\documentclass[aps,prx,nopacs,twocolumn,superscriptaddress,nofootinbib]{revtex4-1}

\usepackage{amsfonts, amsmath, amsthm, amssymb}
\usepackage{bm}
\usepackage{graphicx}
\usepackage{color}
\usepackage[normalem]{ulem}
\usepackage[utf8]{inputenc}
\usepackage[justification=centerlast,singlelinecheck=false, caption=false]{subfig}
\usepackage{hyperref}

\newcommand{\Zt}{\mathbb{Z}_2}
\newcommand{\rr}{\mathbf{r}}
\newcommand{\beq}{\begin{equation}}
\newcommand{\eeq}{\end{equation}}
\newcommand\numberthis{\addtocounter{equation}{1}\tag{\theequation}}

\begin{document}

\title{Quantum phases of two-dimensional $\Zt$ gauge theory \\ coupled to single-component fermion matter}

\author{Umberto Borla}
\affiliation{Physik-Department, Technische Universit\"at M\"unchen, 85748 Garching, Germany}
\affiliation{Munich Center for Quantum Science and Technology (MCQST), 80799 M\"unchen, Germany}
\author{Bhilahari Jeevanesan}
\affiliation{Physik-Department, Technische Universit\"at M\"unchen, 85748 Garching, Germany}
\affiliation{Munich Center for Quantum Science and Technology (MCQST), 80799 M\"unchen, Germany}
\author{Frank Pollmann}
\affiliation{Physik-Department, Technische Universit\"at M\"unchen, 85748 Garching, Germany}
\affiliation{Munich Center for Quantum Science and Technology (MCQST), 80799 M\"unchen, Germany}
\author{Sergej Moroz}
\affiliation{Physik-Department, Technische Universit\"at M\"unchen, 85748 Garching, Germany}
\affiliation{Munich Center for Quantum Science and Technology (MCQST), 80799 M\"unchen, Germany}
\affiliation{Department of Engineering and Physics, Karlstad University, Karlstad, Sweden}

\date{\today}

\begin{abstract}
We investigate the rich quantum phase diagram of Wegner's theory of discrete Ising gauge fields interacting with $U(1)$ symmetric single-component fermion matter hopping on a two-dimensional square lattice. In particular limits the model reduces to (i) pure $\Zt$ even and odd gauge theories, (ii) free fermions in a static background of deconfined $\Zt$ gauge fields, (iii) the kinetic Rokhsar-Kivelson quantum dimer model at a generic dimer filling. We develop a local transformation that maps the lattice gauge theory onto a model of $\Zt$ gauge-invariant spin $1/2$ degrees of freedom. Using the mapping, we perform numerical density matrix renormalization group calculations that corroborate our understanding of the limits identified above. Moreover,  in the absence of the magnetic plaquette term, we reveal signatures of topologically ordered Dirac semimetal and staggered Mott insulator phases at half-filling.  At strong coupling, the lattice gauge theory displays fracton phenomenology with isolated fermions being completely frozen and dimers exhibiting restricted mobility. In that limit, we predict that in the ground state dimers form compact clusters, whose hopping is suppressed exponentially in their size.  We determine the band structure of the smallest clusters numerically using exact diagonalization. The rich phenomenology discussed in this paper can be probed in analog and digital quantum simulators of discrete gauge theories and in Kitaev spin-orbital liquids.
\end{abstract}

\maketitle

\section{\label{sec:intro}Introduction}

\subsection{Background and previous work}

Gauging of symmetries is one of the most successful paradigms of modern physics. Two fundamental theories of the twentieth century, the general theory of relativity \cite{carroll2019spacetime} and the Standard Model of particle interactions \cite{weinberg1995quantum} were conceived by relying on the principle of local gauge invariance. Our faith in gauging is so strong that it is often employed as a guiding principle for frontiers of our knowledge, such as grand-unified theories, string theory and loop quantum gravity. 

Despite all that success, we still do not understand fundamentally why the gauging principle is so useful. Promoting a global symmetry to a local one  is a drastic modification of the original problem: on the one hand the space of states is naively enlarged by the introduction of additional degrees of freedom, but on the other hand a hard gauge constraint, the Gauss law, reduces the space of physical states. Generators of global symmetries and gauge transformations act very differently in the Hilbert space of a quantum system. While the former generically relates different quantum states, the latter is a do-nothing transformation which acts trivially in the physical Hilbert space\footnote{This is true only if we work on a closed space manifold. The situation is more subtle in the presence of a boundary, where the global symmetry that is gauged survives at the edge.}. Gauging just introduces redundancy in our description of Nature. So why is it so inexplicably useful for describing the world around us? Condensed matter physics, where gauge theories emerge in abundance, suggests an answer to this question.
It appears that the gauge description becomes indispensable  to understand deconfined quantum phases of matter which exhibit fractionalization of low-energy excitations, topological order and long-range entanglement \cite{wenbook, Fradkin2013, zeng2019quantum}.

Remarkable progress in our understanding of quantum gauge theories has been made by defining them on a lattice \cite{Wilson1974}. First, the lattice approach provides a unique opportunity to perform strong coupling calculations analytically in a systematic fashion. Moreover, since this method naturally regularizes the field theory under investigation by discretizing the path integral, it is extremely suitable for numerical simulations. In recent years traditional quantum Monte Carlo numerical approaches to lattice gauge theories have been complemented with novel methods such as tensor networks, analog quantum simulators and noisy intermediate-scale quantum computers, for recent reviews see \cite{dalmonte2016lattice, banuls2020review, banuls2019simulating, PhysRevResearch.2.023015}. 

The first example of a lattice gauge theory was discovered by Franz Wegner almost fifty years ago \cite{Wegner1971}. In attempt to generalize the Kramers-Wannier duality to higher dimensional Ising models, he found the most basic lattice gauge model-- the discrete abelian $\Zt$ gauge theory.
The two-dimensional quantum version of the $\Zt$ gauge theory exhibits deconfined and confined regimes  that are separated by a continuous Ising phase transition. It was noticed already by Wegner that the two phases cannot be distinguished by a local Landau order parameter.    It was realized however much later that the deconfined phase exhibits quantum topological order, which manifests itself via a robust four-fold degeneracy of ground states on a torus in the thermodynamic limit.  Being historically the first known example of a model enjoying quantum topological order, the $\Zt$ gauge theory forced a deep paradigm shift in our understanding of possible quantum phases of matter. Remarkably, this model  captures essential low-energy properties of gapped $\Zt$ spin liquids emerging in some strongly-interacting quantum spin models \cite{savary2016quantum}. Although no conclusive experimental realization is known to date \cite{broholm2020quantum}, two-dimensional arrays of Rydberg atoms appear to be a very promising platform to realize highly controllable systems with $\Zt$ topological order \cite{2020arXiv201112295S, 2020arXiv201112310V}.

\begin{figure}
	\includegraphics[width=0.95\linewidth]{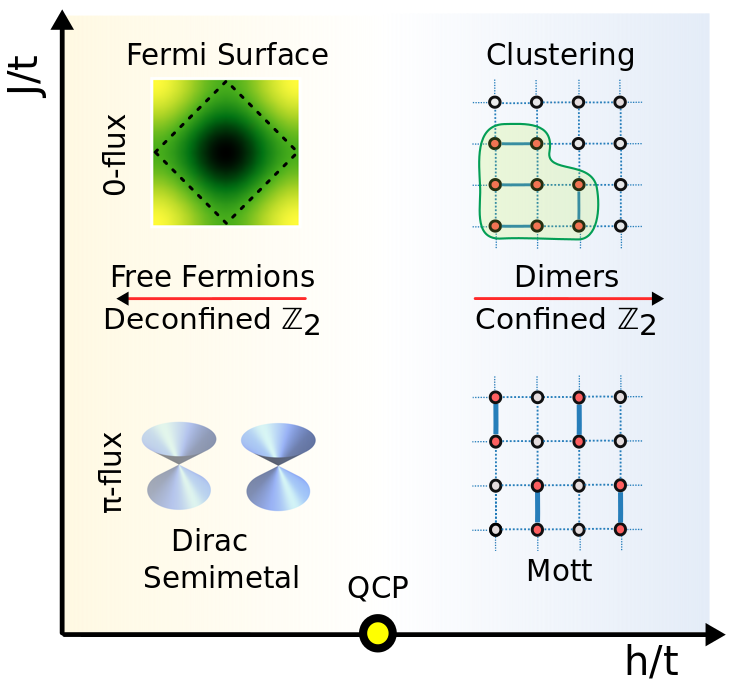}
	\caption{Schematic quantum phase diagram at half-filling: At a weak magnetic coupling compared to the fermionic hopping, $J\ll t$, we find two phases-- the deconfined Dirac semimetal emerging in the $\pi$-flux $\Zt$ gauge field background and the staggered confined Mott insulator. In this paper we present some numerical evidence that the two phases are separated by a single quantum critical point. On the other hand, in the regime $J\gg t$, we find the deconfined Fermi surface and the confined clustered phase. Away from the zero-tension regime $h=0$, the Fermi surface is expected to have a BCS instability towards a p-wave paired superfluid state.}
	\label{fig:hfpt}
\end{figure}
 
What is known about the quantum two-dimensional $\Zt$ gauge theory coupled to dynamical matter fields? In their influential paper \cite{Fradkin1979}, Fradkin and Shenker investigated  Wegner's gauge theory coupled to Ising matter fields. They predicted that the two distinct phases of the pure $\Zt$ gauge theory survive, and that despite apparent quantitative differences, the Higgs and confinement regimes constitute in fact a single phase of matter. Their predictions were confirmed later by quantum Monte Carlo simulations of this model, see for example \cite{tupitsyn2010topological, PhysRevLett.98.070602}.

The fate of the two-dimensional Ising $\Zt$ gauge theory coupled to fermionic matter is now actively studied. The most known example was worked out by Kitaev as a means to solve his celebrated honeycomb model \cite{Kitaev2006}.  An early study of the problem involving $U(1)$ symmetric fermionic matter  was undertaken by Senthil and Fisher in \cite{Senthil2000}, whose motivation stemmed from high-$T_c$ superconductivity. The so-called orthogonal fermions forming a fractionalized non-Fermi liquid \cite{Nandkishore2012} is another example, where the dynamical discrete $\Zt$ gauge fields couple to fermionic (and simultaneously Ising) matter. Substantial progress in our understanding of the quantum phase diagram of a two-component (spinful) Fermi sea interacting with dynamical $\Zt$ gauge fields has been achieved recently due to sign-problem free determinant quantum Monte Carlo studies of Gazit and collaborators \cite{Gazit2017}. 
Contrary to electromagnetism, where equal charges repel, $\Zt$ gauge fields mediate attraction between fermions and trigger \textit{s}-wave fermionic superfluidity, which can be tuned between the weakly-coupled BCS and the strongly-coupled BEC states. In contrast to a conventional two-dimensional \textit{s}-wave superfluid these two regimes are separated by a quantum critical point, where the gauge field undergoes a deconfinement-confinement transition. Moreover, at half-filling a new deconfined phase was discovered where, due to an emergent $\pi$ flux of $\Zt$ gauge fields in the ground state, fermionic matter forms a Dirac semimetal. Surprisingly, Gazit and collaborators identified a single exotic deconfined critical point between the deconfined Dirac phase and the confined BEC phase, instead of a naively anticipated split transition where $U(1)$ global symmetry breaks spontaneously before confinement. This criticality was studied in some detail in \cite{Gazit2018}, and it was argued that it exhibits an emergent global $SO(5)$ symmetry. In addition, in the absence of a confining electric term,  the interpolation between the gauged and ungauged fermionic model discussed above has been studied recently in \cite{konig2019soluble}.  The phase diagram of the Ising gauge theory interacting with gapless fermions, where the Gauss law constraint is not imposed but emerges dynamically, was mapped out in \cite{Grover2016}. Exactly solvable deformations of the two-dimensional Ising gauge theory coupled to fermions were investigated in \cite{Prosko2017, Smith2017}. Spinless fermion matter interacting with dynamical $\Zt$ gauge fields on a cross-linked Creutz-Ising ladder, that  captures some aspects of two-dimensional geometry,  has been studied in \cite{PhysRevX.10.041007}.
The physics of isolated visons in a Fermi sea of $\Zt$-charged anyons was discussed in \cite{2020arXiv201011956P}. Finally, fermionic matter interacting with dynamical $\Zt$ gauge fields has also been actively investigated in one spatial dimension \cite{PhysRevB.84.235148, Frank_2020, borla2020confined, grusdt2020z2, borla2020gauging}.
In summary, recent years enjoyed a surge of theoretical progress towards understanding the physics of gapless $U(1)$ symmetric fermionic matter interacting with dynamical $\Zt$ gauge fields.

A timely motivation to investigate the phase diagram of the Ising gauge theory coupled to $U(1)$ symmetric fermionic matter stems from a remarkable current effort to simulate discrete lattice gauge theories using quantum technologies \cite{banuls2019simulating}. In particular, for the $\Zt$ gauge theory coupled to dynamical matter, a Floquet implementation has been proposed in \cite{Barbiero2018} and proof-of-principle experiments with a two-component mixture of ultracold atoms have been already performed in \cite{Schweizer2019, Gorg2019}. Extension of these experiments to one-dimensional and two-dimensional geometries is expected in the near future. Digital quantum simulations of the Ising gauge theory coupled to fermionic matter were also proposed in \cite{PhysRevLett.118.070501}.

Recently $\Zt$ gauge fields coupled to single-component $U(1)$ fermions hopping on a square lattice have also emerged in deconfined exactly-solvable spin-orbital liquids \cite{PhysRevB.102.201111} that exhibit long-range entanglement of spin and orbital degrees of freedom. In particular, it was shown in \cite{PhysRevB.103.075144} that the simplest so-called $\nu=2$ generalization of the Kitaev spin liquid gives rise to a special case of the $\Zt$ gauge theory studied in our paper, where the electric term that favors confinement is set to zero and the $\Zt$ gauge field is static.

\subsection{Summary of main results}

In this paper we investigate a plethora of quantum phases emerging when two-dimensional single-component (spinless) complex fermions are coupled to Wegner's $\Zt$ gauge theory. To date very little is known about the quantum phase diagram of this basic, but fundamental model. The main reason for that is that in this case the quantum Monte Carlo method pioneered by Gazit and collaborators, which was so successful in the two-component (spinful) case \cite{Gazit2017, Gazit2018, gazit2019fermi}, is plagued with the infamous sign problem. As will become clear in the following,  the model exhibits a rich quantum phase diagram originating from the interplay of gauge and fermionic degrees of freedom. This interplay leads to a broad range of phenomena such as  exotic $p$-wave superfluidity, quantum topological order, restricted mobility, clustering and confinement of quantum matter. 

After introducing the problem in Sec. \ref{sec:model}, we identify in Sec. \ref{sec:limits} four distinct limits, where the model reduces to a simpler and more tractable problem. While the first three limits give rise to well-understood problems, the strong string tension regime exhibits a variety of unexpected features, such a reduced mobility of emergent dimer excitations, the hallmark of the highly-debated fracton phenomenology, and their clustering into large conglomerates. Our particularly surprising analytical finding, derived in Appendix \ref{OneDimerBand},  is the existance of dimer frozen states, which form perfect flat bands. These immobile dimer states emerge from a delicate cancellation of multiple hopping processes and are building blocks of intriguing exact many-body frozen states. 

Contrary to an ordinary fermionic system, the $\Zt$ gauged model studied in this paper has only bosonic gauge-invariant local degrees of freedom in the bulk. In Sec. \ref{sec:spinmap} we eliminate $\Zt$ redundant fermions and introduce a local mapping to rewrite the model in terms of gauge-invariant spin $1/2$ operators. 

Our numerical results are presented in Sec. \ref{sec:Numerics}:
Within the physically-transparent gauge-invariant formulation we perform density matrix renormalization group (DMRG) \cite{PhysRevLett.69.2863, mcculloch2008infinite} calculations. Moreover, we also use exact diagonalization to investigate numerically the physics of the quantum dimer model that emerges in the limit of a strong string tension. First, it is reassuring to observe that all our numerical results fully agree with analytical expectations in the four limiting regimes identified in Sec. \ref{sec:limits}. In particular, in the strong string tension limit in the presence of the magnetic plaquette term, we uncover robust evidence for clustering and determine the band structure of small clusters. Second, we concentrate at the special case of half filling and map out the salient features of the quantum phase diagram that are summarized in Fig. \ref{fig:hfpt}. In particular,  in the absence of the magnetic term, our DMRG simulations in an infinite cylinder geometry unveil signatures of two phases, a topologically ordered Dirac semimetal and a translation symmetry-broken Mott insulator. Using two independent diagnostics- the iDMRG correlation length and the magnetic susceptibility, we reveal strong indications of a direct continuous transition between these two phases on a cylinder of circumference $L_y=4$. It is very surprising that the confinement and spontaneous translation symmetry breaking happen simultaneously. Moreover, the continuous nature of this transition points towards a novel exotic quantum criticality which will be investigated in more detail in a future work.

Finally, in Sec. \ref{sec:conclusions} we draw our conclusions and present some outlook on future work. 

Technical details of our investigation are relegated to appendices.


\section{\label{sec:model}The model}

\begin{figure}
	\includegraphics[width=0.75\linewidth]{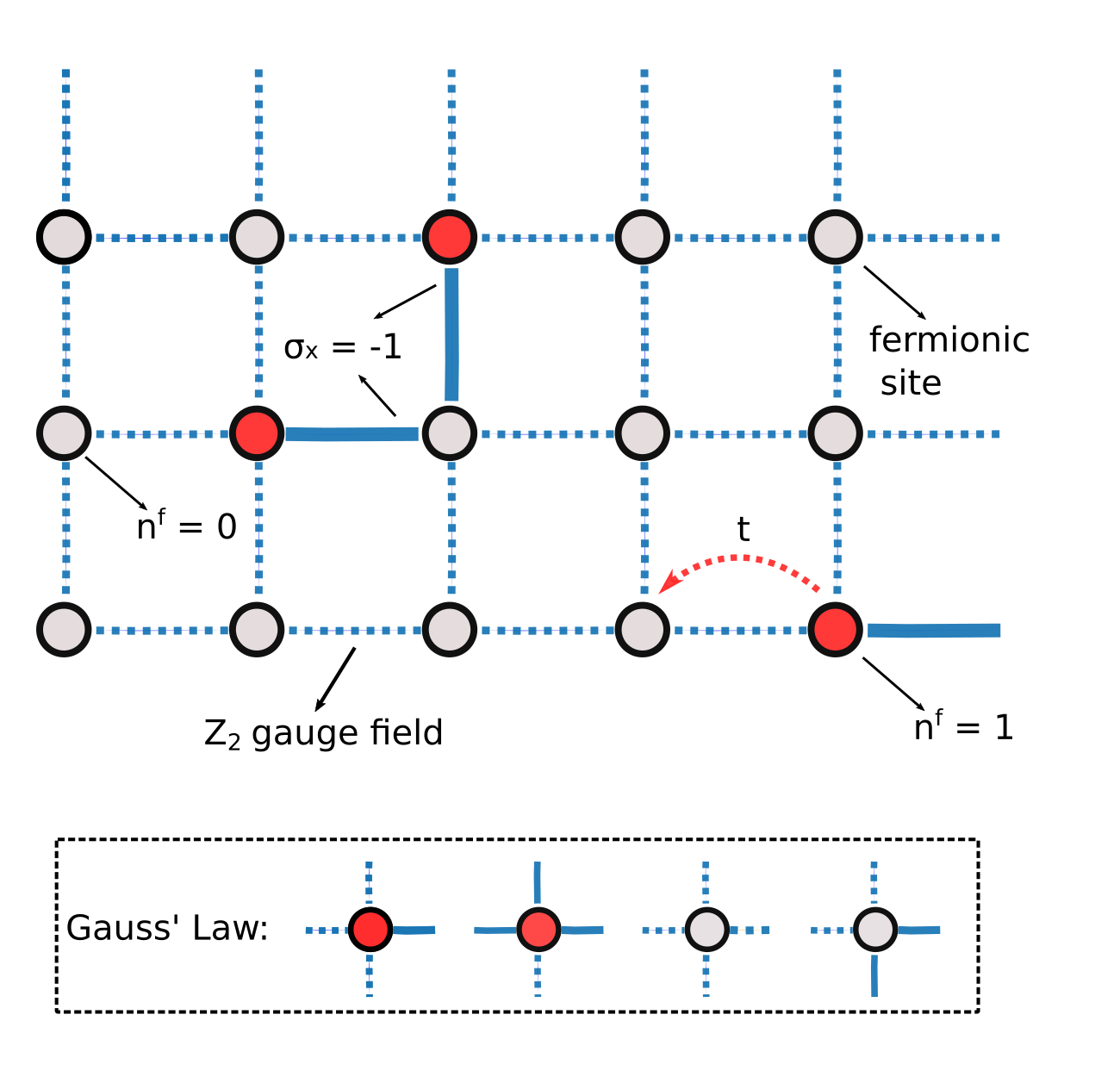}
	\caption{Spinless fermions live on sites, and $\Zt$ gauge fields on links of a two-dimensional square lattice. Occupied sites are denoted in red, empty sites in grey. Dashed and solid blue links correspond to $\sigma^x=1$ and $\sigma^x=-1$ (``electric strings"), respectively.  }
	\label{fig:model}
\end{figure}

In this paper we consider a model of spinless (single-component) fermions hopping on a two-dimensional square lattice coupled to Wegner's $\Zt$ gauge theory. The fermions live on sites of the lattice, while the gauge fields are defined on links as shown in Fig. \ref{fig:model}. The Hamiltonian of the model is given by
\begin{equation} 
H=H_f+H_{\Zt},
\label{eq:H}
\end{equation}
where the fermion Hamiltonian is
 \begin{equation} \label{eq:Hf}
 H_{f} = -t\sum_{ \mathbf{r},\eta} \left( c^\dagger_{\mathbf{r}}\sigma^z_{\mathbf{r},\eta} c_{\mathbf{r}+\eta} + \text{h.c.} \right) 
 -\mu\sum_{\mathbf{r}}c^\dagger_{\mathbf{r}}c_{\mathbf{r}}
 \end{equation} 
and the gauge theory Hamiltonian
\begin{equation} \label{eq:HZ2}
H_{\Zt} = -J \sum_{\mathbf{r}*} \prod_{b\in \square_\mathbf{r}*} \sigma^z_b -h \sum_{\mathbf{r},\eta} \sigma^x_{\mathbf{r},\eta}.
\end{equation} 
Here $\mathbf{r}=(i_x, i_y)$ labels the sites of the lattice and $\eta= \hat x, \hat y$ denotes a unit lattice displacement. In addition, $\square_{\mathbf{r}^*}$ represents a plaquette at position $\mathbf{r}^*$ defined on the dual lattice formed by the centers of the plaquettes. The fermions are minimally coupled to $\Zt$ gauge fields through an appropriate version of the Peierls substitution.

The Hamiltonian \eqref{eq:H} is invariant under a global $U(1)$ symmetry that acts only on the fermionic degrees of freedom as $c_{\rr}\to e^{i\alpha} c_{\rr}$. One can thus introduce the associated $U(1)$ chemical potential $\mu$ to tune the fermionic density $n_{\mathbf{r}}=c^\dagger_{\rr} c_{\rr}$ in the ground state.

The principle underlying a gauge theory is that a global symmetry is promoted to a local one by introducing new degrees of freedom. In this paper, the gauged symmetry is the fermion parity $P=\prod_{\rr} (-1)^{n_{\rr}}$. As a result, the Hamiltonian $H$ is invariant under local $\Zt$ gauge transformations generated by 
\begin{equation}
\label{G}
G_\rr= (-1)^{n_{\rr}} \prod_{b \in +_{\rr}}\sigma^x_b,
\end{equation}
where $\sigma^x$ operators act on the star $+_{\rr}$, i.e., on the links adjacent to the site $\rr$. As a consequence, the Hilbert space separates into a large number of disconnected sectors, determined by the configurations $\left\lbrace G_{\rr} =\pm 1\right\rbrace$.
In the rest of the paper we will work in the sector where $G_{\rr}=1$ for all sites, which corresponds to the gauge theory with no background $\Zt$ charges. In other words, the Hamiltonian \eqref{eq:H} must be diagonalized under the local constraint $G_{\rr}=1$.\footnote{This standard Gauss law is different from the checkerboard Gauss law that emerges dynamically in the spinless case investigated in \cite{Grover2016}.} One important consequence of the gauge constraint is that when defined on a closed surface such as a torus, the fermion number must be necessarily even\footnote{ This follows immediately from taking a product of the generators $G_\rr$ over all sites of the lattice.}, i.e., the physical Hilbert space contains only states with even fermion parity. On the other hand, on open lattices the fate of the global fermion parity $\Zt$ symmetry depends on how the Gauss law is implemented near boundaries. For example, if the lattice terminates everywhere with links, the global fermion parity survives gauging and acts on the boundary links \cite{borla2020gauging}. After taking a product of the Gauss law constraints over all sites, one finds $P=\prod_{b \in \text{edge}} \sigma_b^x$, which is the gauge-invariant t'Hooft loop operator of the $\Zt$ gauge theory traversing the boundary.  As a result, both even and odd fermion numbers are allowed in the physical Hilbert space in that case. 

Better insight into the model can be gained by understanding how it emerges from the more familiar $U(1)$ lattice gauge theory, which describes Maxwell's electrodynamics on a lattice. The elementary building block of the latter gauge theory is the Wilson line operator $e^{iA}$ acting on links, with continuous compact $A\in U(1)$. If one restricts to the $\Zt$ subgroup, $A$ can be either $0$ or $\pi$, and the Wilson line operator $e^{iA}$ can be represented by the Pauli matrix $\sigma^z$ acting in a two-dimensional Hilbert space. The conjugate operator $\sigma_x$ naturally corresponds to $e^{i E}$  in the $U(1)$ gauge theory. The gauge constraint can be interpreted as the $\Zt$ Gauss law for the electric field: the discrete charge at a site equals the number of electric lines going out from that site modulo $2$. Here we defined the electric line as a link state which is the eigenvector of $\sigma^x$ with the eigenvalue $-1$.
The first term in the Hamiltonian \eqref{eq:HZ2} is a $\Zt$ version of the lattice magnetic Hamiltonian, while the second term governs the dynamics of gauge fields and is thus the discrete version of the electric term.  
 The electric term introduces an energy cost for electric lines with $\sigma^x=-1$ and results in an attractive interaction between fermions provided $h>0$. 

The Hamiltonian of interest is invariant under discrete translations and the $D_4$ point group transformations. In addition, the Hamiltonian and the Gauss law are invariant under the following anti-unitary transformation
\beq \label{Tsym}
c_{\rr} \to c_{\rr}, \quad \sigma_{\mathbf{r},\eta}^x\to \sigma_{\mathbf{r},\eta}^x, \quad \sigma_{\mathbf{r},\eta}^y\to -\sigma_{\mathbf{r},\eta}^y, \quad \sigma_{\mathbf{r},\eta}^z\to \sigma_{\mathbf{r},\eta}^z   
\eeq
which realizes time-reversal symmetry in this model.
Next, consider the particle-hole transformation that acts only on the fermions as
$
c_{\rr} \rightarrow (-1)^{{\rr}} c_{\rr}^{\dagger}
$,
where $(-1)^{{\rr}}=(-1)^{i_x+i_y}$. While the hopping term in the Hamiltonian \eqref{eq:Hf} commutes with this transformation, the chemical potential term anti-commutes with it. One thus might naively conclude that the problem has particle-hole symmetry at $\mu=0$. However under this transformation the Gauss law changes sign $G_\rr \rightarrow -G_\rr$ and the even $\Zt$ gauge theory ($G_\rr=1$) transforms onto the odd gauge theory($G_\rr=-1$), where a static $\Zt$ charge occupies every lattice site. As a result, the particle-hole transformation is not a symmetry, but relates the even and odd $\Zt$ gauge theories at chemical potentials $\mu$ and $-\mu$, respectively.

Finally, we notice that the energy spectrum is symmetric under $t\to-t$ since the two cases are related via the unitary transformation generated by $\prod_{\rr, \eta} \sigma_{\rr, \eta}^x$. Similarly, the unitary transformation $\prod_{\rr,\eta} \sigma_{\rr, \eta}^z$ flips the sign of the parameter $h$. As a result, henceforth we will consider only the regime $t, h\ge 0$.

\section{\label{sec:limits}Limiting cases}
In this section we discuss several limiting cases, where the model \eqref{eq:H} reduces to a simpler problem. While the first three limits give rise to a well-understood problems, the strong-tension limit leads to some novel physics of fractonic excitations with restricted mobility and clustering.

\subsection{The limit $\mu\to -\infty$: Even pure $\Zt$ gauge theory} \label{ssec.eZ2}
When the chemical potential is negative and its absolute value much larger than all remaining parameters of the model, there are no fermions in the ground state and the problem \eqref{eq:H} reduces to the even pure $\Zt$ gauge theory defined by the Hamiltonian \eqref{eq:HZ2} with the local Gauss law constraint $\prod_{b \in +_{\rr}}\sigma^x_b=+1$. It is well-known that this theory is dual to the transverse field Ising model (TFIM) and has two distinct phases: the $\Zt$ topologically ordered deconfined phase (that reduces to the toric code at $h=0$) and the featureless confined phase. The two phases are separated by a continuous Ising* quantum phase transition\footnote{In the Ising* theory only local operators that are invariant under the Ising symmetry are allowed.}. Deep in the deconfined phase the excitations are flipped elementary magnetic plaquettes ($\prod_{b\in \square_\mathbf{r}*} \sigma^z_b=-1$), whose energy is $E\sim J$, while deep in the confined regime the excitations are elementary electric loops with $E\sim h$. To excite a pair of $\Zt$ charged fermions, one must invest an energy $E\sim |\mu|\gg J, h$.

\subsection{The limit $\mu\to +\infty$: Odd pure $\Zt$ gauge theory} \label{ssec.oZ2}
If the chemical potential is positive and much larger than all other energy scales, fermions occupy all sites of the lattice in the ground state. Since they cannot hop, they form a static $\Zt$ background which implies that the problem reduces to the odd pure $\Zt$ gauge theory defined by the Hamiltonian \eqref{eq:HZ2} with the local Gauss law constraint $\prod_{b \in +_{\rr}}\sigma^x_b=-1$. This theory naturally arises in the context of quantum dimer models \cite{moessner2001short}.  Adapting the Wegner duality to the odd case, one can show that the model is equivalent to the fully frustrated transverse field Ising model (FFTFIM), where the product of Ising couplings around every elementary plaquette is negative which leads to frustration. Similar to the even pure $\Zt$ gauge theory, there are two quantum phases: the deconfined $\Zt$ topologically ordered phase and the confined phase with no topological order. The physics of these phases, however, is quite different from the even gauge theory \cite{Sachdev_2018}. First,  the topological phase is not featureless, but is symmetry enriched with the $D_8$ non-abelian dihedral symmetry \cite{PhysRevB.84.094419}. Second, the confining phase must break spontaneously lattice translation symmetries and forms a solid. Based on symmetries, one can conclude that the plaquette and columnar translation-broken ordered states are allowed \cite{blankschtein1984fully} with the latter favored by recent quantum Monte Carlo studies \cite{wenzel2012evidence}. Finally, the quantum phase transition separating the two phases exhibits deconfined criticality which in the dual formulation is governed by the XY* conformal field theory \cite{Sachdev_2018}.

\subsection{The limit $h \to 0$: Free fermions in a static $\Zt$ gauge field}
\label{sec:h_0}
The model Hamiltonian \eqref{eq:H} is expressed in terms of gauge non-invariant fermion operators $c_\rr$, which transform as $G_\rr c_\rr G_\rr^{-1}=-c_\rr$. We show here how gauge-invariant fermions can be introduced through a process of string-attachment. Consider the non-local operator \cite{Prosko2017}
\begin{equation}
f_\rr = c_\rr\,S_\rr^z 
\end{equation}
where $S_\rr^z$ is a semi-infinite string of $\sigma^z$ operators that starts at site $\rr$, but is otherwise arbitrary. It is easy to check that the operator $f_\rr$ defined above is gauge invariant, i.e. $G_{\rr'} f_\rr G_{\rr'}^{-1}=f_\rr$ for any choice of sites $\rr$ and $\rr'$ and for any choice of the string $S_\rr^z$. 

The fermionic part of the Hamiltonian can now be rewritten as
\begin{equation} \label{eq:Hff}
 H_{f} = -t\sum_{ \mathbf{r},\eta} \left( f^\dagger_{\mathbf{r}}B_{\mathbf{r},\eta} f_{\mathbf{r}+\eta} + \text{h.c.} \right) 
 -\mu\sum_{\mathbf{r}}f^\dagger_{\mathbf{r}}f_{\mathbf{r}},
 \end{equation} 
where $B_{\mathbf{r},\eta}$ are $\Zt$ gauge-invariant parameters that can have values $\pm 1$\cite{Prosko2017}. One possible convention is to choose horizontal strings that extend to the right of the given fermionic site. In this case one has $B_{\mathbf{r},\eta}=1$ on the horizontal links, while on the vertical links it is possible to express $B_{\mathbf{r},\eta}$ in terms of an infinite horizontal product of magnetic plaquette operators
$B_{\rr, \hat{y}} = \prod_{n\geq 1 }P_{\rr^*+n\hat{x}}\,$.
Importantly, the product of the parameters $B_{\mathbf{r},\eta}$ around a plaquette is the same as the the product of the gauge field $\sigma^z$ around the same plaquette
$\prod_{b \in \square_{\rr^*}}B_b=\prod_{b \in \square_{\rr^*}}{\sigma^z_b}$ and thus the gauge theory Hamiltonian \eqref{eq:HZ2} at $h=0$ reduces to
\beq
H_{\Zt} = -J \sum_{\mathbf{r}*} \prod_{b \in \square_{\rr^*}}B_b.
\eeq

We have therefore rewritten our problem in terms of free $\Zt$ gauge-invariant fermions $f_\rr$ in a background configuration of the fields $B_{\rr, \eta}$, which determine a certain pattern of static $\Zt$ fluxes\footnote{It is worth noting that the attachment of a string $S_\rr^z$ that defines non-local gauge invariant fermions fixes the $\Zt$ gauge redundancy. However, the arbitrariness in the form of this string leads to a new form of redundancy, under which both the  fermions $f_\rr$ and the parameters $B_{\rr,\eta}$ transform non-trivially. 
}.
The ground state of this problem can be found by solving the free fermion problem in different flux backgrounds and choosing the one which minimizes the Hamiltonian $H_f+H_{\Zt}$ \cite{Prosko2017, konig2019soluble}. The half filling case $\mu=0$ is particularly interesting: Lieb's theorem predicts that at $J=0$ a $\pi$-flux configuration is energetically favorable \cite{Lieb94}. In this case, the free-fermion band structure exhibits two Dirac cones at $(k_x,k_y)=(\pm \pi/2,\pi/2)$\footnote{We note that even away from half-filling a $\pi$-flux phase can always be obtained by choosing $J$ to be negative and sufficiently large.}. Any state with a different flux configuration has an energy gap that does not vanish even in the thermodynamic limit, and therefore plaquette excitations always cost a finite amount of energy. In summary, in this phase, $\Zt$ charged fermions form gapless deconfined Dirac excitations, while the $\Zt$ gauge fields are in the topologically ordered phase. 

On the other hand, for $J\gg t$  a zero-flux state is preferred, since every flipped plaquette comes with a large energetic penalty. The transition from a $\pi$-flux phase at small $J$ to a zero-flux phase at large $J$ can happen either sharply or through a series of intermediate configurations. Since each plaquette operator commutes with the Hamiltonian in this regime and thus can only take the values $\pm 1$, such intermediate configurations must necessarily break translational invariance. In this case, the average flux over the extended unit cell takes fractional values (in units of $\pi$), as illustrated in Fig \ref{fig:stripes}. In Appendix \ref{app:flux} we investigate the transition between the $\pi$-flux and zero-flux limits on a thin cylinder with circumference $L_y=2$.  

\begin{figure}
\includegraphics[width=0.9\linewidth]{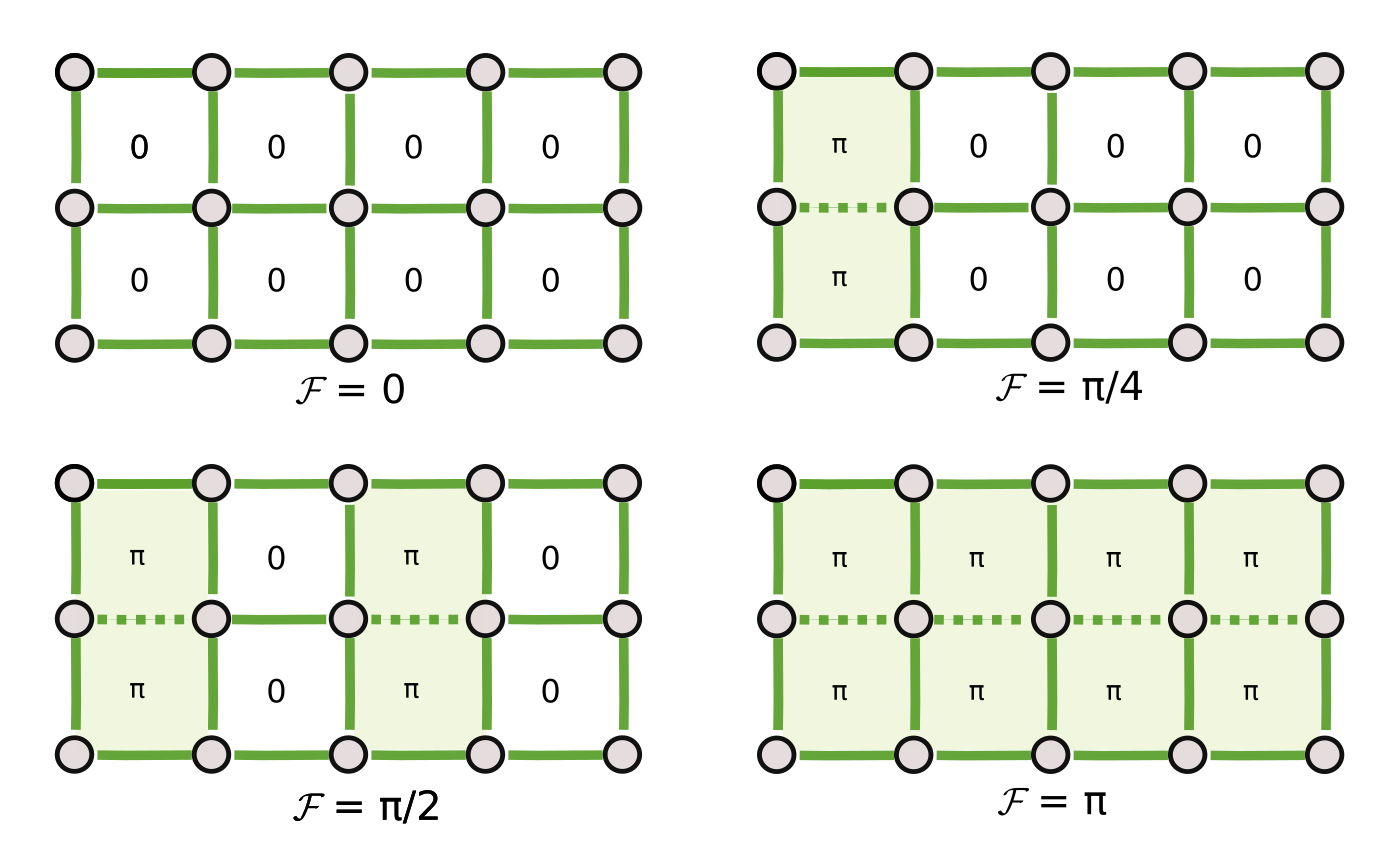}
\caption{Example of hopping configurations that realize vertical stripes of $\pi$-flux plaquettes. The dashed lines denote links with $B_{ij}=-1$ and $\mathcal{F}$ represents the average $\Zt$ flux.}
\label{fig:stripes}
\end{figure}

Finally, we notice that at $h=0$ the model Hamiltonian \eqref{eq:H} commutes with gauge-invariant Wilson loop operators $W=\prod_{b\in \text{loop}} \sigma^z_b$ acting on links forming a closed loop\footnote{On an open lattice terminating with links, gauge-invariant Wilson lines ending on the boundary also commute with the model Hamiltonian.}. As a result, the model enjoys an additional global symmetry. Since this symmetry is generated by operators acting on co-dimension one manifolds, this is a one-form symmetry which is usually referred to as magnetic. Gauged (fermion) parity in a system  with a symmetric ground state under the magnetic (one-form) symmetry implies  symmetry-protected topological (SPT) order \cite{borla2020gauging, Higgs}. On the other hand, spontaneous symmetry breaking of the magnetic one-form symmetry leads to topological order \cite{PhysRevB.99.205139}.

\subsection{The limit $h \to \infty$: Resonating quantum dimers and clustering}
\label{sec:h_inty}
If the coupling $h$ is large, the electric strings become energetically expensive and isolated fermions become immobile \cite{PhysRevResearch.2.013094}. At low energies they must form meson-like dimer states, where pairs of fermions are connected by a unit-length electric string. The dimers can be defined on links of the lattice and are created by the gauge-invariant operators $c^\dagger_{\mathbf{r}}\sigma^z_{\mathbf{r},\eta} c^\dagger_{\mathbf{r}+\eta}$.  Due to the Pauli exclusion principle, any two links that share a site cannot simultaneously host dimers. The dimer limit can be systematically constructed by starting from the classical Hamiltonian
\beq \label{H0d}
\begin{split}
H_0&=-h \sum_{\mathbf{r},\eta} \sigma^x_{\mathbf{r},\eta}-\mu\sum_{\mathbf{r}}c^\dagger_{\mathbf{r}}c_{\mathbf{r}} \\
&=-h \sum_{\mathbf{r},\eta} \sigma^x_{\mathbf{r},\eta}-\frac{\mu}{2}\sum_{\mathbf{r}}\big( 1 -\prod_{b \in +_{\rr}}\sigma^x_b\big), \\
\end{split}
\eeq
where in the second line the Gauss law \eqref{G} was used.
By tuning the chemical potential $\mu\sim h$ one can induce a finite density of dimers in the ground state. All the states with a fixed number of dimers have the same energy, and therefore the spectrum of $H_0$ is highly degenerate. For a fixed number of fermions the first excited states contain one meson of length two. This is separated from the ground state by the gap $2h$.

We will treat now the remaining terms in the Hamiltonian \eqref{eq:H} as small perturbations of the Hamiltonian \eqref{H0d}. While individual fermions cannot move at any order of perturbation theory and thus have fractonic character, dimers acquire dynamics and interact with each other. 

\begin{figure}[th]
	\includegraphics[width=0.8\linewidth]{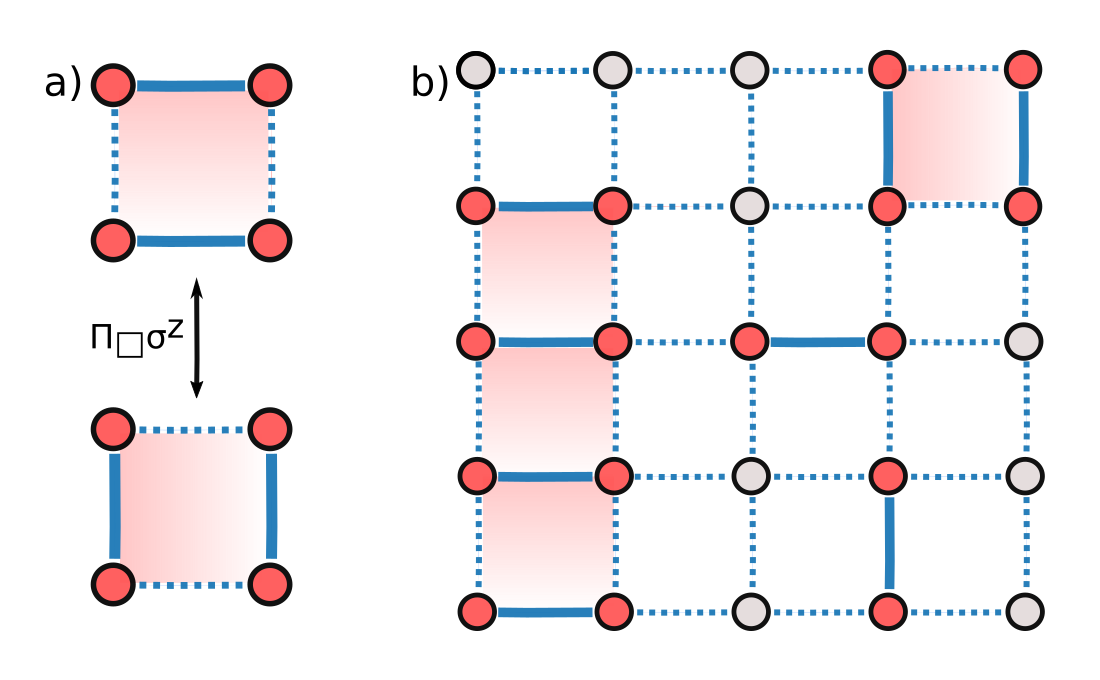}
	\caption{(a) Electric string transitions induced by the magnetic term on a plaquette that is fully occupied with fermions. (b) An example of quantum dimer configuration at partial filling. Highlighted in red are plaquettes fully filled with fermions, where the purely kinetic Rokhsar-Kivelson Hamiltonian resonates electric strings. On the other hand, isolated dimers are not affected by the magnetic term.}
\label{fig:res}
\end{figure}

 At first order in degenerate perturbation theory, only the plaquette term $-J \sum_{\mathbf{r}*} \prod_{b\in \square_\mathbf{r}*} \sigma^z_b$ contributes. On every elementary plaquette fully occupied with fermions, it  induces transitions between two electric string configurations illustrated in Fig. \ref{fig:res}a. Generically this results in a short-range attractive interaction of strength $J$ between the dimers. At full filling, achieved at $\mu\gg h$, the problem reduces to the close-packed, purely kinetic Rokhsar-Kivelson quantum dimer model \cite{PhysRevLett.61.2376}
\begin{eqnarray} 
H_d & =-J\sum\left(|\vcenter{\hbox{\includegraphics[height=0.02\textheight]{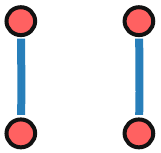}}}\rangle\langle\vcenter{\hbox{\includegraphics[height=0.02\textheight]{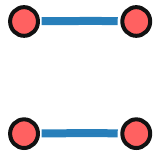}}}|+\text{h.c.}\right).
\end{eqnarray}
The ground state of this Hamiltonian is deeply in the confined regime of the odd $\Zt$ gauge theory described in section \ref{ssec.oZ2}.
At partial filling of fermions, the first-order effective Hamiltonian only resonates electric strings on islands of plaquettes that are fully occupied by fermions, but does not act on isolated dimers, see Fig \ref{fig:res} b. For this reason, configurations where fermions are grouped in clusters are energetically favored.

The fermion hopping term in the Hamiltonian \eqref{eq:H} starts to contribute only at second order of the degenerate perturbation theory. It generates anisotropic short-range dimer hopping processes. At this order of perturbation theory, dimers cannot hop in the direction perpendicular to the electric string and thus exhibit restricted mobility. Contrary to dipoles in some theories of fractons \cite{fractons, pretko2020fracton}, where restricted mobility follows from symmetries, here it is imposed energetically. In addition, second order processes give rise to short-range repulsive interactions between dimers. The energy scale of both these effects is of order $t^2/h$ and thus generically is much smaller than the first-order attraction of strength $J$, discussed above.

In summary, at finite fermion filling for $J\ne 0$ one expects the lowest energy states to be bound clusters of fermions that fully occupy as many plaquettes as possible. In Appendix \ref{app:ClusterProof} we provide a proof of this clustering phenomenon. These clusters are very heavy because they move in one piece by hopping all the dimers in the cluster. For a cluster composed of $n_d$ dimers, this process is of the order $t^{2 n_d}$ in the perturbation theory which leads to an extremely small energy-level splitting.
On the other hand, in the absence of the plaquette term, we expect the relevant degrees of freedom to be hard-core dimers with short-range repulsive interactions, which is  qualitatively similar to the one-dimensional version of the theory that was investigated in \cite{borla2020confined, PhysRevLett.124.207602, PhysRevResearch.2.013094}. We study the problem of dimers at partial filling in some detail both at $J=0$ and $J\ne 0$ in Sec. \ref{sec:Numerics}. 


\section{Mapping to gauge-invariant spin model} \label{sec:spinmap}
The Hamiltonian \eqref{eq:H} is expressed in terms of gauge non-invariant redundant degrees of freedom. 
In this section we demonstrate how the model can be mapped onto a spin $1/2$ model through a local transformation. The advantage of such rewriting is that the new spin degrees of freedom are $\mathbb{Z}_2$ gauge-invariant and thus act directly within the physical Hilbert space whose dimension scales with the number of lattice links $N_l$ as $2^{N_l}$. As a result, there is no  $\mathbb{Z}_2$ gauge redundancy in the spin formulation.

First, we introduce the Majorana operators 
\begin{equation}
\gamma_\rr = c^{\dagger}_\rr+c_\rr^{\vphantom{\dagger}},\,\,\,\,\,\,\,\,\,\,\,\,
\tilde{\gamma}_\rr = i(c^{\dagger}_\rr-c_\rr^{\vphantom{\dagger}}).
\end{equation}
In terms of these variables the Gauss law reads 
\begin{equation}
i \tilde{\gamma}_\rr\gamma_\rr=\prod_{b \in +_{\rr}}\sigma^x_b.
\label{eq:gauss_majorana}
\end{equation}
Out of the original $\mathbb{Z}_2$ gauge fields and Majorana variables, it is possible to construct new gauge-invariant Pauli operators as follows
\begin{align*}
X_{\rr,\eta}&=\sigma^x_{\rr,\eta},\\
Z_{\rr,\hat{x}} &= -i\tilde{\gamma}_\rr \sigma^z_{\rr,\hat{x}}\gamma_{\rr+\hat{x}}\sigma^x_{\rr+\hat{x},-\hat{y}},\\ \numberthis \label{eq:mapping}
Z_{\rr,\hat{y}} &= -i\tilde{\gamma}_\rr \sigma^z_{\rr,\hat{y}}\gamma_{\rr+\hat{y}}\sigma^x_{\rr,\hat{x}}.
\end{align*} 

Note that the factors  $\sigma^x$ in the definition of $Z_{\rr,\eta}$ are needed to ensure that the new spin operators not only satisfy the Pauli algebra on a given link, but also always commute on neighboring links. We also note that this choice is not unique: different arrangements of $\sigma^x$ are possible, including non-symmetric ones where two factors of $\sigma^x$ appear in the definition of $Z_{\rr,\hat{x}}$, but no $\sigma^x$ is needed for $Z_{\rr,\hat{y}}$ (or vice-versa). 
It is worth mentioning that such kind of mapping is possible in any dimension, generically requiring some factors of $\sigma^x$ in the definition of $Z_{\rr,\eta}$  on properly chosen neighboring links. The only exception is one spatial dimension, where no extra operators $\sigma^x$ are needed \cite{PhysRevB.87.041105, radicevic2018spin, borla2020confined, borla2020gauging}.

In the following we will present how different terms of the Hamiltonian \eqref{eq:H} transform under the mapping \eqref{eq:mapping}. In the end, we will collect everything together and express the full Hamiltonian in the gauge-invariant spin formulation. 

\begin{figure}[t]
	\includegraphics[width=0.8\linewidth]{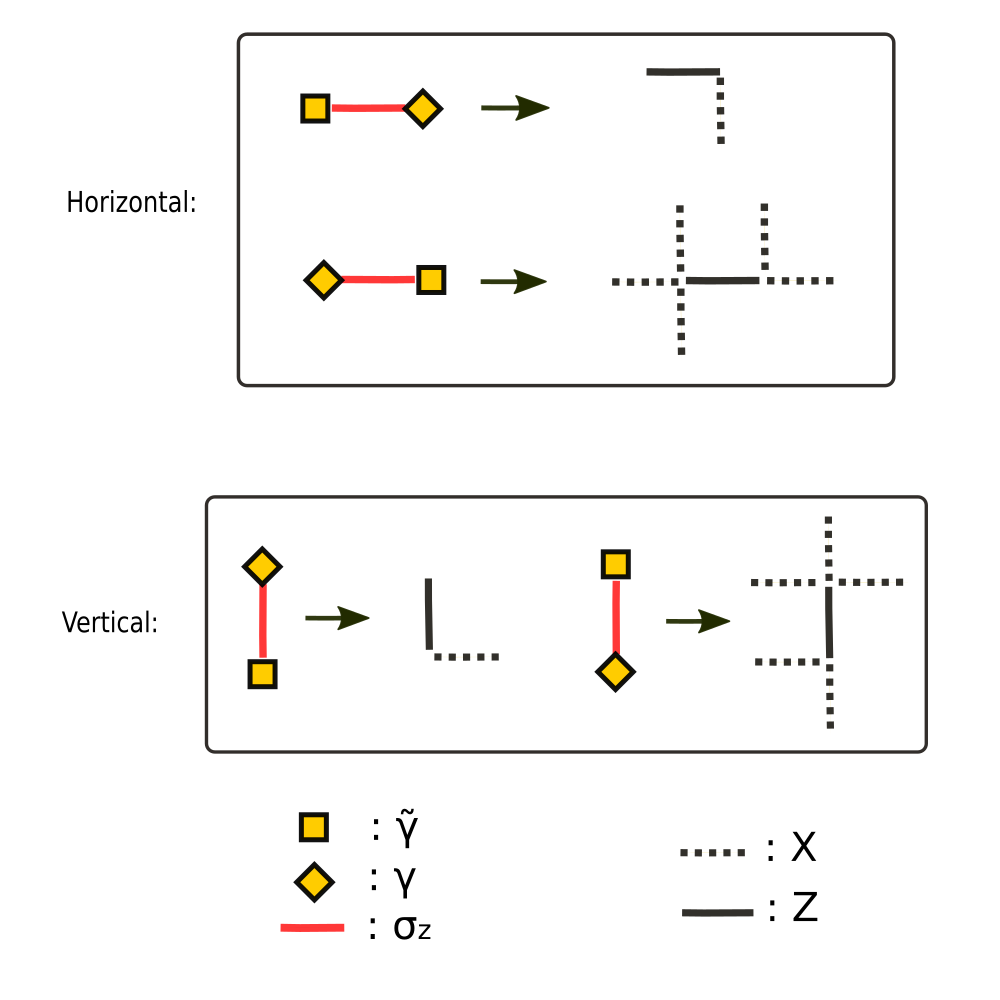}
	\caption{Illustration of the hopping terms under the mapping \eqref{eq:mapping} for horizontal and vertical links.}
\label{fig:hopping_mapping}
\end{figure}
We start with the hopping term, which can be readily expressed using the Majorana operators 
\begin{align*}
&-\left(c^\dagger_{\mathbf{r}}\sigma^z_{\mathbf{r},\eta} c_{\mathbf{r}+\eta} + \text{h.c.}\right)=\\
\numberthis \label{eq:hopping_m}
&\frac{1}{2}\left(i \tilde{\gamma}_{\rr} \sigma^z_{\rr,\eta} \gamma_{\rr+\eta}-i \gamma_{\rr} \sigma^z_{\rr,\eta} \tilde{\gamma}_{\rr+\eta} \right).
\end{align*}
Under the change of variables \eqref{eq:mapping}, the first term in Eq. \eqref{eq:hopping_m} can be immediately rewritten as
\begin{align*}
i \tilde{\gamma}_{\rr} \sigma^z_{\rr,\hat{x}} \gamma_{\rr+\hat{x}}& =  -Z_{\rr,\hat{x}}\,X_{\rr+\hat{x},-\hat{y}}, \numberthis \label{eq:hopping_1}\\
i \tilde{\gamma}_{\rr} \sigma^z_{\rr,\hat{y}} \gamma_{\rr+\hat{y}}& =  -Z_{\rr,\hat{y}}\,X_{\rr,\hat{x}}.
\end{align*}
In the second term in Eq. \eqref{eq:hopping_m}, however, the Majorana operators appear in the reversed order, and so a straightforward mapping is not possible. To remedy this we notice that since all operators are understood to act only on states in the physical Hilbert space, the identity can be inserted on the right of each expression in the form $\mathsf{1}=G_{\rr_1}\dots G_{\rr_k}$. This can be done for an arbitrary choice of sites $\rr_1, \dots \rr_k$ because the Gauss law is enforced on every site independently. Hence on horizontal links one has
\begin{align*}
i \gamma_{\rr} \sigma^z_{\rr,\hat{x}} \tilde{\gamma}_{\rr+\hat{x}}&=i \gamma_{\rr} \sigma^z_{\rr,\hat{x}} \tilde{\gamma}_{\rr+\hat{x}}G_\rr G_{\rr+\hat{x}}  \\
&= i \tilde{\gamma}_{\rr} \sigma^z_{\rr,\hat{x}} \gamma_{\rr+\hat{x}}\prod_{b \in +_{\rr}}X_{b} \prod_{b' \in +_{\rr+\hat{x}}}X_{b'} \\
&=-Z_{\rr,\hat{x}}\prod_{\mu\in \{l\}}X_{\rr,\mu}, \numberthis \label{eq:hopping_2}
\end{align*}
where the last product is over a set of five $X$ operators on the links determined by the displacements $\{-\hat{x},2\hat{x},\hat{y},-\hat{y},\hat{x}+\hat{y}\}$. Similarly, one can rewrite the vertical hopping in terms of the new spin variables only. Both cases are illustrated in Fig. \ref{fig:hopping_mapping}. Combining both contributions, the hopping term \eqref{eq:hopping_m} on horizontal links can be rewritten in the form
\beq \label{eq:projector_hopping_h}
-\underbrace{\frac{1}{2} \left( 1-\prod_{b \in +_{\rr}}X_{b} \prod_{b' \in +_{\rr+\hat{x}}}X_{b'}\right) }_{\mathcal{P}_{\rr,\hat{x}}} Z_{\rr,\hat{x}}\,X_{\rr+\hat{x},-\hat{y}},
\eeq
and analogously for the vertical hopping we find
\beq \label{eq:projector_hopping_v}
-\underbrace{\frac{1}{2} \left( 1-\prod_{b \in +_{\rr}}X_{b} \prod_{b' \in +_{\rr+\hat{y}}}X_{b'}\right)}_{\mathcal{P}_{\rr,\hat{y}}} Z_{\rr,\hat{y}}\,X_{\rr,\hat{x}}. 
\eeq
This compact form has a nice interpretation in terms of the original fermionic theory: due to the Gauss law the factor $\mathcal{P}_{\rr,\hat{\eta}}$ is a projector that annihilates all states with equal fermion parity on sites $\rr$ and $\rr+\hat{\eta}$. This means that the hopping is only possible if one of the sites hosts a fermion particle and the other one does not, as it should be to avoid double occupancy or creation of pairs of particles out of the vacuum. A generic linear combination of the terms \eqref{eq:hopping_1} and \eqref{eq:hopping_2} does not have this property, and would correspond to a $U(1)$ non-conserving fermionic model with anomalous terms of the form $c^\dagger_{\mathbf{r}}\sigma^z_{\mathbf{r},\eta} c^{\dagger}_{\mathbf{r}+\eta} + \text{h.c.}$
Physically, in the spin formulation the operator $Z$ is responsible for the hopping process, as it swaps the fermion parities on the two neighboring sites. The operators $X$, on the other hand, keep track of the fact that the hopping particles are fermions. 
This can be seen explicitly by exchanging two identical particles and verifying that in the process the statistical phase of $\pi$ is acquired, see Appendix \ref{appa} for details. 

Inclusion of the chemical potential term is easy.
Indeed, the Gauss law yields immediately 
\begin{equation}
n_{\rr} = \frac{1-\prod_{b\in +_\mathbf{r}} X}{2},
\end{equation} 
and so up to a constant the chemical potential term maps onto the star operator.

\begin{figure}[t]
	\includegraphics[width=0.9\linewidth]{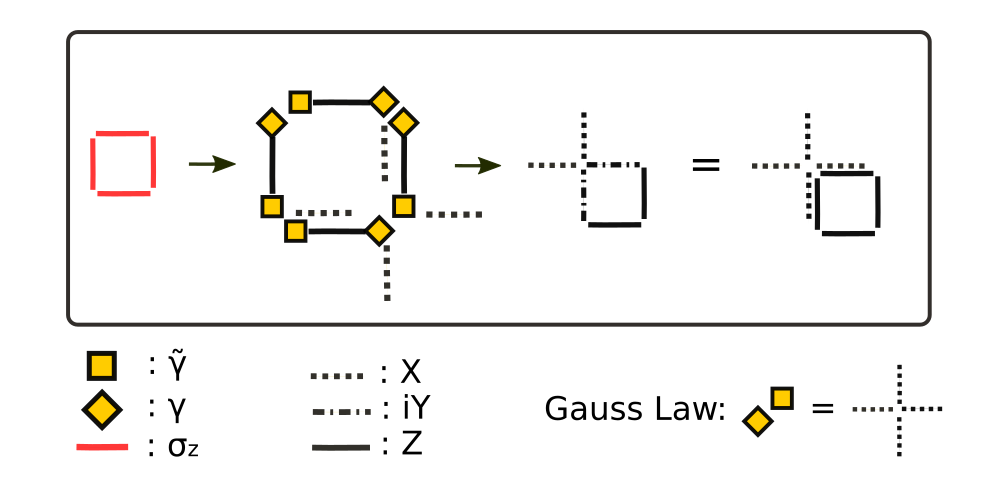}
	\caption{Elementary plaquette $P_{\rr^*}=\prod_{b\in \square_\mathbf{r}*} \sigma^z_b$ under the mapping \eqref{eq:mapping}.  When two $\gamma$ ($\tilde{\gamma}$) appear on the same site, they square to one. On the other hand, when $\gamma$ and $\tilde{\gamma}$ appear, instead, they can be replaced with a star operator as a consequence of the Gauss law.}
\label{fig:plaq_mapping}
\end{figure}

With the help of the gauge constraint the mapping can also be applied to the plaquette term of the Hamiltonian. Each $\sigma_z$ transforms into a combination of $Z$, $X$ and Majorana operators. When taking the product of $\sigma_z$ operators around an elementary plaquette, the Majorana operators on two out of the four vertices square to one, while on the remaining vertices one is left with the product of the Majorana operators $\gamma$ and $\tilde{\gamma}$. Such a product can be replaced with a star operator as a consequence of the Gauss law \eqref{eq:gauss_majorana}, leading eventually to the six-spins term 
\begin{equation}
P_{\rr^*}=\prod_{b\in \square_\mathbf{r}*} \sigma^z_b=- Z_{\rr,\hat{x}}Z_{\rr+\hat{x},\hat{y}}Y_{\rr,\hat{y}}Y_{\rr+\hat{y},\hat{x}}X_{\rr+\hat{y},-\hat{x}}X_{\rr+\hat{y},\hat{y}}
\label{eq:plaq_mapping}
\end{equation}
as shown in Fig. \ref{fig:plaq_mapping}.
This can also be written as
\begin{equation}
P_{\rr^*}=\prod_{b\in \square_\mathbf{r}*} Z \prod_{b\in +_\mathbf{r}} X,
\label{eq:plaq_mapping_2}
\end{equation}
where here $\rr^*$ labels the square of the dual lattice on the bottom-right of the vertex $\rr$. Interestingly, the elementary plaquette operator $P_{\rr^*}$ of the original model \eqref{eq:H} maps onto the plaquette-star composite operator in the spin formulation.
\newline

Finally, the electric term of the Ising gauge theory maps trivially because
$\sigma^x_{\rr,\eta}=X_{\rr,\eta}$.

Combining everything together, the full Hamiltonian \eqref{eq:H} of the $\mathbb{Z}_2$ gauge theory coupled to spinless fermions maps onto the following local spin $1/2$ model
\begin{align*}
H=&-t \sum_{\rr} \left( Z_{\rr,\hat{x}}\,X_{\rr+\hat{x},-\hat{y}}\,\,\mathcal{P}_{\rr,\hat{x}}+Z_{\rr,\hat{y}}\,X_{\rr,\hat{x}}\,\,\mathcal{P}_{\rr,\hat{y}}\right) \\
& -\frac{\mu}{2}\sum_{\rr}\left(1-\prod_{b\in +_\mathbf{r}}X_{b} \right) \\
&-  J \sum_{\rr^*}\prod_{b\in \square_\mathbf{r}*} Z \prod_{b\in +_\mathbf{r}} X -h \sum_{\rr,\eta}X_{\rr,\eta} \numberthis \label{eq:mapped_H} 
\end{align*}
with the projectors $\mathcal{P}_{\rr,\hat{x}}$, $\mathcal{P}_{\rr,\hat{y}}$ defined in Eqs. \eqref{eq:projector_hopping_h} and \eqref{eq:projector_hopping_v}. The time-reversal symmetry \eqref{Tsym} is realized as complex conjugation
\beq
X_{\rr,\eta} \to X_{\rr,\eta}, \qquad Y_{\rr,\eta} \to -Y_{\rr,\eta}, \qquad Z_{\rr,\eta} \to Z_{\rr,\eta}.
\eeq Notice moreover that in the spin
formulation the $U(1)$ particle number symmetry is not onsite. 

Before moving on, we compare the mapping introduced in this section with closely related two-dimensional bosonization mappings introduced in \cite{wosiek1982local, chen2018exact, bochniak2020model, PhysRevResearch.2.023353, 2020arXiv201011956P, rao2020theory}. In the latter case, one starts with a two-dimensional (ungauged) fermionic problem and maps it onto a $\mathbb{Z}_2$-gauged spin 1/2 model. Here on the other hand, our starting point is the gauged fermionic theory \eqref{eq:H} which maps to the unconstrained spin model \eqref{eq:mapped_H}. Similar mapping to the one developed here appeared recently in \cite {shirley2020fractonic}. Related ideas were also introduced in \cite{PhysRevB.98.075119} in order to trade fermionic matter for hard-core bosons in gauge theories which contain $\Zt$ as a normal subgroup.
 
In Appendix \ref{fermspin} we describe how to express the gauge-invariant fermionic correlation function $\langle c^\dagger_{\rr} \prod_{b \in \textit{l}} \sigma^z_b c_{\rr'} \rangle$ in the spin formulation.

Finally, consider the local gauge-invariant operators $b^\dagger_{\mathbf{r}, \eta}= c^\dagger_{\mathbf{r}} \sigma^z_{\mathbf{r},\eta} c^\dagger_{\mathbf{r}+\eta}$ that create dimer states localized at two neighboring sites. In the thermodynamic limit, the expectation values of these operators in the ground state serve as paired superfluid order parameters. Moreover, in the limit $h\to\infty$ such dimers become physical low-energy degrees of freedom, see Sec. \ref{sec:h_inty}. In Appendix \ref{app:dimers} we show that in the gauge-invariant spin formulation the dimer creation operators can be expressed as
\beq
\begin{split}
b^\dagger_{\mathbf{r}, \hat x}&=Z_{\rr,\hat{x}}\,X_{\rr+\hat{x},-\hat{y}}  \Pi_{\rr,\hat{x}}, \\
b^\dagger_{\mathbf{r}, \hat y}&=Z_{\rr,\hat{y}}\,X_{\rr,\hat{x}}  \Pi_{\rr,\hat{y}},
\end{split}
\eeq
where 
\beq
\Pi_{\rr,\eta}=\frac 1 4 \big(1+\prod_{b \in +_{\rr}}X_{b} + \prod_{b' \in +_{\rr+\eta}}X_{b'}+\prod_{b \in +_{\rr}}X_{b} \prod_{b' \in +_{\rr+\eta}}X_{b'} \big)
\eeq
is a projector operator on simultaneously unoccupied sites $\rr$ and $\rr+\eta$.


\section{\label{sec:Numerics} Numerical results}
\subsection{DMRG}
Techniques based on matrix product states (MPS) have been successfully employed to obtain ground states of two-dimensional many-body Hamiltonians for finite geometries. In particular, any two-dimensional system that is finite (or periodic) in one direction (that we will label $y$), i.e. a ladder or a cylinder, can be mapped to an equivalent one-dimensional system with long-range couplings by choosing an appropriate ordering of the physical sites. Such problems can then be tackled using the DMRG algorithm or its infinite version iDMRG (where the thermodynamic limit is taken in the $x$ direction only) that further reduces the computational cost. The limitations are nonetheless severe: even for gapped Hamiltonians, for which DMRG is most efficient, the bond dimension $\chi$ needed to reach a given accuracy scales exponentially with the circumference of the cylinder $L_y$. At finite fermion density, moreover, the studied lattice gauge theory is expected to exhibit a number of exotic gapless phases. This limits the viability of DMRG to even smaller cylinders, and the physical properties of the system are therefore heavily affected by finite-size effects. As a consequence, a reliable study of the thermodynamic limit is not possible for these phases within this method. Nonetheless several interesting features can be appreciated already for small circumferences $L_y$. Which of these features survive in the two-dimensional thermodynamic limit is an open question that will be the subject of future investigations.

In this work we use the tensor network Python library TenPy \cite{hauschild2018efficient} to perform DMRG simulations in infinite and finite cylinder geometries.
We emphasize that the mapping described in Sec. \ref{sec:spinmap} is crucial to perform  numerical simulations of the Hamiltonian \eqref{eq:H} efficiently. The elimination of all gauge redundancy allows us to operate directly in a physical Hilbert space, which guarantees a substantial performance improvement.

\subsubsection{Scans of the $h$-$\mu$ plane: entanglement entropy and phase boundaries.}
The mutual entanglement entropy $S$ under a bipartition of a gapped system obeys the area law, meaning that it is proportional to the area of the cut that divides the system in two parts. For a bipartition of an infinite cylinder, this implies that $S$ scales linearly with the circumference of the cylinder.  This is to be contrasted with gapless systems, for which the entanglement entropy diverges in the thermodynamic limit due to quantum correlations over arbitrarily long length scales. Therefore, peaks in the entanglement entropy reliably detect boundaries between gapped phases. When using iDMRG a correlation length $\xi$ is determined by the second largest eigenvalue of the MPS transfer matrix, and can also be used to detect gapless critical points. 

\begin{figure*}[ht]
	\includegraphics[width=0.98\linewidth]{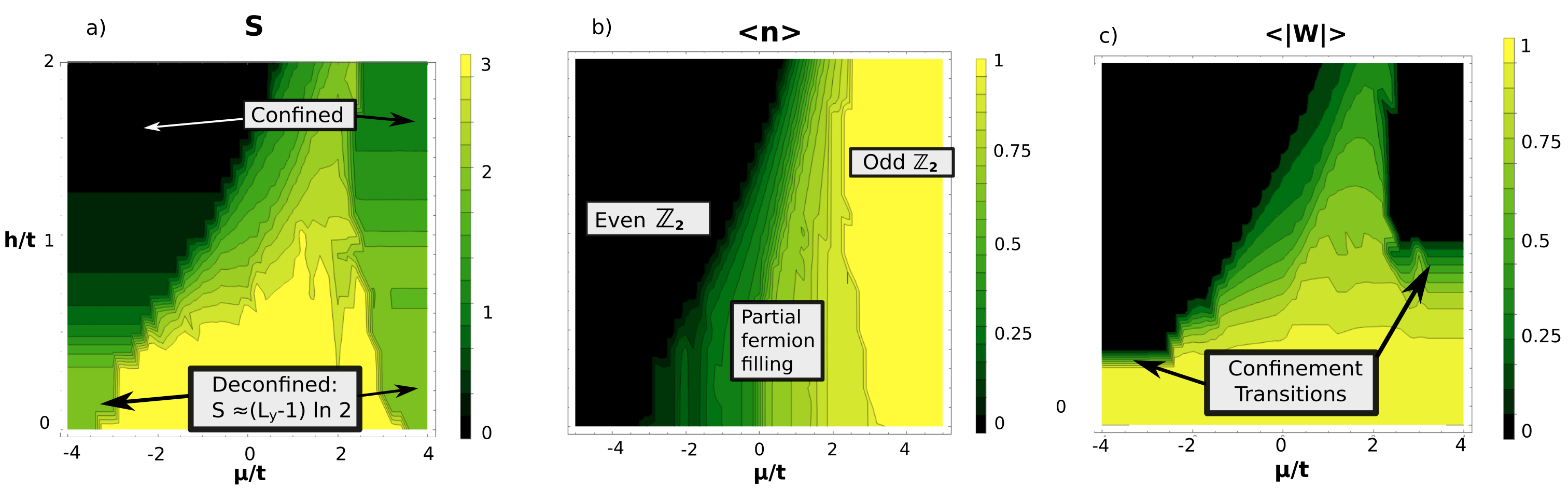}
	\caption{iDMRG scans of the $\mu$-$h$ plane at $J/t=1$ for a cylinder of circumference $L_y=4$: We show (a) the entanglement entropy $S$, (b) the average fermion occupation and (c) the expectation value of the non-contractible Wegner-Wilson loop that winds around the cylinder. The empty ( $\Zt$ even theory) and fully filled ($\Zt$ odd theory) regimes are separated by a partially filled region that has a form of a wedge centered around $h=\mu$. The bond dimension $\chi$ was set to $300$.
	}
\label{fig:contourplots}
\end{figure*} 

\begin{figure}
\centering
\subfloat[]{\includegraphics[width=0.48\linewidth]{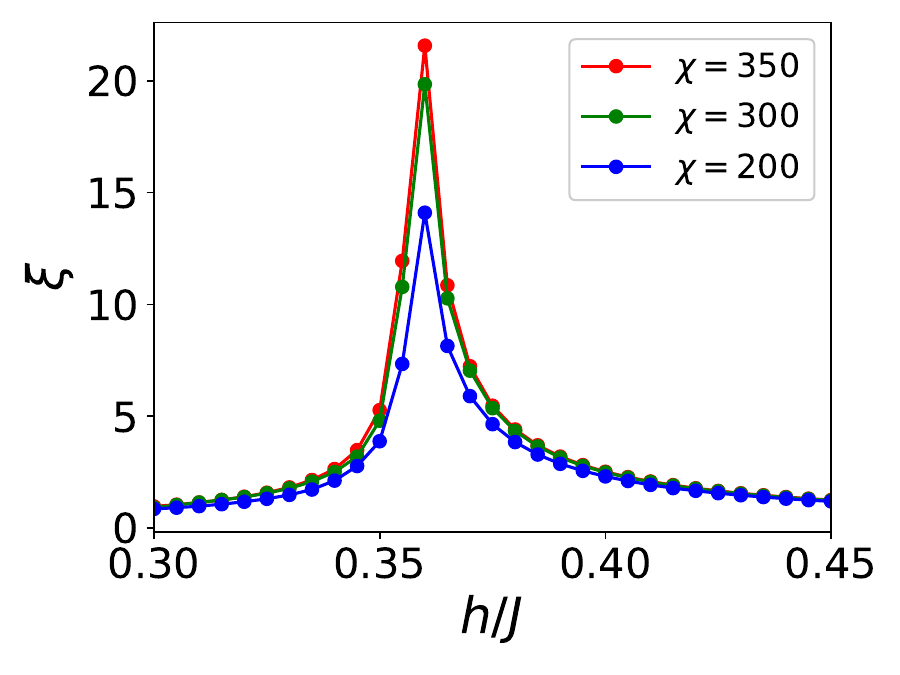}}
\subfloat[]{\includegraphics[width=0.48\linewidth]{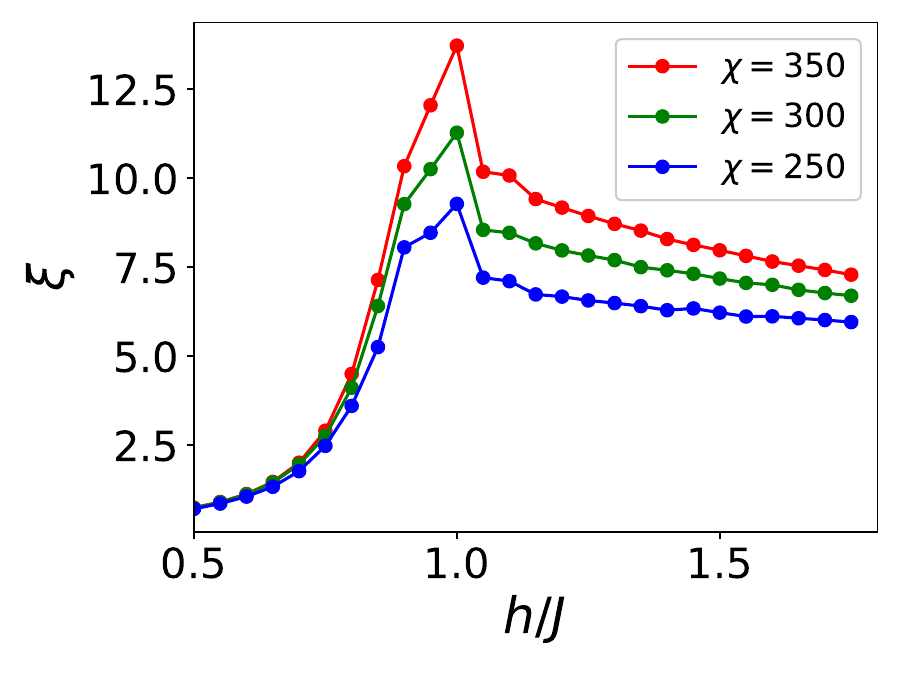}}
\caption{Correlation length $\xi$ as a function of the electric coupling $h$, calculated with iDMRG on cylinders with circumference $L_y=4$: Peaks in the correlation length reveal quantum critical points separating different gapped phases. We show the transition between the deconfined and confined phase for the (a) even and (b) odd $\mathbb{Z}_2$ gauge theories, corresponding to the limits $\mu=\mp \infty$ of our model.}
\label{fig:corr_z2}
\end{figure}

In Fig. \ref{fig:contourplots} we show iDMRG scans of the $h$-$\mu$ plane at fixed values of the magnetic coupling $J$ for a cylinder of circumference $L_y=4$. The gapped even and odd gauge theory regimes are clearly visible for large negative and positive values of the chemical potential $\mu$, respectively. Either of these regimes exhibits deconfined and confined phases described in Secs. \ref{ssec.eZ2} and \ref{ssec.oZ2}. Our DMRG calculations clearly indicate presence of $\Zt$ topological order in both deconfined limits, but not in the confined regimes. While the ground state in the confined regime of the $\Zt$ even case respects lattice translation symmetry, our DMRG results in the odd $\Zt$ regime are consistent with columnar ordering found in \cite{wenzel2012evidence}. The quantum transitions between the deconfined and confined phases can be detected numerically already for small system sizes, see Fig. \ref{fig:corr_z2}. Away from the pure gauge theory limits, for moderate values of the chemical potential an extended region with partially filled fermion occupation can be observed in Fig. \ref{fig:contourplots}. This has the shape of a wedge centered around the line $h=\mu$. This has a clear interpretation in the large $h$ limit: on this line dimers of unit length cost no energy, as a flipped electric string is exactly compensated by the chemical potential. In this limit, there is an emerging energy scale $t^2/h$ that determines the region where a partial occupation of dimers is possible. Large deviations of $\mu$ from this range of values force zero or complete occupancy. Another observable of interest on a cylinder is the expectation value of the non-contractible Wegner-Wilson loop $W$ which we plot in Fig. \ref{fig:contourplots}. In the pure even and odd $\Zt$ gauge theories, expectation value of this operator is (up to sign) unity  deep in the topologically ordered deconfined phase, while dropping abruptly to zero in the trivial (confined) phase. In these cases it can be seen as a non-local order parameter for topological order in the limit $L_y\to\infty$. 

\subsubsection{Topological Order and deconfinement at $h=0$}
\label{ssec:h0}
Deep in the deconfined phase of the $\mathbb{Z}_2$ gauge theory (toric code limit) the entanglement entropy under a bipartition into two half-infinite cylinders is given by 
\begin{equation}
S_{\mathbb{Z}_2}=(L_y-1)\log 2
\end{equation}
with the size-independent  \textit{topological entanglement entropy} $\gamma$ being equal to $-\log 2$. Our iDMRG results for the entanglement entropy plotted in Fig. \ref{fig:contourplots} confirm that the model  reduces to pure even and odd $\Zt$ gauge theories in the $\mu = \mp \infty$ limits.

Moreover, as argued in Sec. \ref{sec:h_0}, at $h=0$ and half-filling we expect a $\pi$-flux phase with a pair of deconfined Dirac fermions carrying $\mathbb{Z}_2$ charge, as long as $J\ll t$ (including the case $J\leq 0$). The entanglement entropy of such system is expected to split
\begin{equation}
S = S_{f}+S_{\mathbb{Z}_2},
\label{eq:S_free_gauge}
\end{equation}
with  $S_f$ being the entanglement entropy of free fermions hopping in the $\pi$-flux background. Our numerical results, presented in the Appendix \ref{appDMRG}, confirm the prediction \eqref{eq:S_free_gauge}.

\subsubsection{Large $h$ limit at $J=0$}
\label{ssec:num_large_h}
\begin{figure}[t]
	\includegraphics[width=\linewidth]{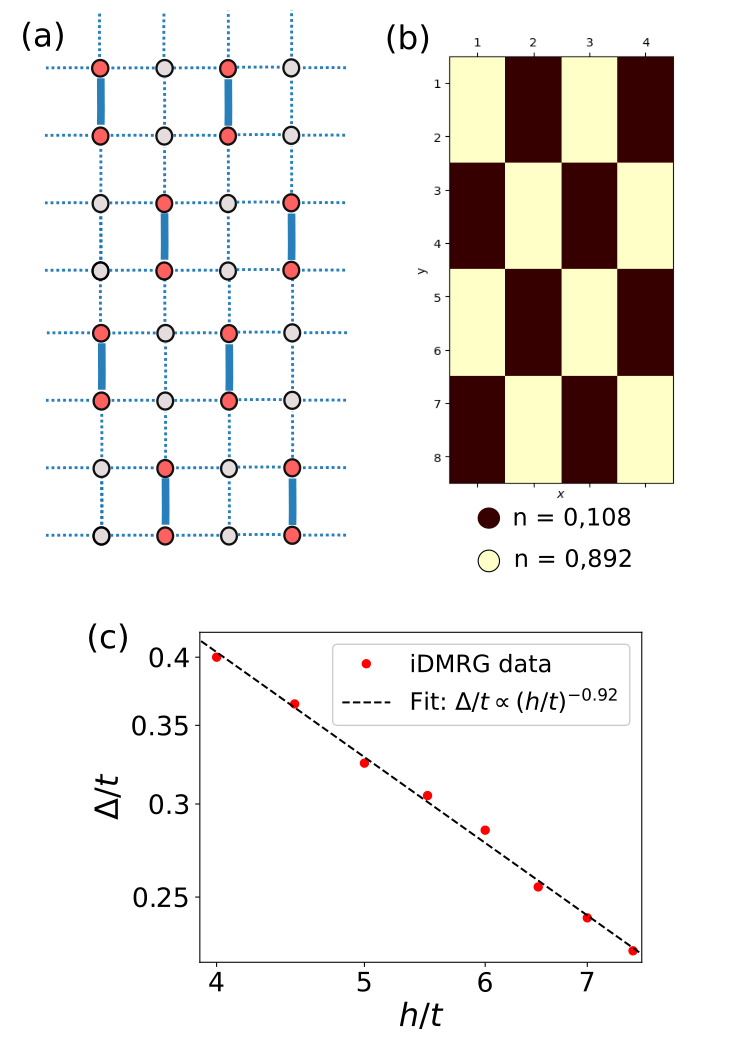}
	\caption{(a) Schematic illustration of a  staggered dimer Mott state at half filling. For each dimer, the six sites directly next to it are empty, minimizing the energy penalty due to the repulsion between different dimers. (b) iDMRG results for the local fermion density on a cylinder of circumference $L_y=8$ at $h/t=\mu/t=4$ and $J=0$. The arrangement of fermions into a staggered pattern of dimers is clearly visible. (c) Mott gap as a function of $h$. For large values of $h$ the gap scales approximately as $t^2/h$, consistent with the second order perturbation theory emergent energy scale.}
\label{fig:mott}
\end{figure}
When $h \gg J, t$ the effective degrees of freedom of the system are bosonic dimers of unit length. We can therefore consider an effective Hamiltonian that operates in the reduced Hilbert space spanned by dimer states. At $J=0$ the leading contributions are hopping terms and short-range repulsive interactions which are both of order $t^2/h$, as explained in Sec. \ref{sec:h_inty}. At commensurate fillings the repulsion can potentially stabilize a Mott-insulating state of dimers. In particular, at half filling the staggered Mott pattern shown in Fig. \ref{fig:mott} is a natural candidate, since this arrangement minimizes the inter-dimer repulsion. To test this prediction, we obtain the ground state wave function on cylinders of circumference up to $L_y=8$ using iDMRG. The results for the fermion density, shown in Fig. \ref{fig:mott}, confirm that the guess is indeed correct for the largest system that we could probe. The Mott gap can be estimated by tuning the chemical potential away from the value $\mu=h$ until the filling deviates from $1/2$. By doing this, we find that the gap decreases as $h$ is increased, and at large $h$ it is proportional to the emergent energy scale $t^2/h$, as shown in Fig. \ref{fig:mott}c.


\subsubsection{Topologically ordered Dirac semimetal to Mott insulator transition at $J=0$}
\begin{figure}
\includegraphics[width=0.9\linewidth]{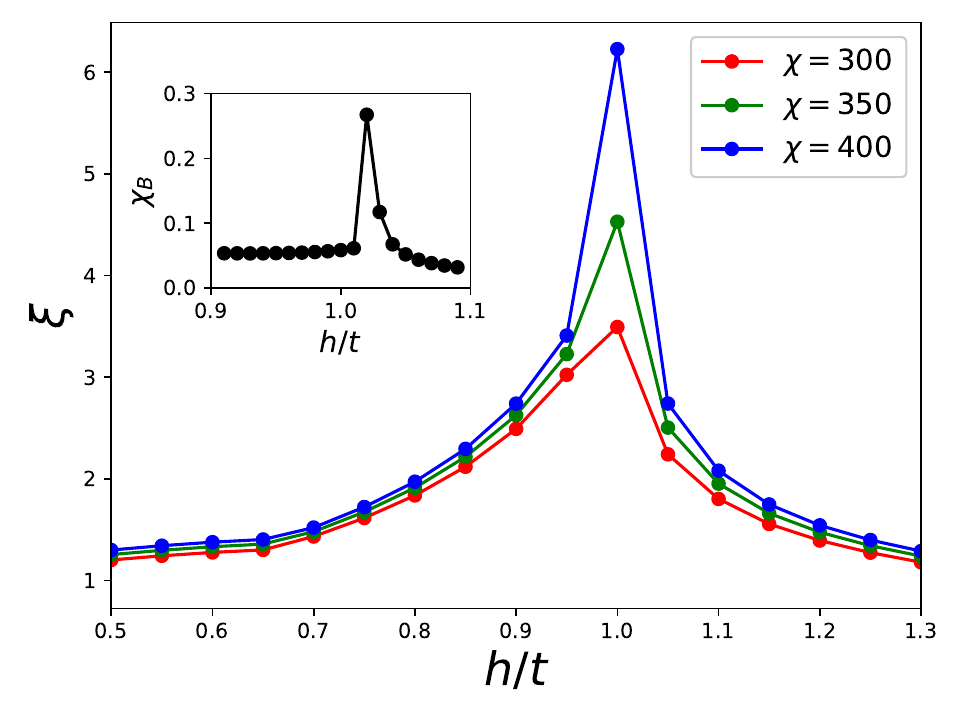}
\caption{Correlation length $\xi$ as a function of the electric coupling $h$, calculated with iDMRG on a cylinder with circumference $L_y=4$ for $J=0$. The system is kept at half filling by setting $\mu=h$. The peak in the correlation length is a signature of a quantum phase transition between a topologically ordered $\pi$-flux Dirac phase and a translational symmetry-breaking staggered Mott phase of dimers. The inset shows the magnetic susceptibility $\chi_B$ that exhibits a peak at the transition.}
\label{fig:MDtrans}
\end{figure}

For $J=0$ at half filling the Mott ground state that exists at large $h\approx \mu$ breaks translational symmetry spontaneously. In the opposite limit $h\rightarrow 0$, the free Dirac fermions coexist with topologically ordered Ising gauge fields, and we expect this phase to be stable with respect to a small finite $h$ perturbation.
On a cylinder of finite $L_y$ both these phases are gapped and thus have finite entanglement entropy $S$ that follows the area law. The two phases must be separated by at least one quantum phase transition. The nature of this transition is of great interest, as it is not clear \textit{a priori} that confinement of $\mathbb{Z}_2$ charges and  spontaneous symmetry breaking of translational symmetry happen simultaneously. From iDMRG simulations on a cylinder with $L_y=4$, plotted in Fig. \ref{fig:MDtrans} we detect a single phase transition at $J=0$ taking place along the half-filling line $h=\mu$. In the gauge sector, we find that the transition exhibits a peak in the $\mathbb{Z}_2$ magnetic susceptibility $\chi_B = \partial \overline{\langle P_{\rr^*} \rangle} /\partial h$, where $P_{\rr^*}$ is the plaquette operator \eqref{eq:plaq_mapping_2} and the overbar denotes the average over a unit cell. This peak is shown in the inset of Fig. \ref{fig:MDtrans}.

\subsubsection{Comparison with the cross-linked ladder study \cite{PhysRevX.10.041007}}
\begin{figure}[h]
	\includegraphics[width=0.6\linewidth]{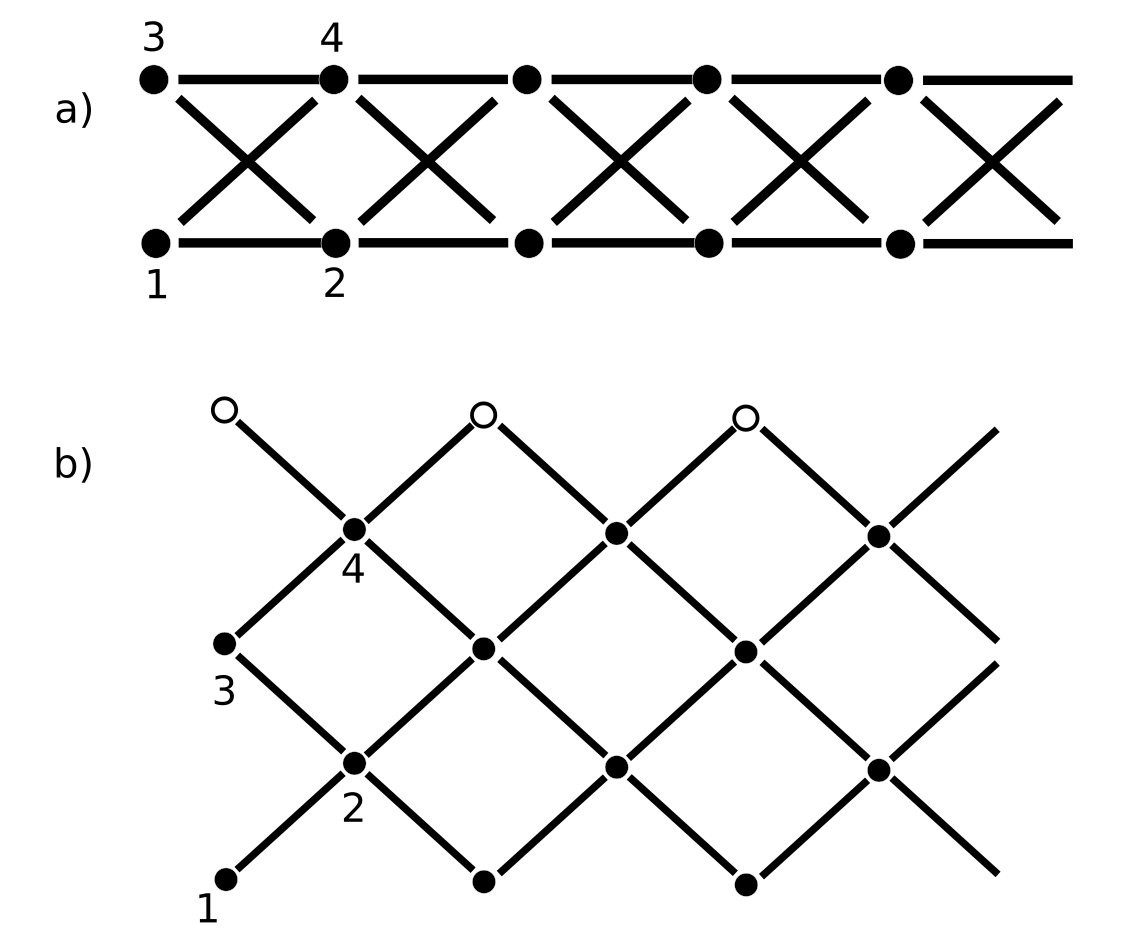}
	\caption{ A cross-linked ladder (a) is equivalent to a rhombic lattice (b) on a cylinder of circumference $L_y=4$. In (b), empty circles must be identified with corresponding sites in the lower row.}
\label{fig:cll}
\end{figure}

Here we will compare and contrast our DMRG findings at $J=0$ on cylinders with square lattice geometry with the recent study of the $\Zt$ gauge theory coupled to spineless fermionic matter on a cross-linked Creutz-Ising ladder \cite{PhysRevX.10.041007}. As illustrated in Fig. \ref{fig:cll}, such a ladder can be equivalently represented as a rhombic lattice on a cylinder with $L_y=4$. Although in the thermodynamic limit the rhombic arrangement is completely equivalent to the square lattice, the situation is very different on a cylinder of circumference $L_y=4$. For $h=J=0$ at half-filling ($\mu=0$), just as on a square lattice, $\Zt$ gauge fields on a cross-linked lattice form a $\pi$ flux background in the ground state due to Lieb's theorem. According to Creutz \cite{creutz1999end}, spinless fermions in such background form an SPT ordered phase with protected edge modes. This SPT order and corresponding edge modes were also identified in \cite{PhysRevX.10.041007} for the $\Zt$ gauged problem even at finite values of the electric coupling $h$. This should be contrasted with the square lattice model studied here, where in cylinder geometry the fermionic bands in the $\pi$-flux background appear to be topologically trivial and thus no SPT ordered state emerges. Therefore we found no evidence for edge modes in DMRG simulations on a finite cylinder. Moreover, at finite $h$ the entanglement spectrum of this problem does not exhibit topological degeneracies that are characteristic to SPT phases \cite{PhysRevB.85.075125}. In addition, the authors of \cite{PhysRevX.10.041007} argued that $\Zt$ gauge fields are deconfined and topological order survives in the cross-linked ladder at arbitrary values $h$. As we have seen above, this is not true in our problem, where at sufficiently large electric coupling the gauge fields confine and a staggered Mott insulator emerges at $J=0$. We believe it will be interesting to employ the mapping developed in Sec \ref{sec:spinmap} to investigate the fate of the SPT and topological orders on rhombic lattices on cylinders of circumferences larger than $L_y=4$.

\subsection{\label{sec:qd} Exact diagonalization: clustering of quantum dimers}
Our discussion in Sec. \ref{sec:h_inty} showed that in the large $h$ limit individual fermions become immobile and at low energies the model \eqref{eq:H} reduces to a quantum dimer model, where two neighboring fermions connected by an electric string form one dimer. In this problem the number of dimers in the ground state can be changed by tuning the chemical potential $\mu$. Here we investigate the resulting quantum dimer model by means of an exact diagonalization study on finite lattices with periodic boundary conditions and a fixed number of dimers.
  
\subsubsection{Zero fermion hopping} We take the magnetic coupling $J>0$ and first analyze the case where the fermion hopping $t$ is strictly zero. As argued in Sec. \ref{sec:h_inty}, in this regime the dynamics of dimers is governed by the kinetic Rokhsar-Kivelson Hamiltonian 
\begin{eqnarray} \label{RKN}
H_d & =-J\sum\left(|\vcenter{\hbox{\includegraphics[height=0.02\textheight]{term1.pdf}}}\rangle\langle\vcenter{\hbox{\includegraphics[height=0.02\textheight]{term2.pdf}}}|+\text{h.c.}\right).
\end{eqnarray}
\begin{figure}[h]
	\includegraphics[width=0.9\linewidth]{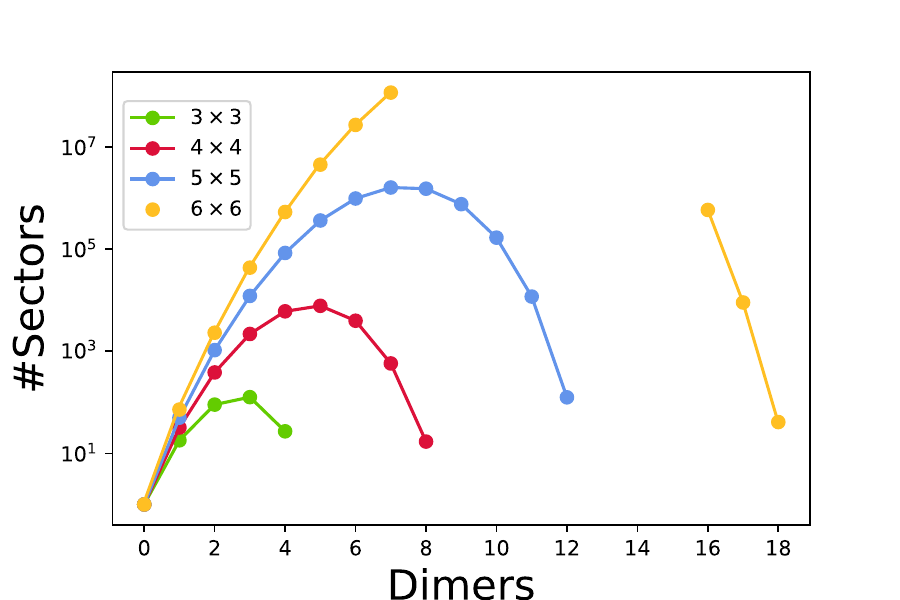}
	\caption{ The number of disconnected subsectors of the Hilbert space as a function of the number of dimers for four different system sizes. }
\label{fig:sectors}
\end{figure} 

We find that the full Hilbert space of the quantum dimer model splits into many disconnected sectors.
The logarithmic plot in Fig. \ref{fig:sectors} shows the number of these sectors as a function of dimer number for several different lattice sizes. Within each sector the states are connected in the sense that one can start with one state and act repeatedly with local terms of the Hamiltonian \eqref{RKN} to produce any other state in the same sector. 
The curves show a steep increase at low fillings. This is intuitively clear, since on a nearly empty lattice the dimers can be placed with hardly any obstructions. Thus the total number of dimer configurations increases exponentially with the dimer number $n_\text{d}$. More precisely, on a square $L\times L$ lattice a dimer can be placed in $L^2$ positions in two orientations (vertical and horizontal). Therefore the number of configurations has a leading term that scales as $(2L^2)^{n_\text{d}}/n_\text{d}!$. In this sparsely populated limit, most of these states give rise to sectors of size one, whenever the dimers cannot resonate. As a consequence the number of sectors is roughly the same as the number of states.

As $n_\text{d}$ increases further, the dimers start to obstruct each other, which leads, first, to a decrease in the rate of growth of the curves in Fig. \ref{fig:sectors} and eventually to their decline. By starting from the fully packed system and decreasing the number of dimers, we can obtain the asymptotics near the right end of the curves. If a dimer is removed, two monomer holes are introduced into the system, each can be placed in $\sim L^2$ positions. This results in an asymptotic formula for the number of states near full packing:  $C \times  (L^2)^{n_\text{h}}/n_\text{h}!$, were $C$ is the number of states for the fully packed case and $n_\text{h}$ is the number of monomer holes. 

For the fully packed case, our results for excitation energies and the number of dimer configurations were compared with Leung et al. \cite{PhysRevB.54.12938}, who studied the Rokhsar-Kivelson model with a potential energy $V$ term. We find complete agreement with their results at $V=0$. 

\begin{figure*}[hbt]
\includegraphics[width=0.9\linewidth]{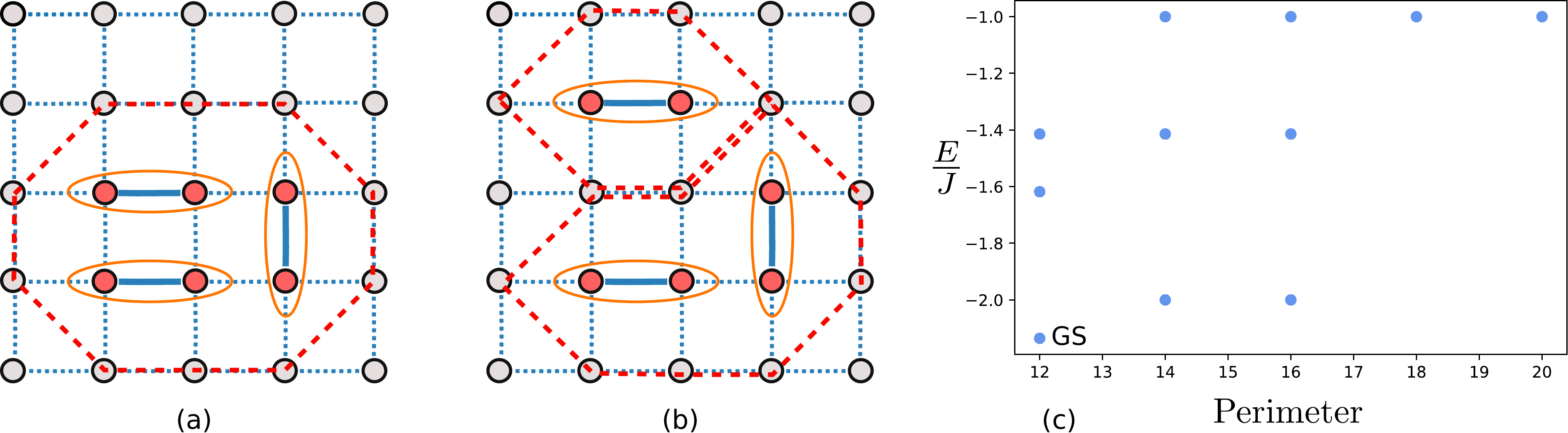}
\caption{Zero fermionic hopping: (a) and (b) The perimeter is defined as the number of sites surrounding the dimers. The red dashed lines are the boundaries. Whenever a site lies on more than one boundary we count it again. Thus the perimeter in (a) is $10$, while the perimeter in (b) is $15$.  (c) Four dimers on a $5\times 5$ lattice: The lowest energy of each sector is plotted together with its perimeter. We observe that the ground state of the Hamiltonian has the smallest perimeter.}
\label{fig:EvsPerim}\label{fig:PerimDef}
\end{figure*}

In the absence of the fermion hopping, the Hamiltonian \eqref{RKN} is only able to change the attachment of the electric lines on fully occupied plaquettes, while the fermions remain frozen in place. 
As illustrated in Fig. \ref{fig:res}b, the effect of the plaquette term is to decrease the energy whenever two dimers can resonate. This leads to a clustering phenomenon, where the dimers in the system energetically prefer to clump together such that they can resonate. In order to detect this numerically, we computed the perimeter of dimer arrangements in several disconnected sectors on a $5\times5$ lattice populated with four dimers. The perimeter is a conserved quantity that is defined as shown in Fig. \ref{fig:PerimDef} (a,b).  We find that a compact packing is preferred, i.e.,  the absolute ground state has the lowest perimeter, see Fig. \ref{fig:EvsPerim} (c). 

Presence of an exponential number of disconnected subsectors in the Hilbert space of the problem governed by the Hamiltonian \eqref{RKN} originates from $\Zt$ gauge invariance of the original lattice gauge theory. Indeed, since we consider here the case with $t=0$, the fermions are completely frozen and thus the problem reduces to a pure $\Zt$ gauge theory in a background of static gauge charges. By construction, the resulting $\Zt$ gauge invariant theory has an extensive number of local conservation laws and thus its Hamiltonian cannot connect subsectors characterized by different configurations of $\Zt$ background charges.  Similar physics arises in the strict confinement limit of the one-dimensional version of this lattice gauge theory studied in \cite{PhysRevLett.124.207602}. The phenomenon discussed here appears to be qualitatively different to the Hilbert space fragmentation introduced in \cite{PhysRevX.10.011047, PhysRevB.101.174204} in the context of dipole-conserving one-dimensional models. While in our case \emph{local} $\Zt$ gauge invariance is responsible for the fracture of the Hilbert space, what ensures the Hilbert space fragmentation of \cite{PhysRevX.10.011047, PhysRevB.101.174204} are  \emph{non-local} statistically localized integrals of motions that were introduced in \cite{PhysRevB.101.125126}.

\subsubsection{Finite fermion hopping}
As discussed in Sec. \ref{sec:h_inty}, a finite fermion hopping $t$ results in second-order processes that allow the dimers to acquire kinetic energy. These transitions, however, are subleading compared to the first-order resonance term, since the energy cost of hopping a single dimer is of the order $t^2/h$, which is much smaller than $J$ in the large $h$ limit. We add now these second-order processes resulting in the following quantum dimer Hamiltonian 
\begin{eqnarray} \label{qdh}
H_d & =&-J\sum\left(|\vcenter{\hbox{\includegraphics[height=0.02\textheight]{term1.pdf}}}\rangle\langle\vcenter{\hbox{\includegraphics[height=0.02\textheight]{term2.pdf}}}|+\text{h.c.}\right)-t_\text{d}\sum \bigg( |\vcenter{\hbox{\includegraphics[height=0.005\textheight]{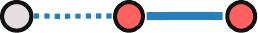}}}\rangle\langle\vcenter{\hbox{\includegraphics[height=0.005\textheight]{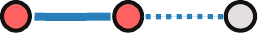}}}|\nonumber \\
&&+|\vcenter{\hbox{\includegraphics[height=0.025\textheight]{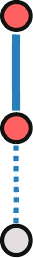}}}\rangle\langle\vcenter{\hbox{\includegraphics[height=0.025\textheight]{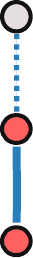}}}|-|\vcenter{\hbox{\includegraphics[height=0.02\textheight]{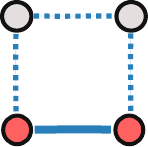}}}\rangle\langle\vcenter{\hbox{\includegraphics[height=0.02\textheight]{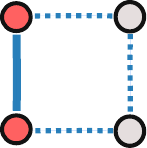}}}|+|\vcenter{\hbox{\includegraphics[height=0.02\textheight]{term7.pdf}}}\rangle\langle\vcenter{\hbox{\includegraphics[height=0.02\textheight]{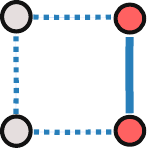}}}|+|\vcenter{\hbox{\includegraphics[height=0.02\textheight]{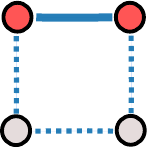}}}\rangle\langle\vcenter{\hbox{\includegraphics[height=0.02\textheight]{term8.pdf}}}|\nonumber\\
&&-|\vcenter{\hbox{\includegraphics[height=0.02\textheight]{term10.pdf}}}\rangle\langle\vcenter{\hbox{\includegraphics[height=0.02\textheight]{term9.pdf}}}|+\text{h.c.}\bigg).
\end{eqnarray}
The coupling $t_\text{d}$ is positive and equals to $t^2/(2h)$.
Here dimers do not hop perpendicularly to the electric string. These perpendicular transitions are allowed in the original model, but appear only at fourth order in the perturbation theory. Thus they scale as $t^4/h^3$ and are neglected here. 
The relative signs between the hopping processes stem from a careful consideration of fermionic statistics.
Note that the Hamiltonian \eqref{qdh} is a simplified version of the complete second-order perturbative effective Hamiltonian because it does not contain the short-ranged repulsion interactions between dimers that are also generated at this order. Nevertheless, this model provides some insight into the hopping mechanism of clusters and emergence of weakly dispersing energy eigenstates of the studied gauge theory in the large $h$ limit.

For $t_\text{d} \neq 0$ the ground state degeneracy is partially lifted and splits into Bloch bands. The band structure for a single dimer is calculated in Appendix \ref{OneDimerBand}. Since the dimers live on links, each unit cell of the square lattice can be occupied by either a horizontal or a vertical dimer, resulting in two energy bands. We find that the lower band is flat, while the upper band has cosine dispersion. The immobile localized excitations forming the flat band are constructed explicitly in Appendix \ref{OneDimerBand}, where they are found to be a superposition of one dimer states around a plaquette. The flatness of the lower band implies that the single dimer groundstate is macroscopically degenerate.
\begin{figure*}[hbt]
\includegraphics[width=0.9\linewidth]{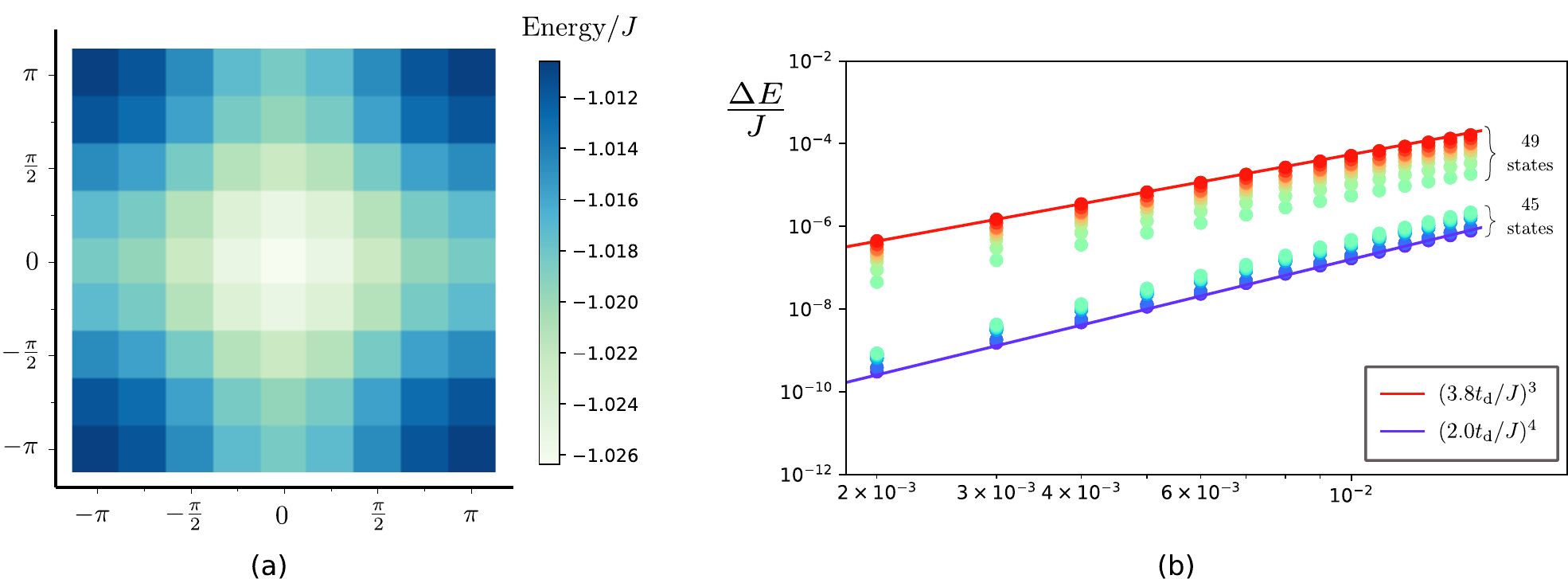}
\caption{Finite fermionic hopping: (a) Excitation energies of the lowest band of the Hamiltonian \eqref{qdh} with $t_\text{d} = 0.05 J$ on an $8\times 8$ lattice with the cluster composed of two dimers, plotted in the first Brillouin zone. The ground state (center) is a zero momentum eigenstate. (b) Double logarithmic plot of the lowest two bands of excitation energies of the Hamiltonian \eqref{qdh}, after subtracting the ground state energy, on a $7\times 7$ lattice for a cluster composed of three dimers. The excitation energies of the two lowest bands grow as  $\propto t^4_\text{d}/J^3$ and $\propto t^3_\text{d}/J^2$ for $t_\text{d} \ll J$. } 
\label{fig:EnMomentum}\label{fig:diffs_7_3}
\end{figure*} 

We first study the physics of the smallest cluster. Specifically, we perform an exact diagonalization calculation for two dimers on a periodic $8\times 8$ lattice.  In Fig. \ref{fig:EnMomentum} (a) the degenerate energy levels of the lowest band have been resolved into momentum eigenstates with wave numbers  $\bm k = (2\pi n/L ,2\pi m/L)$, where $n,m=0,\dots,L-1$.  The ground state has zero momentum. Qualitatively, we observe that the band resembles a simple cosine band of an elementary particle hopping on a square lattice. Consistent with expectations from perturbation theory, we find that the bandwidth scales as $ t_{\text{d}}^2/J$ as $t_{\text{d}}\to 0$.

In order to explore the hopping of larger clusters as a function of $t_\text{d}$, we numerically diagonalized a system of three dimers on a periodic $7\times7$ lattice, see Fig.\ref{fig:diffs_7_3} (b). Here the excited levels are measured from the respective ground state at a given value of $t_\text{d}$, i.e. the plot shows the energy differences $\Delta E \equiv E_{i}(t_\text{d})-E_{0}(t_\text{d})$.  
At $t_\text{d}=0$ the ground state is $98$-fold degenerate. This degeneracy is expected, since the cluster can be placed in $7\times 7$ positions and, due to its rectangular shape, can be oriented either horizontally or vertically. At finite $t_\text{d}$, the degeneracy is partially lifted and the ground state becomes $4$-fold degenerate, with momentum wavevectors $\bm k = (0,\pm  6 \pi /7), (\pm  6 \pi /7,0)$. Due to the four-fold rotation symmetry of the lattice some excited states retain their $4$-fold or $8$-fold degeneracy. The four ground states together with the lowest $45$ excitations make up the first energy band. 
The same figure also shows a power-law fit to the lowest and highest excitation of the first and second band as a function of $t_\text{d}/J$.  The second band has an energy dependence that scales as $(t_\text{d}/J)^3$. This is consistent with our expectation that the bandwidth of a cluster composed of $n_{\text{d}}$ dimers should scale as $\propto t_\text{d}^{n_{\text{d}}}$ in the limit $t_\text{d}\to 0$.
In contrast, the energies in the lowest band scale for $t_\text{d} \ll J$ as the power-law  $(t_\text{d}/J)^4$. 


\section{\label{sec:conclusions}Conclusions and outlook}

In this paper we started to explore the rich and intricate quantum many-body physics of single-component fermion matter coupled to Wegner's $\Zt$ gauge theory in two spatial dimensions. The zero- and fully-filled regimes reduce to the well-understood even and odd versions of the pure $\Zt$ gauge theory, respectively. At partial filling we developed analytical understanding of the $h=0$ and $h\to\infty$ limits. Employing the mapping of this lattice gauge theory to the unconstrained local spin 1/2 model developed in Sec. \ref{sec:spinmap}, we have performed iDMRG simulations which fully support our analytical understanding. In addition, with iDMRG we investigated the half-filled case in the absence of the magnetic coupling $J$, where on an infinite cylinder of circumference $L_y=4$ we identified salient signatures of the topologically ordered Dirac semimetal and the translation symmetry-broken Mott quantum phases separated by a quantum phase transition. In addition, in the strongly-coupled $h\to\infty$ limit and at finite $J$ we have studied the clustering of fermions and its anomalously slow hopping by means of exact diagonalization.

One important aspect that was not addressed in this paper is the nature of spontaneous symmetry breaking of the global $U(1)$ particle number symmetry in this model. At $h>0$ the Ising gauge field mediates attraction between fermions which should result in Cooper pairing. Given that the fermionic matter is single-component, we naturally anticipate $p$-wave orbital pair superfluid to be formed in some region of the phase diagram. Is the non-chiral $p_x/ p_y$ or the time-reversal breaking chiral $p_x+ip_y$ order is preferred? The clustering phenomenon found in this paper in the limit $h\to\infty$ is expected to inhibit condensation of $p$-wave Cooper pairs. How does clustering destroy pair superfluidity? The answers to all these interesting questions are hidden in expectation values and correlation functions of the dimer operators $b^\dagger_{\mathbf{r}, \eta}= c^\dagger_{\mathbf{r}} \sigma_{\mathbf{r},\eta} c^\dagger_{\mathbf{r}+\eta}$ with $\eta= \hat x, \hat y$. Numerical simulations that we have undertaken so far are too anisotropic to draw definite conclusions about the nature of the superfluid order in the two-dimensional thermodynamic limit. 

The fate of $p$-wave superfluidity in the two-dimensional thermodynamic limit could be clarified by finding and implementing more efficient ways of simulating this lattice gauge theory. It is an open question whether this model can be investigated with a sign-problem-free quantum Monte Carlo method. As an alternative, it would be exciting to study this problem with two-dimensional tensor networks, for example using the iPEPS algorithm \cite{PhysRevLett.101.250602, orus2014practical, Cirac:2020obd}. Experimentally, extensions of the Floquet-engineering schemes developed in \cite{Barbiero2018, Schweizer2019, Gorg2019} to two-dimensional lattices are desirable and can shed fresh light on this rich model, for a recent proposal see also \cite{wei2020floquet}. Finally, digital simulations of lattice gauge theories with quantum computers become nowadays reality \cite{banuls2019simulating, banuls2020review}, so they presents an exciting frontier for studies of this lattice gauge theory. For the simulation of the pure $\Zt$ gauge theory a concrete circuit-based digital quantum adiabatic algorithm has already been proposed and tested on a classical computer in \cite{cui2020circuit}. 

In this problem the fermion parity symmetry is completely eliminated by gauging in the bulk of the system, but survives as a global symmetry in the presence of a boundary, where it acts onsite. Remarkably, despite being manifest only near the boundary, such a symmetry can enrich our understanding of the bulk quantum phases \cite{borla2020gauging, Higgs}. It would be useful to work out how this boundary symmetry acts in this model and study its interplay with other global symmetries of the problem. 

It would be interesting to investigate quantum phases of  this lattice gauge theory on different two-dimensional lattices and attempt to generalize the local mapping of Sec. \ref{sec:spinmap} to those geometries.

An exciting frontier is the structure of the energy spectrum far away from the ground state manifold. Quantum scar states are isolated eigenstates in the middle of the energy spectrum of a many-body translation-invariant problem that do not follow the paradigm of the eigenstate thermalization hypothesis \cite{serbyn2020quantum}. Recently, such states were found analytically in the spin model that is equivalent to the one-dimensional $\Zt$ gauge theory coupled to spinless fermions \cite{PhysRevB.101.024306, PhysRevB.101.195131}. Can the methods used in these works be extended to the two-dimensional lattice gauge theory studied in this paper? Ergodicity breaking and disorder-free localization \cite{Smith2017} in the strong confinement limit is another wide open problem.

From the viewpoint of quantum information, our work sheds new light on the stability of the topological toric code with respect to stabilizer violations originating from a finite density of the anyons with the fermionic self-statistics. In the future, it would be interesting to study these questions systematically and extend these investigations to a finite temperature case.

\begin{acknowledgments}
\emph{Acknowledgments:} 
We would like to acknowledge useful discussions with Johannes Feldmeier, Snir Gazit, Johannes Hauschild, Joseph Maciejko, Abhinav Prem, Matthias Punk, Pablo Sala, Urban Seifert, Ruben Verresen, Erez Zohar.
UB, BJ and SM are funded by the Deutsche Forschungsgemeinschaft (DFG, German Research Foundation) under Emmy Noether Programme grant no.~MO 3013/1-1.  FP is funded by the European Research Council (ERC) under the European Unions Horizon 2020 research and innovation program (grant agreement No. 771537). We acknowledge support from the Deutsche Forschungsgemeinschaft (DFG, German Research Foundation) under
Germany's Excellence Strategy-EXC-2111-390814868.
\end{acknowledgments}

\newpage
\begin{widetext}
\appendix

\section{Ground state flux configurations for the $L_y=2$ cylinder at half filling}
\label{app:flux}
As explained in section \ref{sec:h_0} the model at $h=0$ reduces to free fermions in a background of $\mathbb{Z}_2$ magnetic fluxes, determined by a configuration $\left\lbrace B_{\mathbf{r},\eta} \right\rbrace$ of classical link variables. Here we study using iDMRG the ground state of the system at half filling ($\mu=0$) as a function of the magnetic coupling $J$ on a thin infinite cylinder of circumference $L_y=2$. 

Our numerical results demonstrate that a number of intermediate flux phases appears as the dimensionless ratio $J/t$ is tuned, see Fig. \ref{fig:lines}. Within each flux phase, the ground state energy per unit cell is a linear function of $J$, and is given by $E_{gs}(J) = E^{\mathcal{F}}-J\langle\mathcal{P}\rangle$, where $E^{\mathcal{F}}$ is the ground state energy of free fermions in the flux-background $\mathcal{F}$ and $\langle\mathcal{P}\rangle$ is the average value of the plaquette operator within the extended unit cell. The competition between the two terms determines the most favorable configuration of $\mathbb{Z}_2$ fluxes\footnote{In principle, this decomposition of the ground state energy $E_{gs}$ allows to solve the problem analytically. However this involves a cumbersome task of calculating the fermionic band structure for an infinite number of flux configurations $\mathcal{F}$.}. 

Although for larger cylinders the computation of the ground state becomes numerically challenging, our analysis suggests that as $L_y$ is increased the intermediate phases occupy a progressively smaller region of parameter space. In the thermodynamic limit we expect a sharp transition between the $\pi$-flux and the $0$-flux phase, in agreement with the QMC results of \cite{Gazit2017}.

\begin{figure}
   \includegraphics[width=0.5\linewidth]{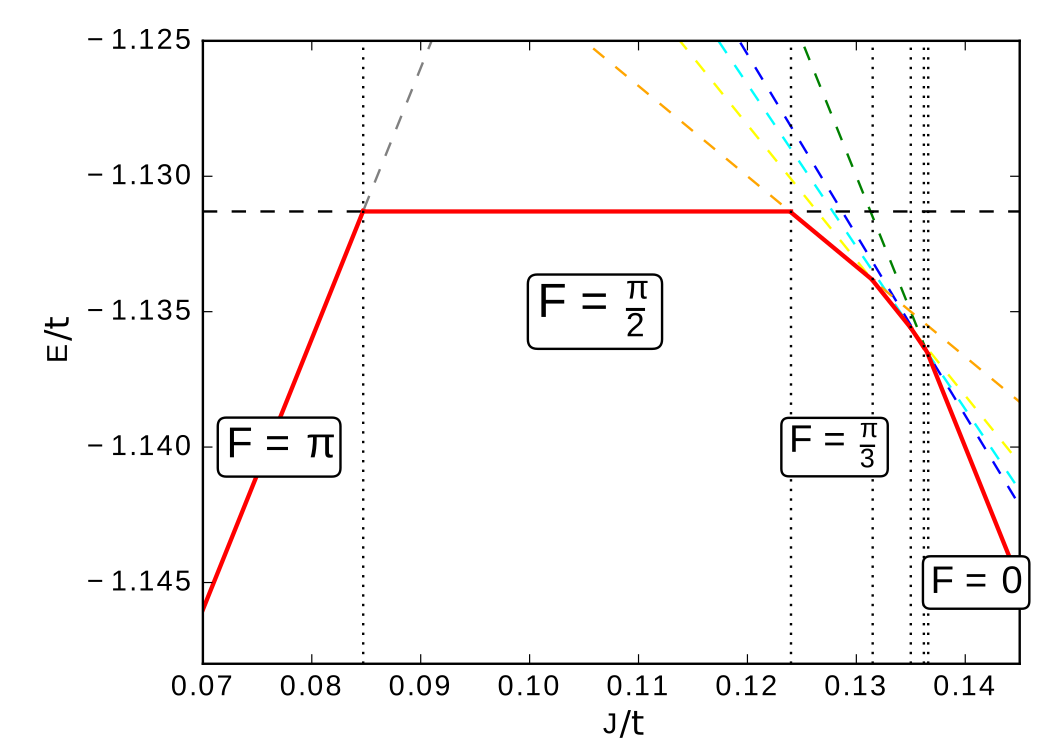}
  \caption{Ground state energy (red) and its average flux $\mathcal{F}$ as a function of $J/t$ on a cylinder with $L_y=2$. Dashed lines denote energies of configurations with different $\Zt$ fluxes.}
	\label{fig:lines}
\end{figure}
\section{\label{app:ClusterProof} Proof of clustering in the kinetic quantum dimer model}
Here we study the phenomenon of clustering in a system governed by the kinetic Rokhsar-Kivelson Hamiltonian
\begin{eqnarray}
H_d & =-J\sum\left(|\vcenter{\hbox{\includegraphics[height=0.02\textheight]{term1.pdf}}}\rangle\langle\vcenter{\hbox{\includegraphics[height=0.02\textheight]{term2.pdf}}}|+\text{h.c.}\right).
\end{eqnarray}
 We begin by considering  two separated islands of dimers as shown in Fig.\ref{fig:Proof}(a), where we highlighted the dimers by drawing rectangles. 
\begin{figure}[h]
\begin{centering}
\includegraphics[scale=0.3]{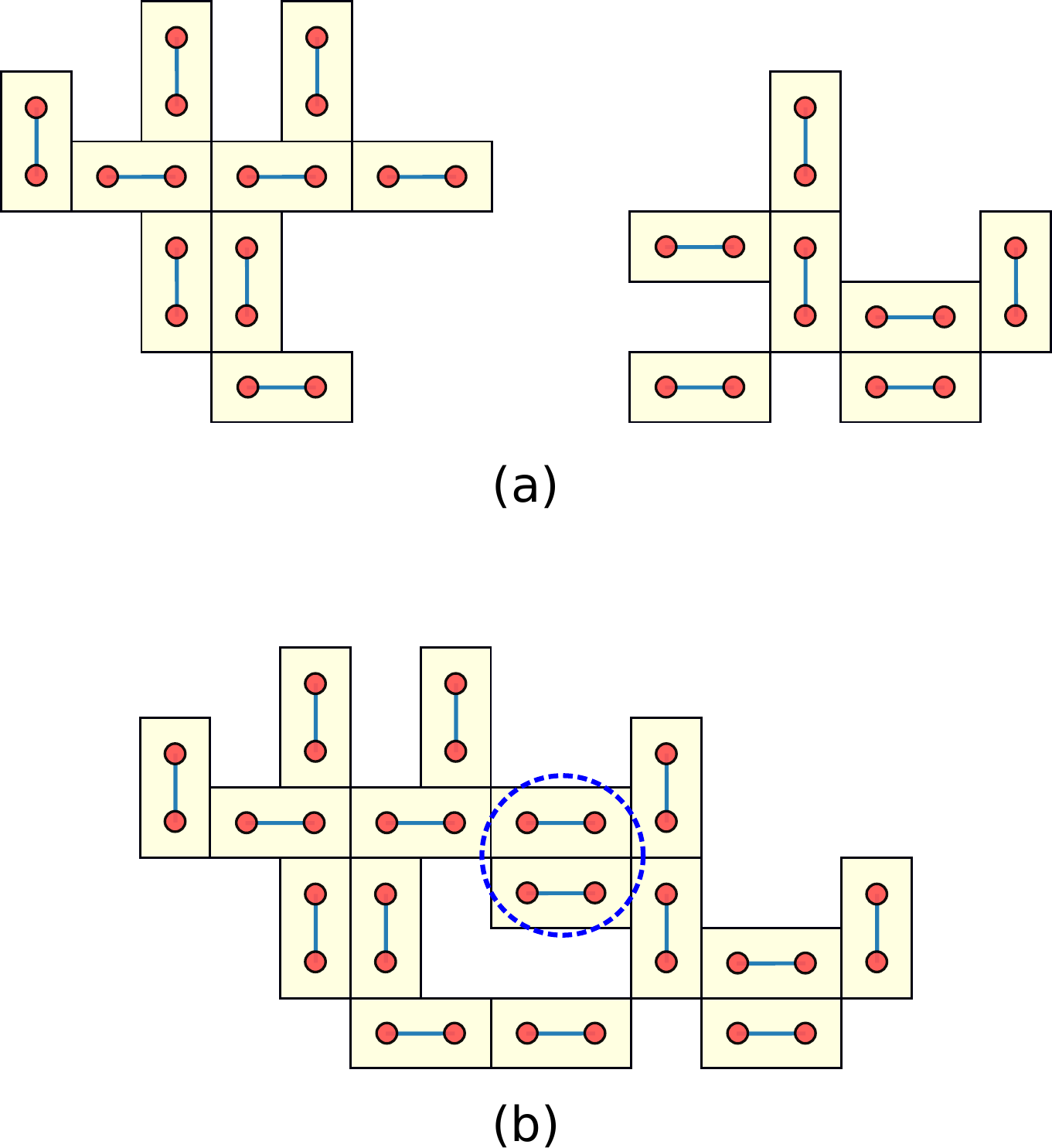}
\par\end{centering}
\caption{(a) State with two separated islands of dimers. Under the action of the Hamiltonian $H_d$ two parallel neighboring dimers resonate. The empty fermion sites are not shown. In (b) the islands are placed next to each other. Now there is one more pair of dimers, circled by a dashed line, that resonate under the action of $H_d$. }
\label{fig:Proof}
\end{figure}
Although the fermions cannot move, the electric lines can resonate under the action of the Hamiltonian. Here we prove that it is always energetically favorable for the islands to cluster together such that electric lines can resonate between them.  The precise shapes of islands do not matter for this proof. 

Let us denote by $S$ the set of all states that can be obtained by acting with the local terms of the Hamiltonian on Fig. \ref{fig:Proof}(a). When the islands are brought together as in  Fig. \ref{fig:Proof}(b), one retains
all the states from $S$. Moreover, new states appear under the action of the local terms in $H_d$. We denote this set of states by $S'$. They are generated by flipping dimers
in the connected region, as for example the circled dimer pair in Fig. \ref{fig:Proof}(b).  The new Hamiltonian has a larger Hilbert space of dimension $|S|+|S'|$. Next to the matrix elements between the states within $S$ and $S'$, the Hamiltonian will now also connect some states in $S$ with states
in $S'$. Hence the problem of the connected islands is governed by a Hamiltonian of the block form 
\begin{eqnarray}
H'=\left(\begin{array}{cc}
H & A\\
A^{T} & B
\end{array}\right).
\end{eqnarray}
\\
The Hamiltonian matrix has dimensions $(|S|+|S'|)\times(|S|+|S'|)$, with the block $A$ of size $|S|\times|S'|$ and the block B of size $|S'|\times|S'|$. We notice that the diagonal entries of $H$ and $H'$ are equal to zero, since the Hamiltonian acting on a state always changes it or yields zero. This fact will become important below.

Intuitively, it is clear that the ground state can hybridize with the new states in the Hilbert space and thereby lower its energy. In the following we provide a proof of this expectation by utilizing the variational principle which  states that the ground state energy $E'_{0}$ of $H'$ satisfies the inequality
\begin{eqnarray}
E'_{0}\leq\frac{\langle\psi|H'|\psi\rangle}{\langle\psi|\psi\rangle},
\end{eqnarray}
for any nonzero vector $|\psi\rangle$. \newline  
With this inequality it is straightforward to show that the ground state energy of $H'$ is not larger than that of $H$. Namely, let $|\psi_0\rangle$ be the ground state of $H$ with energy $E_{0}$, then the stacked vector
\begin{eqnarray}
\left(
\begin{array}{c}
|\psi_0\rangle\\
0
\end{array}
\right)
\end{eqnarray}
can be used as a variational state for $H'$:
\begin{eqnarray}
E_0' \leq
\left(\begin{array}{cc}
\langle \psi_0| &
0
\end{array}\right) 
H'\left(\begin{array}{c}
|\psi_0\rangle\\
0
\end{array}\right)=E_{0}.
\end{eqnarray}
Thus the ground state energy of $H'$ will certainly be smaller or equal to
$E_{0}$. However, it turns out that one can prove a stronger statement: \newline 

\textbf{Statement} The ground state energy of $H'$ will be strictly lower than
the ground state energy of $H$, if the overlap of at least one of the new states with $H'|\text{GS}\rangle$ is non-zero. Here $|\text{GS}\rangle$ is the ground state of $H$.\\
\\

We prove this statement by  starting from $H$ and building $H'$ step by step by adding one row and one column at a time. We show that
at each step the energy will not increase and that at least at one
of the steps the energy must decrease. We start by adding the first
column and first row of $A$ to $H$, such that we have a matrix of
dimension $(|S|+1)\times(|S|+1)$ and $B$ is set to zero because $B_{11} = 0$. Denote
the $i$th column of $A$ by $A_{i}$ and the Hamiltonian at step
$i$ by $H'_{i}$ with the ground state energy $E'_{i0}$. We choose as the variational
wave-function
\begin{eqnarray}
|\psi\rangle=\left(\begin{array}{c}
|\text{GS}\rangle\\
a
\end{array}\right),
\end{eqnarray}
where $|\text{GS}\rangle$ is the normalized ground state of $H$ and $a$
is a real number. Then we have
\begin{align}
\frac{\langle\psi|H'_{1}|\psi\rangle}{\langle\psi|\psi\rangle} & =\frac{E_{0}+a(\langle \text{GS}|A_{1}\rangle+\langle A_{1}|\text{GS}\rangle)}{1+a^{2}} \label{eq:en_fct_of_a}\\
H'_{1} & =\left(\begin{array}{cc}
H & A_{1}\\
A_{1}^{T} & 0
\end{array}\right).
\end{align}
We are writing $A_{1}$ as a vector, since it has the correct dimensions.
If we now minimize \eqref{eq:en_fct_of_a} with the choice 
\begin{eqnarray}
a = \frac{\sqrt{ {E_0}^2+{c_1}^2}-{E_0}}{{c_1}}
\end{eqnarray}
where
\begin{eqnarray}
c_{1}\equiv\langle \text{GS}|A_{1}\rangle+\langle A_{1}|\text{GS}\rangle,
\end{eqnarray}
we have
\begin{eqnarray}
E'_{1,0}\leq\frac{\langle\psi|H_{1}'|\psi\rangle}{\langle\psi|\psi\rangle}=\frac{1}{2} \left(-\sqrt{{E_0}^2+{c_1}^2}+{E_0}\right)\leq E_{0},
\end{eqnarray}
since $E_{0}$ is negative. If $c_1\neq0$ the strict inequality $E'_{1,0}<E_{0}$
holds. To summarize, we have shown that adding column $A_{1}$ and
row $A_{1}^{\dagger}$ decreases the ground state energy, provided
$c_{1}$ is not zero. Now we have a matrix of shape similar to the
one before. In the next step we add column and row $A_{2}$, $A_{2}^{T}$, once more $B_{22} = 0$.
We will find the inequality
\begin{eqnarray}
E'_{2,0}\leq E'_{1,0}.
\end{eqnarray}
By iterating this argument a further $|S'|-2$ times we prove that
\begin{eqnarray}
E'_{0}\equiv E'_{|S'|,0}\leq E_{0},
\end{eqnarray}
i.e. the ground state energy of $H'$ is lower or equal to that of $H$. For the ground state energy of $H'$ to be strictly lower than $H$,
one of the $c_{i}$ has to be nonzero. This is the case, provided
that the ground state of $H$ has a finite overlap with at least one of the
states in $S'$. 

\section{Fermionic statistics in the gauge-invariant spin model} \label{appa}
In the original formulation of Sec. \ref{sec:model}, the $\sigma^z$ operator in the hopping term of the Hamiltonian \eqref{eq:Hf} assigns a phase of $0$ or $\pi$ to the hopping amplitude of the fermion, depending on the state of the gauge field on that link. When a fermion is carried all the way around a closed loop $\mathcal{C}$, it picks up the Aharonov-Bohm phase given by the operator $e^{i \hat{\Phi}}=\prod_{\mathcal{C}}\sigma^z$. This is related by Stokes' theorem to the total $\mathbb{Z}_2$ magnetic flux piercing the surface enclosed by $\mathcal{C}$. 

Here we show how the hopping operators \eqref{eq:projector_hopping_h} and \eqref{eq:projector_hopping_v} encode in the gauge-invariant spin formulation the fermionic statistics of the $\mathbb{Z}_2$ charges.
\begin{figure}[h]
	\includegraphics[width=0.5\linewidth]{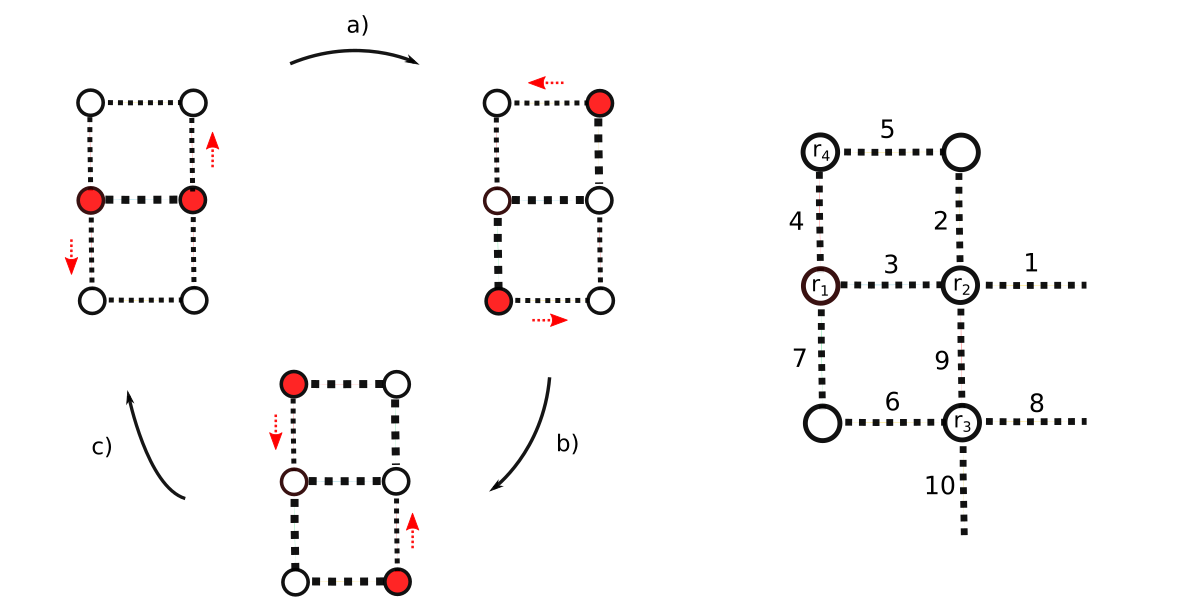}
	\caption{Left: The positions of two identical $\mathbb{Z}_2$ charges (red blobs) are exchanged by successively applying the hopping operators \eqref{eq:projector_hopping_h} and \eqref{eq:projector_hopping_v}. Right: the notation adopted in Eq. \eqref{Bflux}.}
\label{fig:statistics}
\end{figure}
In general, one expects that after an exchange of two identical $\mathbb{Z}_2$-charged fermionic particles the initial state evolves as
\begin{equation}
|\psi_0\rangle \longrightarrow e^{i({\alpha+ \hat{\Phi}})} |\psi_0\rangle,
\end{equation}
where $\alpha=\pi$ is the fermion statistical phase while $\hat{\Phi}$ is the operator that measures the phase acquired due to the magnetic flux as explained above.

For a generic initial state of two neighboring particles, the braiding process shown in Fig. \ref{fig:statistics} can be represented as the action of a braiding operator $\mathcal{B}$ constructed by combining the appropriate hoppings \eqref{eq:projector_hopping_h} and \eqref{eq:projector_hopping_v}:  $|\psi_0\rangle \longrightarrow \mathcal{B} \,|\psi_0\rangle$. Following the notation of Fig. \ref{fig:statistics} for the labeling of the relevant links and sites, one has
\beq \label{Bflux}
\begin{split} 
\mathcal{B}&=Z_9X_8\,Z_4X_3\,Z_6X_{10}\,Z_5X_2\,Z_7X_6\,Z_2X_1 \\
&=-Z_9Z_4Z_6Z_5Z_7Z_2\,X_8X_3X_{10}X_2X_6X_1 \\
&=-\tilde{P}_{\rr_{3}}\tilde{P}_{\rr_{2}}S_{\rr_3}S_{\rr_2},
\end{split}
\eeq
where $\tilde{P}_\rr$ denotes a plaquette operator of $Z$ spins (not of $\sigma^z!$) at the top left of site $\rr$, while $S_\rr = \prod_{b \in +_{\rr}}X_b$. To isolate the contribution from the $\mathbb{Z}_2$ flux, we express $\tilde{P}$ in terms of the operators $X$ and $\sigma^z$ by inverting Eq. \eqref{eq:plaq_mapping_2}. We get
\begin{equation}
\mathcal{B}= -\left( \prod_{\mathcal{C}}\sigma^z \right)S_{\rr_1}S_{\rr_2}S_{\rr_3}S_{\rr_4}= -\prod_{\mathcal{C}}\sigma^z,
\end{equation}
where in the last step we have used the fact that $S_{\rr_1}=S_{\rr_2}=-1$ and $S_{\rr_3}=S_{\rr_4}=1$ for the state $|\psi_0\rangle$. This completes the proof: besides the Bohm-Aharonov phase, an extra minus sign manifests the fermionic nature of the $\mathbb{Z}_2$ charges of the original gauge theory.

\section{Fermion correlator in the gauge-invariant spin formulation}
\label{fermspin}
The bare equal-time fermionic two-point function $\langle c^\dagger_{\rr} c_{\rr'} \rangle$ is not a gauge invariant object, and therefore it cannot have a finite expectation value due to Elitzur's theorem. We now show how its natural gauge-invariant generalization
\begin{equation} \label{fermcorr}
\langle f^\dagger_{\rr} f^{\vphantom{\dagger}}_{\rr'} \rangle=\langle c^\dagger_{\rr} \prod_{b \in \textit{l}} \sigma^z_b c_{\rr'} \rangle
\end{equation}
is expressed in the spin language. For concreteness, we consider sites $\rr$ and $\rr'$ separated in the $x$ direction, such that the Wilson line $\textit{l}$ connecting them is straight and horizontal. Let us see first how an infinite Wilson line transforms, so that first one does not need to worry about the endpoints. In order to use the mapping \eqref{eq:mapping}, we insert at each site crossed by the Wilson line the identity operator in the form $\mathit{I} = \gamma^2 \tilde{\gamma}^2$. One then gets
\begin{equation}
\prod_{b \in \textit{l}} \sigma^z_b = \dots \tilde{\gamma}_{\rr_1}\underbrace{\tilde{\gamma}_{\rr_1}\sigma^z_{\rr_1,\hat{x}}\gamma_{\rr_2}}_{i Z_{\rr_1,\hat{x}}X_{\rr_2,-\hat{y}}}\,\,\underbrace{\gamma_{\rr_2}\tilde{\gamma}_{\rr_2}}_{i S_{\rr_2}}\tilde{\gamma}_{\rr_2}\dots
\end{equation}
where $S_{\rr_i}$ is the star operator at site $\rr_i$. The $X$ operators on horizontal and lower vertical links square to one, so for the infinite Wilson line the mapping takes the simple form
\begin{equation}
\sigma^z_{\rr,\hat{x}} \longrightarrow Z_{\rr,\hat{x}}\,X_{\rr,\hat{y}}.
\end{equation}
Now we explain how to do the mapping for the gauge-invariant fermionic correlators.
At the endpoints, where fermions reside, not all the $X$ operators cancel and therefore the mapping needs to be complemented with the following endpoint rules
\begin{align*}
\gamma_{\rr}&\longrightarrow X_{\rr,-\hat{x}}\,X_{\rr,-\hat{y}}, \\ \numberthis
\tilde{\gamma}_{\rr}&\longrightarrow X_{\rr,\hat{x}}\,X_{\rr,\hat{y}}.
\end{align*}
From these building blocks one can easily reconstruct all fermionic two-point functions. An example is given in Fig. \ref{fig:correlator}.

\begin{figure}[h]
	\includegraphics[width=0.9\linewidth]{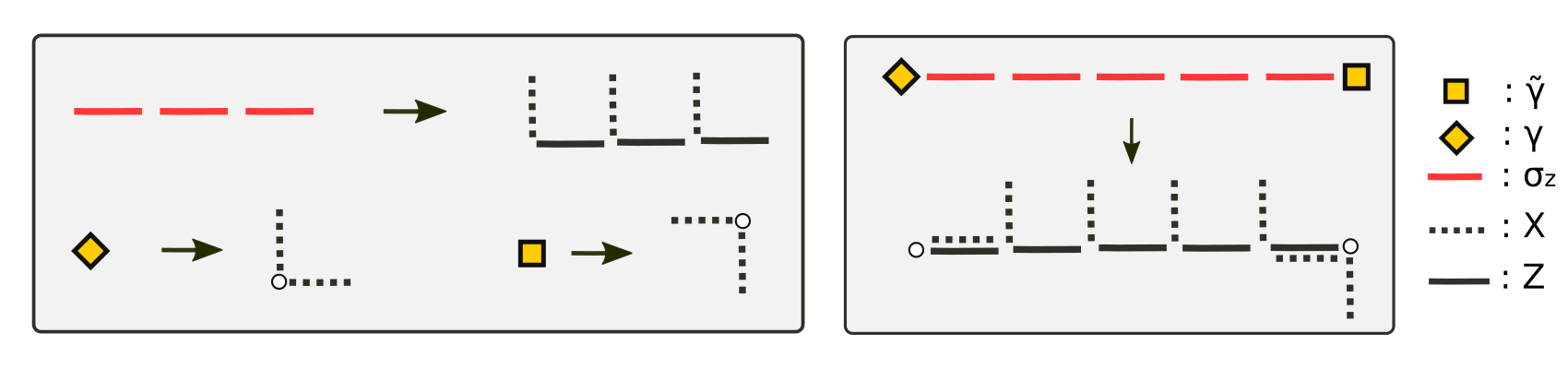}
	\caption{Left panel displays expressions for the Wilson line and endpoint fermionic operators in terms of gauge invariant spin variables using the mapping \eqref{eq:mapping}. From such building blocks one can construct all fermionic two-point functions. As an example, the mapping of the correlator $\tilde{\gamma}\,\sigma^z\dots \gamma$ is shown in the right panel.}
\label{fig:correlator}
\end{figure}



\section{Dimer operators in the gauge-invariant spin formulation}
\label{app:dimers}
In this appendix we show how the gauge-invariant dimer operator $b^\dagger_{\mathbf{r}, \eta}= c^\dagger_{\mathbf{r}} \sigma_{\mathbf{r},\eta} c^\dagger_{\mathbf{r}+\eta}$ can be expressed in terms of gauge-invariant spin operators introducing in Sec. \ref{sec:spinmap}. First, we write it in terms of the Majorana operators
\beq \label{ds}
\begin{split}
b^\dagger_{\mathbf{r}, \eta}=&\frac 1 4 (\gamma_{\mathbf{r}}-i \tilde \gamma_{\mathbf{r}}) \sigma^z_{\mathbf{r}, \eta}  (\gamma_{\mathbf{r}+\eta}-i \tilde \gamma_{\mathbf{r}+\eta}) \\
=& \frac {-i} {4}   \left(\tilde \gamma_{\mathbf{r}}\sigma^z_{\mathbf{r}, \eta} \gamma_{\mathbf{r}+\eta}+ \gamma_{\mathbf{r}}\sigma^z_{\mathbf{r}, \eta} \tilde \gamma_{\mathbf{r}+\eta}\right) 
+\frac 1 4 \left( \gamma_{\mathbf{r}}\sigma^z_{\mathbf{r}, \eta} \gamma_{\mathbf{r}+\eta}- \tilde \gamma_{\mathbf{r}}\sigma^z_{\mathbf{r}, \eta} \tilde \gamma_{\mathbf{r}+\eta}\right).
\end{split}
\eeq
Using our calculation of the hopping part of the Hamiltonian from Sec. \ref{sec:spinmap}, we  find that the first bracketed summand above is just
\beq
\begin{split}
\eta&=\hat x: \qquad \frac 1 2 Z_{\rr,\hat{x}}\,X_{\rr+\hat{x},-\hat{y}}\,\, \mathcal{\tilde P}_{\rr,\hat{x}}, \\
\eta&=\hat y: \qquad \frac 1 2 Z_{\rr,\hat{y}}\,X_{\rr,\hat{x}}\,\,\mathcal{\tilde P}_{\rr,\hat{y}}, \\
\end{split}
\eeq
where we introduced 
\beq
\mathcal{\tilde P}_{\rr,\hat{\eta}}=\frac 1 2 \left( 1+\prod_{b \in +_{\rr}}X_{b} \prod_{b' \in +_{\rr+\eta}}X_{b'}\right)
\eeq
 which annihilates states with opposite fermion parity on sites $\rr$ and $\rr+\eta$. The spin representation of the second bracketed summand in Eq. \eqref{ds} can be computed by using the Gauss law, we find
 \beq
 \begin{split}
 \frac 1 4 \left( \gamma_{\mathbf{r}}\sigma^z_{\mathbf{r}, \hat x} \gamma_{\mathbf{r}+\hat x}- \tilde \gamma_{\mathbf{r}}\sigma^z_{\mathbf{r}, \hat x} \tilde \gamma_{\mathbf{r}+\hat x}\right)&=\frac i 2 Y_{\mathbf{r}, \hat x} X_{\mathbf{r}+\hat x, \hat x} X_{\mathbf{r}+\hat x, \hat y}  \mathcal{\tilde P}_{\rr,\hat{x}}, \\
 \frac 1 4 \left( \gamma_{\mathbf{r}}\sigma^z_{\mathbf{r}, \hat y} \gamma_{\mathbf{r}+\hat y}- \tilde \gamma_{\mathbf{r}}\sigma^z_{\mathbf{r}, \hat y} \tilde \gamma_{\mathbf{r}+\hat y}\right)&=\frac i 2 Y_{\mathbf{r}, \hat y} X_{\mathbf{r}, -\hat x} X_{\mathbf{r}, -\hat y}  \mathcal{\tilde P}_{\rr,\hat{x}}.
 \end{split}
 \eeq
Combining everything together we write the dimer creation operators in the spin language
\beq
\begin{split}
b^\dagger_{\mathbf{r}, \hat x}&=\frac 1 2 \big( Z_{\rr,\hat{x}}\,X_{\rr+\hat{x},-\hat{y}}+i Y_{\mathbf{r}, \hat x} X_{\mathbf{r}+\hat x, \hat x} X_{\mathbf{r}+\hat x, \hat y} \big)\,\, \mathcal{\tilde P}_{\rr,\hat{x}} 
=Z_{\rr,\hat{x}}\,X_{\rr+\hat{x},-\hat{y}} \Pi_{\rr,\hat{x}}, \\
b^\dagger_{\mathbf{r}, \hat y}&=\frac 1 2 \big( Z_{\rr,\hat{y}}\,X_{\rr,\hat{x}}+i Y_{\mathbf{r}, \hat y} X_{\mathbf{r}, -\hat x} X_{\mathbf{r}, -\hat y} \big)\,\, \mathcal{\tilde P}_{\rr,\hat{y}} 
=Z_{\rr,\hat{y}}\,X_{\rr,\hat{x}} \Pi_{\rr,\hat{y}}, \\
\end{split}
\eeq
where
\beq
\Pi_{\rr,\eta}=\frac 1 4 \big(1+\prod_{b \in +_{\rr}}X_{b} + \prod_{b' \in +_{\rr+\eta}}X_{b'}+\prod_{b \in +_{\rr}}X_{b} \prod_{b' \in +_{\rr+\eta}}X_{b'} \big)
\eeq
is a projector on simultaneously unoccupied sites $\rr$ and $\rr+\eta$.
The annihilation operators of dimers can be obtained by hermitian conjugation
\beq
\begin{split}
b_{\mathbf{r}, \hat x}&=Z_{\rr,\hat{x}}\,X_{\rr+\hat{x},-\hat{y}} \tilde \Pi_{\rr,\hat{x}}, \\
b_{\mathbf{r}, \hat y}&=Z_{\rr,\hat{y}}\,X_{\rr,\hat{x}} \tilde \Pi_{\rr,\hat{y}}, \\
\end{split}
\eeq
where
\beq
\tilde \Pi_{\rr,\eta}=\frac 1 4 \big(1-\prod_{b \in +_{\rr}}X_{b} - \prod_{b' \in +_{\rr+\eta}}X_{b'}+\prod_{b \in +_{\rr}}X_{b} \prod_{b' \in +_{\rr+\eta}}X_{b'} \big)
\eeq
is a projector on simultaneously occupied sites $\rr$ and $\rr+\eta$.


\section{Entanglement entropy at $h=0$} \label{appDMRG}
We present here numerical evidence of the phase discussed in Secs. \ref{sec:h_0} and \ref{ssec:h0}, where free Dirac fermions coexist with deconfined $\mathbb{Z}_2$ gauge fields. In particular, we present here results for the entanglement entropy $S$ under a bipartition on an infinite cylinder. At $h=0$, according to Eq. \eqref{eq:S_free_gauge} we expect it to be a simple sum of the fermionic entropy $S_f$ and the gauge field contribution $S_{\mathbb{Z}_2}$. In the thermodynamic limit the entropy $S_f$ is unbounded because spinless fermions in a $\pi$-flux background form two Dirac cones that are at neutrality point at half filling. On the other hand, in a cylinder geometry quantization of momentum in the $y$-direction implies the existence of a finite size gap and the resulting entanglement entropy $S_f$ is therefore finite\footnote{To be more precise, for a given $L_y$, whether a gap is present or not depends on the boundary conditions chosen for the fermions (periodic or antiperiodic). The latter can be interpreted as a the presence of a $\pi$ flux threading the cylinder. We observe that one of the two sectors remains gapless while the other one is gapped. Our numerics show that the absolute ground state always belongs to the gapped sector.}. In Table \ref{tab:table1} we show that the formula \eqref{eq:S_free_gauge} works very well for cylinders of size up to $L_y = 6$. This is a clear signature of a topologically ordered Dirac semimetal phase around half filling at $J\ll t$.

\begin{table}[h!]
  \begin{center}
    \begin{tabular}{l|c|c|c|r} 
      $L_y$ & $\chi$ & $S_f+ S_{\mathbb{Z}_2}$ & $S$ & Rel. Error \\
      \hline
      2 & 400 & 1.03972 & 1,03972 & 0.00\\
      4 & 1000 & 3.04080 & 3.03225 & $\approx 0.28 \% $\\
      6 & 2000 & 5.05664 & 4.93008 & $\approx 2.5 \% $\\
    \end{tabular}
  \end{center}
    \caption{Comparison between the entanglement entropy of our model at $J=h=0$ at half filling with the predicted result $S = S_{f}+S_{\mathbb{Z}_2}$. The entropy $S_f$ for hopping fermions in the $\pi$-flux background is computed numerically with iDMRG, with an error that is negligible compared to the one of $S$. The gauge contribution $S_{\mathbb{Z}_2}=(L_y-1)\log 2$.}
    \label{tab:table1}
\end{table}

\section{Band structure for a single dimer}
\label{OneDimerBand}

Consider the case of a single dimer hopping in the quantum dimer model \eqref{qdh}. In this problem the magnetic $J$ term is not relevant, and thus it is sufficient to consider the dimer hopping term only. The unit cell of this problem is the smallest plaquette of the square lattice.
In this cell there are two possible dimer configurations, horizontal and vertical. We denote the creation operators of these states in the unit cell with site $\bm r$ by the operators $c^{\dagger}(\bm{r})$ and $d^{\dagger}(\bm{r})$, respectively.
The dimensionless Hamiltonian takes the form 
\begin{eqnarray}
H/t_{\text{d}}= & -\sum_{\bm{r}}\left[-d^{\dagger}(\bm{r})+d^{\dagger}(\bm{r}+\hat x)+d^{\dagger}(\bm{r}-\hat y)-d^{\dagger}(\bm{r}+\hat x-\hat y)\right]c(\bm{r})+\text{h.c.} \nonumber\\
 & -\sum_{\bm{r}}c^{\dagger}(\bm{r}+\hat x)c(\bm{r})+\text{h.c.}-\sum_{\bm{r}}d^{\dagger}(\bm{r}+\hat y)d(\bm{r})+\text{h.c.}
\end{eqnarray}
 After Fourier transforming, we can write this Hamiltonian as 
\begin{eqnarray}
H/t_{\text{d}}= & -\sum_{\bm{k}}\left(\begin{array}{c}
c_{\bm{k}}^{\dagger}\\
d_{\bm{k}}^{\dagger}
\end{array}\right)\left(\begin{array}{cc}
2\cos k_{x} & -1+e^{-ik_{x}}+e^{ik_{y}}-e^{-i(k_{x}-k_{y})}\\
-1+e^{ik_{x}}+e^{-ik_{y}}-e^{i(k_{x}-k_{y})} & 2\cos k_{y}
\end{array}\right)\left(\begin{array}{c}
c_{\bm{k}}\\
d_{\bm{k}}
\end{array}\right).\label{eq:matH}
\end{eqnarray}
For an $L\times L$ lattice with periodic boundary conditions we have
\begin{eqnarray}
\bm{k} & =&(k_{x},k_{y})=\left(\frac{2\pi}{L}n,\frac{2\pi}{L}m \right)\\
n,m & =&0,\dots,L-1.
\end{eqnarray}
The energy is found by diagonalizing the matrix in (\ref{eq:matH}), resulting in two bands.
The lowest band
\begin{eqnarray}
\epsilon_{\bm{k}}^{(1)}=-2t_{\text{d}},
\end{eqnarray}
is flat, implying a macroscopic ground state degeneracy.
The second band has cosine dispersion
\begin{eqnarray}
\epsilon_{\bm{k}}^{(2)}=2t_{\text{d}}\left(1-\cos k_{x}-\cos k_{y}\right).
\end{eqnarray}

The flatness of the lower band has striking consequences. To start, it implies that it is possible to create localized immobile excitations. We can construct such a frozen excitation explicitly as the following linear combination of the four dimer states around a single plaquette
\begin{eqnarray}
F^{\dagger}(\bm r) \equiv  \frac{1}{2}\left[d^{\dagger}(\bm{r})-d^{\dagger}(\bm{r}+\hat x)-c^{\dagger}(\bm{r})+c^{\dagger}(\bm{r}+\hat y)\right].
\end{eqnarray}
The state $|\psi(\bm r)\rangle \equiv F^{\dagger}(\bm r) |0\rangle$ created by this operator can be represented visually as
\begin{eqnarray}
|\psi(\bm r)\rangle  = \frac{1}{2}\left[|\vcenter{\hbox{\includegraphics[height=0.02\textheight]{term8.pdf}}}\rangle
- |\vcenter{\hbox{\includegraphics[height=0.02\textheight]{term9.pdf}}}\rangle
- |\vcenter{\hbox{\includegraphics[height=0.02\textheight]{term7.pdf}}}\rangle
+ |\vcenter{\hbox{\includegraphics[height=0.02\textheight]{term10.pdf}}}\rangle\right].
\end{eqnarray}
When the Hamiltonian acts on this  state, the relative signs of its hopping terms, which have their origin in fermionic statistics, lead to a cancellation of hopping processes that would move the dimer away from the plaquette. Thus the dimer remains localized on this plaquette forever.
As a consequence, it is possible to write down a many-body frozen state by creating well-separated frozen plaquette excitations. Since these dimers do not spread, they cannot interact and thereby are exact eigenstates of the Hamiltonian.
\end{widetext}

\bibliography{library}

\begin{thebibliography}{80}%
\makeatletter
\providecommand \@ifxundefined [1]{%
 \@ifx{#1\undefined}
}%
\providecommand \@ifnum [1]{%
 \ifnum #1\expandafter \@firstoftwo
 \else \expandafter \@secondoftwo
 \fi
}%
\providecommand \@ifx [1]{%
 \ifx #1\expandafter \@firstoftwo
 \else \expandafter \@secondoftwo
 \fi
}%
\providecommand \natexlab [1]{#1}%
\providecommand \enquote  [1]{``#1''}%
\providecommand \bibnamefont  [1]{#1}%
\providecommand \bibfnamefont [1]{#1}%
\providecommand \citenamefont [1]{#1}%
\providecommand \href@noop [0]{\@secondoftwo}%
\providecommand \href [0]{\begingroup \@sanitize@url \@href}%
\providecommand \@href[1]{\@@startlink{#1}\@@href}%
\providecommand \@@href[1]{\endgroup#1\@@endlink}%
\providecommand \@sanitize@url [0]{\catcode `\\12\catcode `\$12\catcode
  `\&12\catcode `\#12\catcode `\^12\catcode `\_12\catcode `\%12\relax}%
\providecommand \@@startlink[1]{}%
\providecommand \@@endlink[0]{}%
\providecommand \url  [0]{\begingroup\@sanitize@url \@url }%
\providecommand \@url [1]{\endgroup\@href {#1}{\urlprefix }}%
\providecommand \urlprefix  [0]{URL }%
\providecommand \Eprint [0]{\href }%
\providecommand \doibase [0]{http://dx.doi.org/}%
\providecommand \selectlanguage [0]{\@gobble}%
\providecommand \bibinfo  [0]{\@secondoftwo}%
\providecommand \bibfield  [0]{\@secondoftwo}%
\providecommand \translation [1]{[#1]}%
\providecommand \BibitemOpen [0]{}%
\providecommand \bibitemStop [0]{}%
\providecommand \bibitemNoStop [0]{.\EOS\space}%
\providecommand \EOS [0]{\spacefactor3000\relax}%
\providecommand \BibitemShut  [1]{\csname bibitem#1\endcsname}%
\let\auto@bib@innerbib\@empty
\bibitem [{\citenamefont {Carroll}(2019)}]{carroll2019spacetime}%
  \BibitemOpen
  \bibfield  {author} {\bibinfo {author} {\bibfnamefont {S.~M.}\ \bibnamefont
  {Carroll}},\ }\href@noop {} {\emph {\bibinfo {title} {Spacetime and
  geometry}}}\ (\bibinfo  {publisher} {Cambridge University Press},\ \bibinfo
  {year} {2019})\BibitemShut {NoStop}%
\bibitem [{\citenamefont {Weinberg}(1995)}]{weinberg1995quantum}%
  \BibitemOpen
  \bibfield  {author} {\bibinfo {author} {\bibfnamefont {S.}~\bibnamefont
  {Weinberg}},\ }\href@noop {} {\emph {\bibinfo {title} {The quantum theory of
  fields}}},\ Vol.~\bibinfo {volume} {2}\ (\bibinfo  {publisher} {Cambridge
  university press},\ \bibinfo {year} {1995})\BibitemShut {NoStop}%
\bibitem [{\citenamefont {Wen}(2004)}]{wenbook}%
  \BibitemOpen
  \bibfield  {author} {\bibinfo {author} {\bibfnamefont {X.}~\bibnamefont
  {Wen}},\ }\href {https://books.google.com/books?id=llnlrfdR4YgC} {\emph
  {\bibinfo {title} {Quantum Field Theory of Many-Body Systems}}},\ Oxford
  Graduate Texts\ (\bibinfo  {publisher} {OUP Oxford},\ \bibinfo {year}
  {2004})\BibitemShut {NoStop}%
\bibitem [{\citenamefont {Fradkin}(2013)}]{Fradkin2013}%
  \BibitemOpen
  \bibfield  {author} {\bibinfo {author} {\bibfnamefont {E.}~\bibnamefont
  {Fradkin}},\ }\href
  {http://books.google.com/books?hl=en\&lr=\&id=x7\_6MX4ye\_wC\&oi=fnd\&pg=PR11\&dq=Field+Theories+of+Condensed+Matter+Physics\&ots=OTMzLv0\_tI\&sig=iT7am\_ANJcpca8PXm7hBPm7Dl1E}
  {\emph {\bibinfo {title} {{Field Theories of Condensed Matter Physics}}}}\
  (\bibinfo  {publisher} {Cambridge University Press},\ \bibinfo {year}
  {2013})\BibitemShut {NoStop}%
\bibitem [{\citenamefont {Zeng}\ \emph {et~al.}(2019)\citenamefont {Zeng},
  \citenamefont {Chen}, \citenamefont {Zhou},\ and\ \citenamefont
  {Wen}}]{zeng2019quantum}%
  \BibitemOpen
  \bibfield  {author} {\bibinfo {author} {\bibfnamefont {B.}~\bibnamefont
  {Zeng}}, \bibinfo {author} {\bibfnamefont {X.}~\bibnamefont {Chen}}, \bibinfo
  {author} {\bibfnamefont {D.-L.}\ \bibnamefont {Zhou}}, \ and\ \bibinfo
  {author} {\bibfnamefont {X.-G.}\ \bibnamefont {Wen}},\ }\href@noop {} {\emph
  {\bibinfo {title} {Quantum Information Meets Quantum Matter: From Quantum
  Entanglement to Topological Phases of Many-Body Systems}}}\ (\bibinfo
  {publisher} {Springer},\ \bibinfo {year} {2019})\BibitemShut {NoStop}%
\bibitem [{\citenamefont {Wilson}(1974)}]{Wilson1974}%
  \BibitemOpen
  \bibfield  {author} {\bibinfo {author} {\bibfnamefont {K.~G.}\ \bibnamefont
  {Wilson}},\ }\href {\doibase 10.1103/PhysRevD.10.2445} {\bibfield  {journal}
  {\bibinfo  {journal} {Phys. Rev. D}\ }\textbf {\bibinfo {volume} {10}},\
  \bibinfo {pages} {2445} (\bibinfo {year} {1974})}\BibitemShut {NoStop}%
\bibitem [{\citenamefont {Dalmonte}\ and\ \citenamefont
  {Montangero}(2016)}]{dalmonte2016lattice}%
  \BibitemOpen
  \bibfield  {author} {\bibinfo {author} {\bibfnamefont {M.}~\bibnamefont
  {Dalmonte}}\ and\ \bibinfo {author} {\bibfnamefont {S.}~\bibnamefont
  {Montangero}},\ }\href@noop {} {\bibfield  {journal} {\bibinfo  {journal}
  {Contemporary Physics}\ }\textbf {\bibinfo {volume} {57}},\ \bibinfo {pages}
  {388} (\bibinfo {year} {2016})}\BibitemShut {NoStop}%
\bibitem [{\citenamefont {Ba{\~n}uls}\ and\ \citenamefont
  {Cichy}(2020)}]{banuls2020review}%
  \BibitemOpen
  \bibfield  {author} {\bibinfo {author} {\bibfnamefont {M.~C.}\ \bibnamefont
  {Ba{\~n}uls}}\ and\ \bibinfo {author} {\bibfnamefont {K.}~\bibnamefont
  {Cichy}},\ }\href@noop {} {\bibfield  {journal} {\bibinfo  {journal} {Reports
  on Progress in Physics}\ }\textbf {\bibinfo {volume} {83}},\ \bibinfo {pages}
  {024401} (\bibinfo {year} {2020})}\BibitemShut {NoStop}%
\bibitem [{\citenamefont {Ba{\~n}uls}\ \emph {et~al.}(2020)\citenamefont
  {Ba{\~n}uls}, \citenamefont {Blatt}, \citenamefont {Catani}, \citenamefont
  {Celi}, \citenamefont {Cirac}, \citenamefont {Dalmonte}, \citenamefont
  {Fallani}, \citenamefont {Jansen}, \citenamefont {Lewenstein}, \citenamefont
  {Montangero}, \citenamefont {Muschik}, \citenamefont {Reznik}, \citenamefont
  {Rico}, \citenamefont {Tagliacozzo}, \citenamefont {Van~Acoleyen},
  \citenamefont {Verstraete}, \citenamefont {Wiese}, \citenamefont {Wingate},
  \citenamefont {Zakrzewski},\ and\ \citenamefont
  {Zoller}}]{banuls2019simulating}%
  \BibitemOpen
  \bibfield  {author} {\bibinfo {author} {\bibfnamefont {M.~C.}\ \bibnamefont
  {Ba{\~n}uls}}, \bibinfo {author} {\bibfnamefont {R.}~\bibnamefont {Blatt}},
  \bibinfo {author} {\bibfnamefont {J.}~\bibnamefont {Catani}}, \bibinfo
  {author} {\bibfnamefont {A.}~\bibnamefont {Celi}}, \bibinfo {author}
  {\bibfnamefont {J.~I.}\ \bibnamefont {Cirac}}, \bibinfo {author}
  {\bibfnamefont {M.}~\bibnamefont {Dalmonte}}, \bibinfo {author}
  {\bibfnamefont {L.}~\bibnamefont {Fallani}}, \bibinfo {author} {\bibfnamefont
  {K.}~\bibnamefont {Jansen}}, \bibinfo {author} {\bibfnamefont
  {M.}~\bibnamefont {Lewenstein}}, \bibinfo {author} {\bibfnamefont
  {S.}~\bibnamefont {Montangero}}, \bibinfo {author} {\bibfnamefont {C.~A.}\
  \bibnamefont {Muschik}}, \bibinfo {author} {\bibfnamefont {B.}~\bibnamefont
  {Reznik}}, \bibinfo {author} {\bibfnamefont {E.}~\bibnamefont {Rico}},
  \bibinfo {author} {\bibfnamefont {L.}~\bibnamefont {Tagliacozzo}}, \bibinfo
  {author} {\bibfnamefont {K.}~\bibnamefont {Van~Acoleyen}}, \bibinfo {author}
  {\bibfnamefont {F.}~\bibnamefont {Verstraete}}, \bibinfo {author}
  {\bibfnamefont {U.-J.}\ \bibnamefont {Wiese}}, \bibinfo {author}
  {\bibfnamefont {M.}~\bibnamefont {Wingate}}, \bibinfo {author} {\bibfnamefont
  {J.}~\bibnamefont {Zakrzewski}}, \ and\ \bibinfo {author} {\bibfnamefont
  {P.}~\bibnamefont {Zoller}},\ }\href {\doibase 10.1140/epjd/e2020-100571-8}
  {\bibfield  {journal} {\bibinfo  {journal} {The European Physical Journal D}\
  }\textbf {\bibinfo {volume} {74}},\ \bibinfo {pages} {165} (\bibinfo {year}
  {2020})}\BibitemShut {NoStop}%
\bibitem [{\citenamefont {Davoudi}\ \emph {et~al.}(2020)\citenamefont
  {Davoudi}, \citenamefont {Hafezi}, \citenamefont {Monroe}, \citenamefont
  {Pagano}, \citenamefont {Seif},\ and\ \citenamefont
  {Shaw}}]{PhysRevResearch.2.023015}%
  \BibitemOpen
  \bibfield  {author} {\bibinfo {author} {\bibfnamefont {Z.}~\bibnamefont
  {Davoudi}}, \bibinfo {author} {\bibfnamefont {M.}~\bibnamefont {Hafezi}},
  \bibinfo {author} {\bibfnamefont {C.}~\bibnamefont {Monroe}}, \bibinfo
  {author} {\bibfnamefont {G.}~\bibnamefont {Pagano}}, \bibinfo {author}
  {\bibfnamefont {A.}~\bibnamefont {Seif}}, \ and\ \bibinfo {author}
  {\bibfnamefont {A.}~\bibnamefont {Shaw}},\ }\href {\doibase
  10.1103/PhysRevResearch.2.023015} {\bibfield  {journal} {\bibinfo  {journal}
  {Phys. Rev. Research}\ }\textbf {\bibinfo {volume} {2}},\ \bibinfo {pages}
  {023015} (\bibinfo {year} {2020})}\BibitemShut {NoStop}%
\bibitem [{\citenamefont {Wegner}(1971)}]{Wegner1971}%
  \BibitemOpen
  \bibfield  {author} {\bibinfo {author} {\bibfnamefont {F.~J.}\ \bibnamefont
  {Wegner}},\ }\href {\doibase 10.1063/1.1665530} {\bibfield  {journal}
  {\bibinfo  {journal} {Journal of Mathematical Physics}\ }\textbf {\bibinfo
  {volume} {12}},\ \bibinfo {pages} {2259} (\bibinfo {year}
  {1971})}\BibitemShut {NoStop}%
\bibitem [{\citenamefont {Savary}\ and\ \citenamefont
  {Balents}(2016)}]{savary2016quantum}%
  \BibitemOpen
  \bibfield  {author} {\bibinfo {author} {\bibfnamefont {L.}~\bibnamefont
  {Savary}}\ and\ \bibinfo {author} {\bibfnamefont {L.}~\bibnamefont
  {Balents}},\ }\href@noop {} {\bibfield  {journal} {\bibinfo  {journal}
  {Reports on Progress in Physics}\ }\textbf {\bibinfo {volume} {80}},\
  \bibinfo {pages} {016502} (\bibinfo {year} {2016})}\BibitemShut {NoStop}%
\bibitem [{\citenamefont {Broholm}\ \emph {et~al.}(2020)\citenamefont
  {Broholm}, \citenamefont {Cava}, \citenamefont {Kivelson}, \citenamefont
  {Nocera}, \citenamefont {Norman},\ and\ \citenamefont
  {Senthil}}]{broholm2020quantum}%
  \BibitemOpen
  \bibfield  {author} {\bibinfo {author} {\bibfnamefont {C.}~\bibnamefont
  {Broholm}}, \bibinfo {author} {\bibfnamefont {R.}~\bibnamefont {Cava}},
  \bibinfo {author} {\bibfnamefont {S.}~\bibnamefont {Kivelson}}, \bibinfo
  {author} {\bibfnamefont {D.}~\bibnamefont {Nocera}}, \bibinfo {author}
  {\bibfnamefont {M.}~\bibnamefont {Norman}}, \ and\ \bibinfo {author}
  {\bibfnamefont {T.}~\bibnamefont {Senthil}},\ }\href@noop {} {\bibfield
  {journal} {\bibinfo  {journal} {Science}\ }\textbf {\bibinfo {volume} {367}}
  (\bibinfo {year} {2020})}\BibitemShut {NoStop}%
\bibitem [{\citenamefont {{Samajdar}}\ \emph {et~al.}(2020)\citenamefont
  {{Samajdar}}, \citenamefont {{Ho}}, \citenamefont {{Pichler}}, \citenamefont
  {{Lukin}},\ and\ \citenamefont {{Sachdev}}}]{2020arXiv201112295S}%
  \BibitemOpen
  \bibfield  {author} {\bibinfo {author} {\bibfnamefont {R.}~\bibnamefont
  {{Samajdar}}}, \bibinfo {author} {\bibfnamefont {W.~W.}\ \bibnamefont
  {{Ho}}}, \bibinfo {author} {\bibfnamefont {H.}~\bibnamefont {{Pichler}}},
  \bibinfo {author} {\bibfnamefont {M.~D.}\ \bibnamefont {{Lukin}}}, \ and\
  \bibinfo {author} {\bibfnamefont {S.}~\bibnamefont {{Sachdev}}},\ }\href@noop
  {} {\bibfield  {journal} {\bibinfo  {journal} {arXiv:2011.12295}\ } (\bibinfo
  {year} {2020})}\BibitemShut {NoStop}%
\bibitem [{\citenamefont {{Verresen}}\ \emph {et~al.}(2020)\citenamefont
  {{Verresen}}, \citenamefont {{Lukin}},\ and\ \citenamefont
  {{Vishwanath}}}]{2020arXiv201112310V}%
  \BibitemOpen
  \bibfield  {author} {\bibinfo {author} {\bibfnamefont {R.}~\bibnamefont
  {{Verresen}}}, \bibinfo {author} {\bibfnamefont {M.~D.}\ \bibnamefont
  {{Lukin}}}, \ and\ \bibinfo {author} {\bibfnamefont {A.}~\bibnamefont
  {{Vishwanath}}},\ }\href@noop {} {\bibfield  {journal} {\bibinfo  {journal}
  {arXiv:2011.12310}\ } (\bibinfo {year} {2020})}\BibitemShut {NoStop}%
\bibitem [{\citenamefont {Fradkin}\ and\ \citenamefont
  {Shenker}(1979)}]{Fradkin1979}%
  \BibitemOpen
  \bibfield  {author} {\bibinfo {author} {\bibfnamefont {E.}~\bibnamefont
  {Fradkin}}\ and\ \bibinfo {author} {\bibfnamefont {S.~H.}\ \bibnamefont
  {Shenker}},\ }\href {\doibase 10.1103/PhysRevD.19.3682} {\bibfield  {journal}
  {\bibinfo  {journal} {Phys. Rev. D}\ }\textbf {\bibinfo {volume} {19}},\
  \bibinfo {pages} {3682} (\bibinfo {year} {1979})}\BibitemShut {NoStop}%
\bibitem [{\citenamefont {Tupitsyn}\ \emph {et~al.}(2010)\citenamefont
  {Tupitsyn}, \citenamefont {Kitaev}, \citenamefont {Prokofiev},\ and\
  \citenamefont {Stamp}}]{tupitsyn2010topological}%
  \BibitemOpen
  \bibfield  {author} {\bibinfo {author} {\bibfnamefont {I.}~\bibnamefont
  {Tupitsyn}}, \bibinfo {author} {\bibfnamefont {A.}~\bibnamefont {Kitaev}},
  \bibinfo {author} {\bibfnamefont {N.}~\bibnamefont {Prokofiev}}, \ and\
  \bibinfo {author} {\bibfnamefont {P.}~\bibnamefont {Stamp}},\ }\href@noop {}
  {\bibfield  {journal} {\bibinfo  {journal} {Physical Review B}\ }\textbf
  {\bibinfo {volume} {82}},\ \bibinfo {pages} {085114} (\bibinfo {year}
  {2010})}\BibitemShut {NoStop}%
\bibitem [{\citenamefont {Trebst}\ \emph {et~al.}(2007)\citenamefont {Trebst},
  \citenamefont {Werner}, \citenamefont {Troyer}, \citenamefont {Shtengel},\
  and\ \citenamefont {Nayak}}]{PhysRevLett.98.070602}%
  \BibitemOpen
  \bibfield  {author} {\bibinfo {author} {\bibfnamefont {S.}~\bibnamefont
  {Trebst}}, \bibinfo {author} {\bibfnamefont {P.}~\bibnamefont {Werner}},
  \bibinfo {author} {\bibfnamefont {M.}~\bibnamefont {Troyer}}, \bibinfo
  {author} {\bibfnamefont {K.}~\bibnamefont {Shtengel}}, \ and\ \bibinfo
  {author} {\bibfnamefont {C.}~\bibnamefont {Nayak}},\ }\href {\doibase
  10.1103/PhysRevLett.98.070602} {\bibfield  {journal} {\bibinfo  {journal}
  {Phys. Rev. Lett.}\ }\textbf {\bibinfo {volume} {98}},\ \bibinfo {pages}
  {070602} (\bibinfo {year} {2007})}\BibitemShut {NoStop}%
\bibitem [{\citenamefont {Kitaev}(2006)}]{Kitaev2006}%
  \BibitemOpen
  \bibfield  {author} {\bibinfo {author} {\bibfnamefont {A.}~\bibnamefont
  {Kitaev}},\ }\href {\doibase 10.1016/j.aop.2005.10.005} {\bibfield  {journal}
  {\bibinfo  {journal} {Ann. Phys. (N. Y).}\ }\textbf {\bibinfo {volume}
  {321}},\ \bibinfo {pages} {2} (\bibinfo {year} {2006})}\BibitemShut {NoStop}%
\bibitem [{\citenamefont {Senthil}\ and\ \citenamefont
  {Fisher}(2000)}]{Senthil2000}%
  \BibitemOpen
  \bibfield  {author} {\bibinfo {author} {\bibfnamefont {T.}~\bibnamefont
  {Senthil}}\ and\ \bibinfo {author} {\bibfnamefont {M.~P.~A.}\ \bibnamefont
  {Fisher}},\ }\href {\doibase 10.1103/PhysRevB.62.7850} {\bibfield  {journal}
  {\bibinfo  {journal} {Phys. Rev. B}\ }\textbf {\bibinfo {volume} {62}},\
  \bibinfo {pages} {7850} (\bibinfo {year} {2000})}\BibitemShut {NoStop}%
\bibitem [{\citenamefont {Nandkishore}\ \emph {et~al.}(2012)\citenamefont
  {Nandkishore}, \citenamefont {Metlitski},\ and\ \citenamefont
  {Senthil}}]{Nandkishore2012}%
  \BibitemOpen
  \bibfield  {author} {\bibinfo {author} {\bibfnamefont {R.}~\bibnamefont
  {Nandkishore}}, \bibinfo {author} {\bibfnamefont {M.~A.}\ \bibnamefont
  {Metlitski}}, \ and\ \bibinfo {author} {\bibfnamefont {T.}~\bibnamefont
  {Senthil}},\ }\href {\doibase 10.1103/PhysRevB.86.045128} {\bibfield
  {journal} {\bibinfo  {journal} {Phys. Rev. B}\ }\textbf {\bibinfo {volume}
  {86}},\ \bibinfo {pages} {045128} (\bibinfo {year} {2012})}\BibitemShut
  {NoStop}%
\bibitem [{\citenamefont {Gazit}\ \emph {et~al.}(2017)\citenamefont {Gazit},
  \citenamefont {Randeria},\ and\ \citenamefont {Vishwanath}}]{Gazit2017}%
  \BibitemOpen
  \bibfield  {author} {\bibinfo {author} {\bibfnamefont {S.}~\bibnamefont
  {Gazit}}, \bibinfo {author} {\bibfnamefont {M.}~\bibnamefont {Randeria}}, \
  and\ \bibinfo {author} {\bibfnamefont {A.}~\bibnamefont {Vishwanath}},\
  }\href@noop {} {\bibfield  {journal} {\bibinfo  {journal} {Nature Physics}\
  }\textbf {\bibinfo {volume} {13}},\ \bibinfo {pages} {484} (\bibinfo {year}
  {2017})}\BibitemShut {NoStop}%
\bibitem [{\citenamefont {Gazit}\ \emph {et~al.}(2018)\citenamefont {Gazit},
  \citenamefont {Assaad}, \citenamefont {Sachdev}, \citenamefont {Vishwanath},\
  and\ \citenamefont {Wang}}]{Gazit2018}%
  \BibitemOpen
  \bibfield  {author} {\bibinfo {author} {\bibfnamefont {S.}~\bibnamefont
  {Gazit}}, \bibinfo {author} {\bibfnamefont {F.~F.}\ \bibnamefont {Assaad}},
  \bibinfo {author} {\bibfnamefont {S.}~\bibnamefont {Sachdev}}, \bibinfo
  {author} {\bibfnamefont {A.}~\bibnamefont {Vishwanath}}, \ and\ \bibinfo
  {author} {\bibfnamefont {C.}~\bibnamefont {Wang}},\ }\href {\doibase
  10.1073/pnas.1806338115} {\bibfield  {journal} {\bibinfo  {journal}
  {Proceedings of the National Academy of Sciences}\ }\textbf {\bibinfo
  {volume} {115}},\ \bibinfo {pages} {E6987} (\bibinfo {year}
  {2018})}\BibitemShut {NoStop}%
\bibitem [{\citenamefont {K\"onig}\ \emph {et~al.}(2020)\citenamefont
  {K\"onig}, \citenamefont {Coleman},\ and\ \citenamefont
  {Tsvelik}}]{konig2019soluble}%
  \BibitemOpen
  \bibfield  {author} {\bibinfo {author} {\bibfnamefont {E.~J.}\ \bibnamefont
  {K\"onig}}, \bibinfo {author} {\bibfnamefont {P.}~\bibnamefont {Coleman}}, \
  and\ \bibinfo {author} {\bibfnamefont {A.~M.}\ \bibnamefont {Tsvelik}},\
  }\href {\doibase 10.1103/PhysRevB.102.155143} {\bibfield  {journal} {\bibinfo
   {journal} {Phys. Rev. B}\ }\textbf {\bibinfo {volume} {102}},\ \bibinfo
  {pages} {155143} (\bibinfo {year} {2020})}\BibitemShut {NoStop}%
\bibitem [{\citenamefont {Assaad}\ and\ \citenamefont
  {Grover}(2016)}]{Grover2016}%
  \BibitemOpen
  \bibfield  {author} {\bibinfo {author} {\bibfnamefont {F.~F.}\ \bibnamefont
  {Assaad}}\ and\ \bibinfo {author} {\bibfnamefont {T.}~\bibnamefont
  {Grover}},\ }\href {\doibase 10.1103/PhysRevX.6.041049} {\bibfield  {journal}
  {\bibinfo  {journal} {Phys. Rev. X}\ }\textbf {\bibinfo {volume} {6}},\
  \bibinfo {pages} {041049} (\bibinfo {year} {2016})}\BibitemShut {NoStop}%
\bibitem [{\citenamefont {Prosko}\ \emph {et~al.}(2017)\citenamefont {Prosko},
  \citenamefont {Lee},\ and\ \citenamefont {Maciejko}}]{Prosko2017}%
  \BibitemOpen
  \bibfield  {author} {\bibinfo {author} {\bibfnamefont {C.}~\bibnamefont
  {Prosko}}, \bibinfo {author} {\bibfnamefont {S.-P.}\ \bibnamefont {Lee}}, \
  and\ \bibinfo {author} {\bibfnamefont {J.}~\bibnamefont {Maciejko}},\ }\href
  {\doibase 10.1103/PhysRevB.96.205104} {\bibfield  {journal} {\bibinfo
  {journal} {Phys. Rev. B}\ }\textbf {\bibinfo {volume} {96}},\ \bibinfo
  {pages} {205104} (\bibinfo {year} {2017})}\BibitemShut {NoStop}%
\bibitem [{\citenamefont {Smith}\ \emph {et~al.}(2017)\citenamefont {Smith},
  \citenamefont {Knolle}, \citenamefont {Kovrizhin},\ and\ \citenamefont
  {Moessner}}]{Smith2017}%
  \BibitemOpen
  \bibfield  {author} {\bibinfo {author} {\bibfnamefont {A.}~\bibnamefont
  {Smith}}, \bibinfo {author} {\bibfnamefont {J.}~\bibnamefont {Knolle}},
  \bibinfo {author} {\bibfnamefont {D.~L.}\ \bibnamefont {Kovrizhin}}, \ and\
  \bibinfo {author} {\bibfnamefont {R.}~\bibnamefont {Moessner}},\ }\href
  {\doibase 10.1103/PhysRevLett.118.266601} {\bibfield  {journal} {\bibinfo
  {journal} {Phys. Rev. Lett.}\ }\textbf {\bibinfo {volume} {118}},\ \bibinfo
  {pages} {266601} (\bibinfo {year} {2017})}\BibitemShut {NoStop}%
\bibitem [{\citenamefont {Gonz\'alez-Cuadra}\ \emph {et~al.}(2020)\citenamefont
  {Gonz\'alez-Cuadra}, \citenamefont {Tagliacozzo}, \citenamefont
  {Lewenstein},\ and\ \citenamefont {Bermudez}}]{PhysRevX.10.041007}%
  \BibitemOpen
  \bibfield  {author} {\bibinfo {author} {\bibfnamefont {D.}~\bibnamefont
  {Gonz\'alez-Cuadra}}, \bibinfo {author} {\bibfnamefont {L.}~\bibnamefont
  {Tagliacozzo}}, \bibinfo {author} {\bibfnamefont {M.}~\bibnamefont
  {Lewenstein}}, \ and\ \bibinfo {author} {\bibfnamefont {A.}~\bibnamefont
  {Bermudez}},\ }\href {\doibase 10.1103/PhysRevX.10.041007} {\bibfield
  {journal} {\bibinfo  {journal} {Phys. Rev. X}\ }\textbf {\bibinfo {volume}
  {10}},\ \bibinfo {pages} {041007} (\bibinfo {year} {2020})}\BibitemShut
  {NoStop}%
\bibitem [{\citenamefont {{Pozo}}\ \emph {et~al.}(2020)\citenamefont {{Pozo}},
  \citenamefont {{Rao}}, \citenamefont {{Chen}},\ and\ \citenamefont
  {{Sodemann}}}]{2020arXiv201011956P}%
  \BibitemOpen
  \bibfield  {author} {\bibinfo {author} {\bibfnamefont {O.}~\bibnamefont
  {{Pozo}}}, \bibinfo {author} {\bibfnamefont {P.}~\bibnamefont {{Rao}}},
  \bibinfo {author} {\bibfnamefont {C.}~\bibnamefont {{Chen}}}, \ and\ \bibinfo
  {author} {\bibfnamefont {I.}~\bibnamefont {{Sodemann}}},\ }\href@noop {}
  {\bibfield  {journal} {\bibinfo  {journal} {arXiv:2010.11956}\ } (\bibinfo
  {year} {2020})}\BibitemShut {NoStop}%
\bibitem [{\citenamefont {Lai}\ and\ \citenamefont
  {Motrunich}(2011)}]{PhysRevB.84.235148}%
  \BibitemOpen
  \bibfield  {author} {\bibinfo {author} {\bibfnamefont {H.-H.}\ \bibnamefont
  {Lai}}\ and\ \bibinfo {author} {\bibfnamefont {O.~I.}\ \bibnamefont
  {Motrunich}},\ }\href {\doibase 10.1103/PhysRevB.84.235148} {\bibfield
  {journal} {\bibinfo  {journal} {Phys. Rev. B}\ }\textbf {\bibinfo {volume}
  {84}},\ \bibinfo {pages} {235148} (\bibinfo {year} {2011})}\BibitemShut
  {NoStop}%
\bibitem [{\citenamefont {Frank}\ \emph {et~al.}(2020)\citenamefont {Frank},
  \citenamefont {Huffman},\ and\ \citenamefont {Chandrasekharan}}]{Frank_2020}%
  \BibitemOpen
  \bibfield  {author} {\bibinfo {author} {\bibfnamefont {J.}~\bibnamefont
  {Frank}}, \bibinfo {author} {\bibfnamefont {E.}~\bibnamefont {Huffman}}, \
  and\ \bibinfo {author} {\bibfnamefont {S.}~\bibnamefont {Chandrasekharan}},\
  }\href {\doibase 10.1016/j.physletb.2020.135484} {\bibfield  {journal}
  {\bibinfo  {journal} {Physics Letters B}\ }\textbf {\bibinfo {volume}
  {806}},\ \bibinfo {pages} {135484} (\bibinfo {year} {2020})}\BibitemShut
  {NoStop}%
\bibitem [{\citenamefont {Borla}\ \emph {et~al.}(2020)\citenamefont {Borla},
  \citenamefont {Verresen}, \citenamefont {Grusdt},\ and\ \citenamefont
  {Moroz}}]{borla2020confined}%
  \BibitemOpen
  \bibfield  {author} {\bibinfo {author} {\bibfnamefont {U.}~\bibnamefont
  {Borla}}, \bibinfo {author} {\bibfnamefont {R.}~\bibnamefont {Verresen}},
  \bibinfo {author} {\bibfnamefont {F.}~\bibnamefont {Grusdt}}, \ and\ \bibinfo
  {author} {\bibfnamefont {S.}~\bibnamefont {Moroz}},\ }\href@noop {}
  {\bibfield  {journal} {\bibinfo  {journal} {Phys. Rev. Lett.}\ }\textbf
  {\bibinfo {volume} {124}},\ \bibinfo {pages} {120503} (\bibinfo {year}
  {2020})}\BibitemShut {NoStop}%
\bibitem [{\citenamefont {Grusdt}\ and\ \citenamefont
  {Pollet}(2020)}]{grusdt2020z2}%
  \BibitemOpen
  \bibfield  {author} {\bibinfo {author} {\bibfnamefont {F.}~\bibnamefont
  {Grusdt}}\ and\ \bibinfo {author} {\bibfnamefont {L.}~\bibnamefont
  {Pollet}},\ }\href@noop {} {\bibfield  {journal} {\bibinfo  {journal}
  {arXiv:2007.13759}\ } (\bibinfo {year} {2020})}\BibitemShut {NoStop}%
\bibitem [{\citenamefont {Borla}\ \emph {et~al.}(2021)\citenamefont {Borla},
  \citenamefont {Verresen}, \citenamefont {Shah},\ and\ \citenamefont
  {Moroz}}]{borla2020gauging}%
  \BibitemOpen
  \bibfield  {author} {\bibinfo {author} {\bibfnamefont {U.}~\bibnamefont
  {Borla}}, \bibinfo {author} {\bibfnamefont {R.}~\bibnamefont {Verresen}},
  \bibinfo {author} {\bibfnamefont {J.}~\bibnamefont {Shah}}, \ and\ \bibinfo
  {author} {\bibfnamefont {S.}~\bibnamefont {Moroz}},\ }\href@noop {}
  {\bibfield  {journal} {\bibinfo  {journal} {SciPost Physics}\ }\textbf
  {\bibinfo {volume} {10}},\ \bibinfo {pages} {148} (\bibinfo {year}
  {2021})}\BibitemShut {NoStop}%
\bibitem [{\citenamefont {Barbiero}\ \emph {et~al.}(2019)\citenamefont
  {Barbiero}, \citenamefont {Schweizer}, \citenamefont {Aidelsburger},
  \citenamefont {Demler}, \citenamefont {Goldman},\ and\ \citenamefont
  {Grusdt}}]{Barbiero2018}%
  \BibitemOpen
  \bibfield  {author} {\bibinfo {author} {\bibfnamefont {L.}~\bibnamefont
  {Barbiero}}, \bibinfo {author} {\bibfnamefont {C.}~\bibnamefont {Schweizer}},
  \bibinfo {author} {\bibfnamefont {M.}~\bibnamefont {Aidelsburger}}, \bibinfo
  {author} {\bibfnamefont {E.}~\bibnamefont {Demler}}, \bibinfo {author}
  {\bibfnamefont {N.}~\bibnamefont {Goldman}}, \ and\ \bibinfo {author}
  {\bibfnamefont {F.}~\bibnamefont {Grusdt}},\ }\href@noop {} {\bibfield
  {journal} {\bibinfo  {journal} {Science advances}\ }\textbf {\bibinfo
  {volume} {5}},\ \bibinfo {pages} {7444} (\bibinfo {year} {2019})}\BibitemShut
  {NoStop}%
\bibitem [{\citenamefont {Schweizer}\ \emph {et~al.}(2019)\citenamefont
  {Schweizer}, \citenamefont {Grusdt}, \citenamefont {Berngruber},
  \citenamefont {Barbiero}, \citenamefont {Demler}, \citenamefont {Goldman},
  \citenamefont {Bloch},\ and\ \citenamefont {Aidelsburger}}]{Schweizer2019}%
  \BibitemOpen
  \bibfield  {author} {\bibinfo {author} {\bibfnamefont {C.}~\bibnamefont
  {Schweizer}}, \bibinfo {author} {\bibfnamefont {F.}~\bibnamefont {Grusdt}},
  \bibinfo {author} {\bibfnamefont {M.}~\bibnamefont {Berngruber}}, \bibinfo
  {author} {\bibfnamefont {L.}~\bibnamefont {Barbiero}}, \bibinfo {author}
  {\bibfnamefont {E.}~\bibnamefont {Demler}}, \bibinfo {author} {\bibfnamefont
  {N.}~\bibnamefont {Goldman}}, \bibinfo {author} {\bibfnamefont
  {I.}~\bibnamefont {Bloch}}, \ and\ \bibinfo {author} {\bibfnamefont
  {M.}~\bibnamefont {Aidelsburger}},\ }\href {\doibase
  10.1038/s41567-019-0649-7} {\bibfield  {journal} {\bibinfo  {journal} {Nature
  Physics}\ } (\bibinfo {year} {2019}),\ 10.1038/s41567-019-0649-7}\BibitemShut
  {NoStop}%
\bibitem [{\citenamefont {G{\"o}rg}\ \emph {et~al.}(2019)\citenamefont
  {G{\"o}rg}, \citenamefont {Sandholzer}, \citenamefont {Minguzzi},
  \citenamefont {Desbuquois}, \citenamefont {Messer},\ and\ \citenamefont
  {Esslinger}}]{Gorg2019}%
  \BibitemOpen
  \bibfield  {author} {\bibinfo {author} {\bibfnamefont {F.}~\bibnamefont
  {G{\"o}rg}}, \bibinfo {author} {\bibfnamefont {K.}~\bibnamefont
  {Sandholzer}}, \bibinfo {author} {\bibfnamefont {J.}~\bibnamefont
  {Minguzzi}}, \bibinfo {author} {\bibfnamefont {R.}~\bibnamefont
  {Desbuquois}}, \bibinfo {author} {\bibfnamefont {M.}~\bibnamefont {Messer}},
  \ and\ \bibinfo {author} {\bibfnamefont {T.}~\bibnamefont {Esslinger}},\
  }\href {\doibase 10.1038/s41567-019-0615-4} {\bibfield  {journal} {\bibinfo
  {journal} {Nature Physics}\ } (\bibinfo {year} {2019}),\
  10.1038/s41567-019-0615-4}\BibitemShut {NoStop}%
\bibitem [{\citenamefont {Zohar}\ \emph {et~al.}(2017)\citenamefont {Zohar},
  \citenamefont {Farace}, \citenamefont {Reznik},\ and\ \citenamefont
  {Cirac}}]{PhysRevLett.118.070501}%
  \BibitemOpen
  \bibfield  {author} {\bibinfo {author} {\bibfnamefont {E.}~\bibnamefont
  {Zohar}}, \bibinfo {author} {\bibfnamefont {A.}~\bibnamefont {Farace}},
  \bibinfo {author} {\bibfnamefont {B.}~\bibnamefont {Reznik}}, \ and\ \bibinfo
  {author} {\bibfnamefont {J.~I.}\ \bibnamefont {Cirac}},\ }\href {\doibase
  10.1103/PhysRevLett.118.070501} {\bibfield  {journal} {\bibinfo  {journal}
  {Phys. Rev. Lett.}\ }\textbf {\bibinfo {volume} {118}},\ \bibinfo {pages}
  {070501} (\bibinfo {year} {2017})}\BibitemShut {NoStop}%
\bibitem [{\citenamefont {Chulliparambil}\ \emph {et~al.}(2020)\citenamefont
  {Chulliparambil}, \citenamefont {Seifert}, \citenamefont {Vojta},
  \citenamefont {Janssen},\ and\ \citenamefont {Tu}}]{PhysRevB.102.201111}%
  \BibitemOpen
  \bibfield  {author} {\bibinfo {author} {\bibfnamefont {S.}~\bibnamefont
  {Chulliparambil}}, \bibinfo {author} {\bibfnamefont {U.~F.~P.}\ \bibnamefont
  {Seifert}}, \bibinfo {author} {\bibfnamefont {M.}~\bibnamefont {Vojta}},
  \bibinfo {author} {\bibfnamefont {L.}~\bibnamefont {Janssen}}, \ and\
  \bibinfo {author} {\bibfnamefont {H.-H.}\ \bibnamefont {Tu}},\ }\href
  {\doibase 10.1103/PhysRevB.102.201111} {\bibfield  {journal} {\bibinfo
  {journal} {Phys. Rev. B}\ }\textbf {\bibinfo {volume} {102}},\ \bibinfo
  {pages} {201111} (\bibinfo {year} {2020})}\BibitemShut {NoStop}%
\bibitem [{\citenamefont {Chulliparambil}\ \emph {et~al.}(2021)\citenamefont
  {Chulliparambil}, \citenamefont {Janssen}, \citenamefont {Vojta},
  \citenamefont {Tu},\ and\ \citenamefont {Seifert}}]{PhysRevB.103.075144}%
  \BibitemOpen
  \bibfield  {author} {\bibinfo {author} {\bibfnamefont {S.}~\bibnamefont
  {Chulliparambil}}, \bibinfo {author} {\bibfnamefont {L.}~\bibnamefont
  {Janssen}}, \bibinfo {author} {\bibfnamefont {M.}~\bibnamefont {Vojta}},
  \bibinfo {author} {\bibfnamefont {H.-H.}\ \bibnamefont {Tu}}, \ and\ \bibinfo
  {author} {\bibfnamefont {U.~F.~P.}\ \bibnamefont {Seifert}},\ }\href
  {\doibase 10.1103/PhysRevB.103.075144} {\bibfield  {journal} {\bibinfo
  {journal} {Phys. Rev. B}\ }\textbf {\bibinfo {volume} {103}},\ \bibinfo
  {pages} {075144} (\bibinfo {year} {2021})}\BibitemShut {NoStop}%
\bibitem [{\citenamefont {Gazit}\ \emph {et~al.}(2020)\citenamefont {Gazit},
  \citenamefont {Assaad},\ and\ \citenamefont {Sachdev}}]{gazit2019fermi}%
  \BibitemOpen
  \bibfield  {author} {\bibinfo {author} {\bibfnamefont {S.}~\bibnamefont
  {Gazit}}, \bibinfo {author} {\bibfnamefont {F.~F.}\ \bibnamefont {Assaad}}, \
  and\ \bibinfo {author} {\bibfnamefont {S.}~\bibnamefont {Sachdev}},\
  }\href@noop {} {\bibfield  {journal} {\bibinfo  {journal} {Physical Review
  X}\ }\textbf {\bibinfo {volume} {10}},\ \bibinfo {pages} {041057} (\bibinfo
  {year} {2020})}\BibitemShut {NoStop}%
\bibitem [{\citenamefont {White}(1992)}]{PhysRevLett.69.2863}%
  \BibitemOpen
  \bibfield  {author} {\bibinfo {author} {\bibfnamefont {S.~R.}\ \bibnamefont
  {White}},\ }\href {\doibase 10.1103/PhysRevLett.69.2863} {\bibfield
  {journal} {\bibinfo  {journal} {Phys. Rev. Lett.}\ }\textbf {\bibinfo
  {volume} {69}},\ \bibinfo {pages} {2863} (\bibinfo {year}
  {1992})}\BibitemShut {NoStop}%
\bibitem [{\citenamefont {McCulloch}(2008)}]{mcculloch2008infinite}%
  \BibitemOpen
  \bibfield  {author} {\bibinfo {author} {\bibfnamefont {I.~P.}\ \bibnamefont
  {McCulloch}},\ }\href@noop {} {\bibfield  {journal} {\bibinfo  {journal}
  {arXiv:0804.2509}\ } (\bibinfo {year} {2008})}\BibitemShut {NoStop}%
\bibitem [{\citenamefont {Moessner}\ \emph {et~al.}(2001)\citenamefont
  {Moessner}, \citenamefont {Sondhi},\ and\ \citenamefont
  {Fradkin}}]{moessner2001short}%
  \BibitemOpen
  \bibfield  {author} {\bibinfo {author} {\bibfnamefont {R.}~\bibnamefont
  {Moessner}}, \bibinfo {author} {\bibfnamefont {S.~L.}\ \bibnamefont
  {Sondhi}}, \ and\ \bibinfo {author} {\bibfnamefont {E.}~\bibnamefont
  {Fradkin}},\ }\href@noop {} {\bibfield  {journal} {\bibinfo  {journal}
  {Physical Review B}\ }\textbf {\bibinfo {volume} {65}},\ \bibinfo {pages}
  {024504} (\bibinfo {year} {2001})}\BibitemShut {NoStop}%
\bibitem [{\citenamefont {Sachdev}(2018)}]{Sachdev_2018}%
  \BibitemOpen
  \bibfield  {author} {\bibinfo {author} {\bibfnamefont {S.}~\bibnamefont
  {Sachdev}},\ }\href {\doibase 10.1088/1361-6633/aae110} {\bibfield  {journal}
  {\bibinfo  {journal} {Reports on Progress in Physics}\ }\textbf {\bibinfo
  {volume} {82}},\ \bibinfo {pages} {014001} (\bibinfo {year}
  {2018})}\BibitemShut {NoStop}%
\bibitem [{\citenamefont {Huh}\ \emph {et~al.}(2011)\citenamefont {Huh},
  \citenamefont {Punk},\ and\ \citenamefont {Sachdev}}]{PhysRevB.84.094419}%
  \BibitemOpen
  \bibfield  {author} {\bibinfo {author} {\bibfnamefont {Y.}~\bibnamefont
  {Huh}}, \bibinfo {author} {\bibfnamefont {M.}~\bibnamefont {Punk}}, \ and\
  \bibinfo {author} {\bibfnamefont {S.}~\bibnamefont {Sachdev}},\ }\href
  {\doibase 10.1103/PhysRevB.84.094419} {\bibfield  {journal} {\bibinfo
  {journal} {Phys. Rev. B}\ }\textbf {\bibinfo {volume} {84}},\ \bibinfo
  {pages} {094419} (\bibinfo {year} {2011})}\BibitemShut {NoStop}%
\bibitem [{\citenamefont {Blankschtein}\ \emph {et~al.}(1984)\citenamefont
  {Blankschtein}, \citenamefont {Ma},\ and\ \citenamefont
  {Berker}}]{blankschtein1984fully}%
  \BibitemOpen
  \bibfield  {author} {\bibinfo {author} {\bibfnamefont {D.}~\bibnamefont
  {Blankschtein}}, \bibinfo {author} {\bibfnamefont {M.}~\bibnamefont {Ma}}, \
  and\ \bibinfo {author} {\bibfnamefont {A.~N.}\ \bibnamefont {Berker}},\
  }\href@noop {} {\bibfield  {journal} {\bibinfo  {journal} {Physical Review
  B}\ }\textbf {\bibinfo {volume} {30}},\ \bibinfo {pages} {1362} (\bibinfo
  {year} {1984})}\BibitemShut {NoStop}%
\bibitem [{\citenamefont {Wenzel}\ \emph {et~al.}(2012)\citenamefont {Wenzel},
  \citenamefont {Coletta}, \citenamefont {Korshunov},\ and\ \citenamefont
  {Mila}}]{wenzel2012evidence}%
  \BibitemOpen
  \bibfield  {author} {\bibinfo {author} {\bibfnamefont {S.}~\bibnamefont
  {Wenzel}}, \bibinfo {author} {\bibfnamefont {T.}~\bibnamefont {Coletta}},
  \bibinfo {author} {\bibfnamefont {S.~E.}\ \bibnamefont {Korshunov}}, \ and\
  \bibinfo {author} {\bibfnamefont {F.}~\bibnamefont {Mila}},\ }\href@noop {}
  {\bibfield  {journal} {\bibinfo  {journal} {Physical Review Letters}\
  }\textbf {\bibinfo {volume} {109}},\ \bibinfo {pages} {187202} (\bibinfo
  {year} {2012})}\BibitemShut {NoStop}%
\bibitem [{\citenamefont {Lieb}(1994)}]{Lieb94}%
  \BibitemOpen
  \bibfield  {author} {\bibinfo {author} {\bibfnamefont {E.~H.}\ \bibnamefont
  {Lieb}},\ }\href {\doibase 10.1103/PhysRevLett.73.2158} {\bibfield  {journal}
  {\bibinfo  {journal} {Phys. Rev. Lett.}\ }\textbf {\bibinfo {volume} {73}},\
  \bibinfo {pages} {2158} (\bibinfo {year} {1994})}\BibitemShut {NoStop}%
\bibitem [{\citenamefont {Verresen}\ \emph {et~al.}(2020)\citenamefont
  {Verresen}, \citenamefont {Thorngren}, \citenamefont {Borla}, \citenamefont
  {Moroz},\ and\ \citenamefont {Vishwanath}}]{Higgs}%
  \BibitemOpen
  \bibfield  {author} {\bibinfo {author} {\bibfnamefont {R.}~\bibnamefont
  {Verresen}}, \bibinfo {author} {\bibfnamefont {R.}~\bibnamefont {Thorngren}},
  \bibinfo {author} {\bibfnamefont {U.}~\bibnamefont {Borla}}, \bibinfo
  {author} {\bibfnamefont {S.}~\bibnamefont {Moroz}}, \ and\ \bibinfo {author}
  {\bibfnamefont {A.}~\bibnamefont {Vishwanath}},\ }\href@noop {} {} (\bibinfo
  {year} {2020}),\ \bibinfo {note} {to appear}\BibitemShut {NoStop}%
\bibitem [{\citenamefont {Wen}(2019)}]{PhysRevB.99.205139}%
  \BibitemOpen
  \bibfield  {author} {\bibinfo {author} {\bibfnamefont {X.-G.}\ \bibnamefont
  {Wen}},\ }\href {\doibase 10.1103/PhysRevB.99.205139} {\bibfield  {journal}
  {\bibinfo  {journal} {Phys. Rev. B}\ }\textbf {\bibinfo {volume} {99}},\
  \bibinfo {pages} {205139} (\bibinfo {year} {2019})}\BibitemShut {NoStop}%
\bibitem [{\citenamefont {Pai}\ and\ \citenamefont
  {Pretko}(2020)}]{PhysRevResearch.2.013094}%
  \BibitemOpen
  \bibfield  {author} {\bibinfo {author} {\bibfnamefont {S.}~\bibnamefont
  {Pai}}\ and\ \bibinfo {author} {\bibfnamefont {M.}~\bibnamefont {Pretko}},\
  }\href {\doibase 10.1103/PhysRevResearch.2.013094} {\bibfield  {journal}
  {\bibinfo  {journal} {Phys. Rev. Research}\ }\textbf {\bibinfo {volume}
  {2}},\ \bibinfo {pages} {013094} (\bibinfo {year} {2020})}\BibitemShut
  {NoStop}%
\bibitem [{\citenamefont {Rokhsar}\ and\ \citenamefont
  {Kivelson}(1988)}]{PhysRevLett.61.2376}%
  \BibitemOpen
  \bibfield  {author} {\bibinfo {author} {\bibfnamefont {D.~S.}\ \bibnamefont
  {Rokhsar}}\ and\ \bibinfo {author} {\bibfnamefont {S.~A.}\ \bibnamefont
  {Kivelson}},\ }\href {\doibase 10.1103/PhysRevLett.61.2376} {\bibfield
  {journal} {\bibinfo  {journal} {Phys. Rev. Lett.}\ }\textbf {\bibinfo
  {volume} {61}},\ \bibinfo {pages} {2376} (\bibinfo {year}
  {1988})}\BibitemShut {NoStop}%
\bibitem [{\citenamefont {Nandkishore}\ and\ \citenamefont
  {Hermele}(2019)}]{fractons}%
  \BibitemOpen
  \bibfield  {author} {\bibinfo {author} {\bibfnamefont {R.~M.}\ \bibnamefont
  {Nandkishore}}\ and\ \bibinfo {author} {\bibfnamefont {M.}~\bibnamefont
  {Hermele}},\ }\href {\doibase 10.1146/annurev-conmatphys-031218-013604}
  {\bibfield  {journal} {\bibinfo  {journal} {Annual Review of Condensed Matter
  Physics}\ }\textbf {\bibinfo {volume} {10}},\ \bibinfo {pages} {295}
  (\bibinfo {year} {2019})}\BibitemShut {NoStop}%
\bibitem [{\citenamefont {Pretko}\ \emph {et~al.}(2020)\citenamefont {Pretko},
  \citenamefont {Chen},\ and\ \citenamefont {You}}]{pretko2020fracton}%
  \BibitemOpen
  \bibfield  {author} {\bibinfo {author} {\bibfnamefont {M.}~\bibnamefont
  {Pretko}}, \bibinfo {author} {\bibfnamefont {X.}~\bibnamefont {Chen}}, \ and\
  \bibinfo {author} {\bibfnamefont {Y.}~\bibnamefont {You}},\ }\href@noop {}
  {\bibfield  {journal} {\bibinfo  {journal} {International Journal of Modern
  Physics A}\ }\textbf {\bibinfo {volume} {35}},\ \bibinfo {pages} {2030003}
  (\bibinfo {year} {2020})}\BibitemShut {NoStop}%
\bibitem [{\citenamefont {Yang}\ \emph {et~al.}(2020)\citenamefont {Yang},
  \citenamefont {Liu}, \citenamefont {Gorshkov},\ and\ \citenamefont
  {Iadecola}}]{PhysRevLett.124.207602}%
  \BibitemOpen
  \bibfield  {author} {\bibinfo {author} {\bibfnamefont {Z.-C.}\ \bibnamefont
  {Yang}}, \bibinfo {author} {\bibfnamefont {F.}~\bibnamefont {Liu}}, \bibinfo
  {author} {\bibfnamefont {A.~V.}\ \bibnamefont {Gorshkov}}, \ and\ \bibinfo
  {author} {\bibfnamefont {T.}~\bibnamefont {Iadecola}},\ }\href {\doibase
  10.1103/PhysRevLett.124.207602} {\bibfield  {journal} {\bibinfo  {journal}
  {Phys. Rev. Lett.}\ }\textbf {\bibinfo {volume} {124}},\ \bibinfo {pages}
  {207602} (\bibinfo {year} {2020})}\BibitemShut {NoStop}%
\bibitem [{\citenamefont {Cobanera}\ \emph {et~al.}(2013)\citenamefont
  {Cobanera}, \citenamefont {Ortiz},\ and\ \citenamefont
  {Nussinov}}]{PhysRevB.87.041105}%
  \BibitemOpen
  \bibfield  {author} {\bibinfo {author} {\bibfnamefont {E.}~\bibnamefont
  {Cobanera}}, \bibinfo {author} {\bibfnamefont {G.}~\bibnamefont {Ortiz}}, \
  and\ \bibinfo {author} {\bibfnamefont {Z.}~\bibnamefont {Nussinov}},\ }\href
  {\doibase 10.1103/PhysRevB.87.041105} {\bibfield  {journal} {\bibinfo
  {journal} {Phys. Rev. B}\ }\textbf {\bibinfo {volume} {87}},\ \bibinfo
  {pages} {041105} (\bibinfo {year} {2013})}\BibitemShut {NoStop}%
\bibitem [{\citenamefont {Radicevic}(2018)}]{radicevic2018spin}%
  \BibitemOpen
  \bibfield  {author} {\bibinfo {author} {\bibfnamefont {D.}~\bibnamefont
  {Radicevic}},\ }\href@noop {} {\  (\bibinfo {year} {2018})},\ \Eprint
  {http://arxiv.org/abs/1809.07757} {arXiv:1809.07757} \BibitemShut {NoStop}%
\bibitem [{\citenamefont {Wosiek}(1982)}]{wosiek1982local}%
  \BibitemOpen
  \bibfield  {author} {\bibinfo {author} {\bibfnamefont {J.}~\bibnamefont
  {Wosiek}},\ }\href@noop {} {\bibfield  {journal} {\bibinfo  {journal} {Acta
  Physica Polonica. Series B}\ }\textbf {\bibinfo {volume} {13}},\ \bibinfo
  {pages} {543} (\bibinfo {year} {1982})}\BibitemShut {NoStop}%
\bibitem [{\citenamefont {Chen}\ \emph {et~al.}(2018)\citenamefont {Chen},
  \citenamefont {Kapustin},\ and\ \citenamefont {Radicevic}}]{chen2018exact}%
  \BibitemOpen
  \bibfield  {author} {\bibinfo {author} {\bibfnamefont {Y.-A.}\ \bibnamefont
  {Chen}}, \bibinfo {author} {\bibfnamefont {A.}~\bibnamefont {Kapustin}}, \
  and\ \bibinfo {author} {\bibfnamefont {D.}~\bibnamefont {Radicevic}},\
  }\href@noop {} {\bibfield  {journal} {\bibinfo  {journal} {Annals of
  Physics}\ }\textbf {\bibinfo {volume} {393}},\ \bibinfo {pages} {234}
  (\bibinfo {year} {2018})}\BibitemShut {NoStop}%
\bibitem [{\citenamefont {Bochniak}\ and\ \citenamefont
  {Ruba}(2020)}]{bochniak2020model}%
  \BibitemOpen
  \bibfield  {author} {\bibinfo {author} {\bibfnamefont {A.}~\bibnamefont
  {Bochniak}}\ and\ \bibinfo {author} {\bibfnamefont {B.}~\bibnamefont
  {Ruba}},\ }\href@noop {} {\bibfield  {journal} {\bibinfo  {journal}
  {arXiv:2003.06905}\ } (\bibinfo {year} {2020})}\BibitemShut {NoStop}%
\bibitem [{\citenamefont {Tantivasadakarn}(2020)}]{PhysRevResearch.2.023353}%
  \BibitemOpen
  \bibfield  {author} {\bibinfo {author} {\bibfnamefont {N.}~\bibnamefont
  {Tantivasadakarn}},\ }\href {\doibase 10.1103/PhysRevResearch.2.023353}
  {\bibfield  {journal} {\bibinfo  {journal} {Phys. Rev. Research}\ }\textbf
  {\bibinfo {volume} {2}},\ \bibinfo {pages} {023353} (\bibinfo {year}
  {2020})}\BibitemShut {NoStop}%
\bibitem [{\citenamefont {Rao}\ and\ \citenamefont
  {Sodemann}(2020)}]{rao2020theory}%
  \BibitemOpen
  \bibfield  {author} {\bibinfo {author} {\bibfnamefont {P.}~\bibnamefont
  {Rao}}\ and\ \bibinfo {author} {\bibfnamefont {I.}~\bibnamefont {Sodemann}},\
  }\href@noop {} {\bibfield  {journal} {\bibinfo  {journal} {arXiv:2012.12280}\
  } (\bibinfo {year} {2020})}\BibitemShut {NoStop}%
\bibitem [{\citenamefont {Shirley}(2020)}]{shirley2020fractonic}%
  \BibitemOpen
  \bibfield  {author} {\bibinfo {author} {\bibfnamefont {W.}~\bibnamefont
  {Shirley}},\ }\href@noop {} {\bibfield  {journal} {\bibinfo  {journal}
  {arXiv:2002.12026}\ } (\bibinfo {year} {2020})}\BibitemShut {NoStop}%
\bibitem [{\citenamefont {Zohar}\ and\ \citenamefont
  {Cirac}(2018)}]{PhysRevB.98.075119}%
  \BibitemOpen
  \bibfield  {author} {\bibinfo {author} {\bibfnamefont {E.}~\bibnamefont
  {Zohar}}\ and\ \bibinfo {author} {\bibfnamefont {J.~I.}\ \bibnamefont
  {Cirac}},\ }\href {\doibase 10.1103/PhysRevB.98.075119} {\bibfield  {journal}
  {\bibinfo  {journal} {Phys. Rev. B}\ }\textbf {\bibinfo {volume} {98}},\
  \bibinfo {pages} {075119} (\bibinfo {year} {2018})}\BibitemShut {NoStop}%
\bibitem [{\citenamefont {Hauschild}\ and\ \citenamefont
  {Pollmann}(2018)}]{hauschild2018efficient}%
  \BibitemOpen
  \bibfield  {author} {\bibinfo {author} {\bibfnamefont {J.}~\bibnamefont
  {Hauschild}}\ and\ \bibinfo {author} {\bibfnamefont {F.}~\bibnamefont
  {Pollmann}},\ }\href@noop {} {\bibfield  {journal} {\bibinfo  {journal}
  {SciPost Physics Lecture Notes}\ } (\bibinfo {year} {2018})}\BibitemShut
  {NoStop}%
\bibitem [{\citenamefont {Creutz}(1999)}]{creutz1999end}%
  \BibitemOpen
  \bibfield  {author} {\bibinfo {author} {\bibfnamefont {M.}~\bibnamefont
  {Creutz}},\ }\href@noop {} {\bibfield  {journal} {\bibinfo  {journal}
  {Physical Review Letters}\ }\textbf {\bibinfo {volume} {83}},\ \bibinfo
  {pages} {2636} (\bibinfo {year} {1999})}\BibitemShut {NoStop}%
\bibitem [{\citenamefont {Pollmann}\ \emph {et~al.}(2012)\citenamefont
  {Pollmann}, \citenamefont {Berg}, \citenamefont {Turner},\ and\ \citenamefont
  {Oshikawa}}]{PhysRevB.85.075125}%
  \BibitemOpen
  \bibfield  {author} {\bibinfo {author} {\bibfnamefont {F.}~\bibnamefont
  {Pollmann}}, \bibinfo {author} {\bibfnamefont {E.}~\bibnamefont {Berg}},
  \bibinfo {author} {\bibfnamefont {A.~M.}\ \bibnamefont {Turner}}, \ and\
  \bibinfo {author} {\bibfnamefont {M.}~\bibnamefont {Oshikawa}},\ }\href
  {\doibase 10.1103/PhysRevB.85.075125} {\bibfield  {journal} {\bibinfo
  {journal} {Phys. Rev. B}\ }\textbf {\bibinfo {volume} {85}},\ \bibinfo
  {pages} {075125} (\bibinfo {year} {2012})}\BibitemShut {NoStop}%
\bibitem [{\citenamefont {Leung}\ \emph {et~al.}(1996)\citenamefont {Leung},
  \citenamefont {Chiu},\ and\ \citenamefont {Runge}}]{PhysRevB.54.12938}%
  \BibitemOpen
  \bibfield  {author} {\bibinfo {author} {\bibfnamefont {P.~W.}\ \bibnamefont
  {Leung}}, \bibinfo {author} {\bibfnamefont {K.~C.}\ \bibnamefont {Chiu}}, \
  and\ \bibinfo {author} {\bibfnamefont {K.~J.}\ \bibnamefont {Runge}},\ }\href
  {\doibase 10.1103/PhysRevB.54.12938} {\bibfield  {journal} {\bibinfo
  {journal} {Phys. Rev. B}\ }\textbf {\bibinfo {volume} {54}},\ \bibinfo
  {pages} {12938} (\bibinfo {year} {1996})}\BibitemShut {NoStop}%
\bibitem [{\citenamefont {Sala}\ \emph {et~al.}(2020)\citenamefont {Sala},
  \citenamefont {Rakovszky}, \citenamefont {Verresen}, \citenamefont {Knap},\
  and\ \citenamefont {Pollmann}}]{PhysRevX.10.011047}%
  \BibitemOpen
  \bibfield  {author} {\bibinfo {author} {\bibfnamefont {P.}~\bibnamefont
  {Sala}}, \bibinfo {author} {\bibfnamefont {T.}~\bibnamefont {Rakovszky}},
  \bibinfo {author} {\bibfnamefont {R.}~\bibnamefont {Verresen}}, \bibinfo
  {author} {\bibfnamefont {M.}~\bibnamefont {Knap}}, \ and\ \bibinfo {author}
  {\bibfnamefont {F.}~\bibnamefont {Pollmann}},\ }\href {\doibase
  10.1103/PhysRevX.10.011047} {\bibfield  {journal} {\bibinfo  {journal} {Phys.
  Rev. X}\ }\textbf {\bibinfo {volume} {10}},\ \bibinfo {pages} {011047}
  (\bibinfo {year} {2020})}\BibitemShut {NoStop}%
\bibitem [{\citenamefont {Khemani}\ \emph {et~al.}(2020)\citenamefont
  {Khemani}, \citenamefont {Hermele},\ and\ \citenamefont
  {Nandkishore}}]{PhysRevB.101.174204}%
  \BibitemOpen
  \bibfield  {author} {\bibinfo {author} {\bibfnamefont {V.}~\bibnamefont
  {Khemani}}, \bibinfo {author} {\bibfnamefont {M.}~\bibnamefont {Hermele}}, \
  and\ \bibinfo {author} {\bibfnamefont {R.}~\bibnamefont {Nandkishore}},\
  }\href {\doibase 10.1103/PhysRevB.101.174204} {\bibfield  {journal} {\bibinfo
   {journal} {Phys. Rev. B}\ }\textbf {\bibinfo {volume} {101}},\ \bibinfo
  {pages} {174204} (\bibinfo {year} {2020})}\BibitemShut {NoStop}%
\bibitem [{\citenamefont {Rakovszky}\ \emph {et~al.}(2020)\citenamefont
  {Rakovszky}, \citenamefont {Sala}, \citenamefont {Verresen}, \citenamefont
  {Knap},\ and\ \citenamefont {Pollmann}}]{PhysRevB.101.125126}%
  \BibitemOpen
  \bibfield  {author} {\bibinfo {author} {\bibfnamefont {T.}~\bibnamefont
  {Rakovszky}}, \bibinfo {author} {\bibfnamefont {P.}~\bibnamefont {Sala}},
  \bibinfo {author} {\bibfnamefont {R.}~\bibnamefont {Verresen}}, \bibinfo
  {author} {\bibfnamefont {M.}~\bibnamefont {Knap}}, \ and\ \bibinfo {author}
  {\bibfnamefont {F.}~\bibnamefont {Pollmann}},\ }\href {\doibase
  10.1103/PhysRevB.101.125126} {\bibfield  {journal} {\bibinfo  {journal}
  {Phys. Rev. B}\ }\textbf {\bibinfo {volume} {101}},\ \bibinfo {pages}
  {125126} (\bibinfo {year} {2020})}\BibitemShut {NoStop}%
\bibitem [{\citenamefont {Jordan}\ \emph {et~al.}(2008)\citenamefont {Jordan},
  \citenamefont {Or\'us}, \citenamefont {Vidal}, \citenamefont {Verstraete},\
  and\ \citenamefont {Cirac}}]{PhysRevLett.101.250602}%
  \BibitemOpen
  \bibfield  {author} {\bibinfo {author} {\bibfnamefont {J.}~\bibnamefont
  {Jordan}}, \bibinfo {author} {\bibfnamefont {R.}~\bibnamefont {Or\'us}},
  \bibinfo {author} {\bibfnamefont {G.}~\bibnamefont {Vidal}}, \bibinfo
  {author} {\bibfnamefont {F.}~\bibnamefont {Verstraete}}, \ and\ \bibinfo
  {author} {\bibfnamefont {J.~I.}\ \bibnamefont {Cirac}},\ }\href {\doibase
  10.1103/PhysRevLett.101.250602} {\bibfield  {journal} {\bibinfo  {journal}
  {Phys. Rev. Lett.}\ }\textbf {\bibinfo {volume} {101}},\ \bibinfo {pages}
  {250602} (\bibinfo {year} {2008})}\BibitemShut {NoStop}%
\bibitem [{\citenamefont {Or{\'u}s}(2014)}]{orus2014practical}%
  \BibitemOpen
  \bibfield  {author} {\bibinfo {author} {\bibfnamefont {R.}~\bibnamefont
  {Or{\'u}s}},\ }\href@noop {} {\bibfield  {journal} {\bibinfo  {journal}
  {Annals of Physics}\ }\textbf {\bibinfo {volume} {349}},\ \bibinfo {pages}
  {117} (\bibinfo {year} {2014})}\BibitemShut {NoStop}%
\bibitem [{\citenamefont {Cirac}\ \emph {et~al.}(2020)\citenamefont {Cirac},
  \citenamefont {Perez-Garcia}, \citenamefont {Schuch},\ and\ \citenamefont
  {Verstraete}}]{Cirac:2020obd}%
  \BibitemOpen
  \bibfield  {author} {\bibinfo {author} {\bibfnamefont {I.}~\bibnamefont
  {Cirac}}, \bibinfo {author} {\bibfnamefont {D.}~\bibnamefont {Perez-Garcia}},
  \bibinfo {author} {\bibfnamefont {N.}~\bibnamefont {Schuch}}, \ and\ \bibinfo
  {author} {\bibfnamefont {F.}~\bibnamefont {Verstraete}},\ }\href@noop {} {\
  (\bibinfo {year} {2020})},\ \Eprint {http://arxiv.org/abs/2011.12127}
  {arXiv:2011.12127 [quant-ph]} \BibitemShut {NoStop}%
\bibitem [{\citenamefont {Wei}\ and\ \citenamefont
  {Pengfei}(2020)}]{wei2020floquet}%
  \BibitemOpen
  \bibfield  {author} {\bibinfo {author} {\bibfnamefont {Z.}~\bibnamefont
  {Wei}}\ and\ \bibinfo {author} {\bibfnamefont {Z.}~\bibnamefont {Pengfei}},\
  }\href@noop {} {\bibfield  {journal} {\bibinfo  {journal} {arXiv:2011.01500}\
  } (\bibinfo {year} {2020})}\BibitemShut {NoStop}%
\bibitem [{\citenamefont {Cui}\ \emph {et~al.}(2020)\citenamefont {Cui},
  \citenamefont {Shi},\ and\ \citenamefont {Yang}}]{cui2020circuit}%
  \BibitemOpen
  \bibfield  {author} {\bibinfo {author} {\bibfnamefont {X.}~\bibnamefont
  {Cui}}, \bibinfo {author} {\bibfnamefont {Y.}~\bibnamefont {Shi}}, \ and\
  \bibinfo {author} {\bibfnamefont {J.-C.}\ \bibnamefont {Yang}},\ }\href@noop
  {} {\bibfield  {journal} {\bibinfo  {journal} {Journal of High Energy
  Physics}\ }\textbf {\bibinfo {volume} {2020}},\ \bibinfo {pages} {1}
  (\bibinfo {year} {2020})}\BibitemShut {NoStop}%
\bibitem [{\citenamefont {Serbyn}\ \emph {et~al.}(2020)\citenamefont {Serbyn},
  \citenamefont {Abanin},\ and\ \citenamefont {Papic}}]{serbyn2020quantum}%
  \BibitemOpen
  \bibfield  {author} {\bibinfo {author} {\bibfnamefont {M.}~\bibnamefont
  {Serbyn}}, \bibinfo {author} {\bibfnamefont {D.~A.}\ \bibnamefont {Abanin}},
  \ and\ \bibinfo {author} {\bibfnamefont {Z.}~\bibnamefont {Papic}},\
  }\href@noop {} {\bibfield  {journal} {\bibinfo  {journal} {arXiv:2011.09486}\
  } (\bibinfo {year} {2020})}\BibitemShut {NoStop}%
\bibitem [{\citenamefont {Iadecola}\ and\ \citenamefont
  {Schecter}(2020)}]{PhysRevB.101.024306}%
  \BibitemOpen
  \bibfield  {author} {\bibinfo {author} {\bibfnamefont {T.}~\bibnamefont
  {Iadecola}}\ and\ \bibinfo {author} {\bibfnamefont {M.}~\bibnamefont
  {Schecter}},\ }\href {\doibase 10.1103/PhysRevB.101.024306} {\bibfield
  {journal} {\bibinfo  {journal} {Phys. Rev. B}\ }\textbf {\bibinfo {volume}
  {101}},\ \bibinfo {pages} {024306} (\bibinfo {year} {2020})}\BibitemShut
  {NoStop}%
\bibitem [{\citenamefont {Mark}\ \emph {et~al.}(2020)\citenamefont {Mark},
  \citenamefont {Lin},\ and\ \citenamefont {Motrunich}}]{PhysRevB.101.195131}%
  \BibitemOpen
  \bibfield  {author} {\bibinfo {author} {\bibfnamefont {D.~K.}\ \bibnamefont
  {Mark}}, \bibinfo {author} {\bibfnamefont {C.-J.}\ \bibnamefont {Lin}}, \
  and\ \bibinfo {author} {\bibfnamefont {O.~I.}\ \bibnamefont {Motrunich}},\
  }\href {\doibase 10.1103/PhysRevB.101.195131} {\bibfield  {journal} {\bibinfo
   {journal} {Phys. Rev. B}\ }\textbf {\bibinfo {volume} {101}},\ \bibinfo
  {pages} {195131} (\bibinfo {year} {2020})}\BibitemShut {NoStop}%
\end{thebibliography}%

\end{document}